\documentclass{dmathesis}
\usepackage{fancyhdr}
\usepackage{epsfig}
\usepackage{cite}
\usepackage{graphicx}
\usepackage{amsmath}
\usepackage{theorem}
\usepackage{amssymb}
\usepackage{latexsym}
\usepackage{epic}

\newcounter{ind}

{\theorembodyfont{\rmfamily}}
{\theorembodyfont{\rmfamily}}
{\theorembodyfont{\rmfamily}}
{\theorembodyfont{\rmfamily}}
{\theorembodyfont{\rmfamily}}
{\theorembodyfont{\rmfamily}}

\newcommand{\Tr}{{\rm Tr}}

\newcommand{\gl}{\lambda}
\newcommand{\refe}[1]{Eqn.~(\ref{#1})} 

\newcommand{\be}{\begin{equation}}
\newcommand{\ee}{\end{equation}}
\newcommand{\bea}{\begin{eqnarray}}
\newcommand{\nn}{\nonumber}
\newcommand{\eea}{\end{eqnarray}}
\newcommand{\ti}{\widetilde}
\usepackage{slashed}
\usepackage{braket}

\begin{document}

\pagenumbering{roman}

\setcounter{page}{1}

\newpage

\thispagestyle{empty}
\begin{center}
  \vspace*{1cm}
  {\Huge \bf Gauge Mediated Supersymmetry Breaking in Five Dimensions}

  \vspace*{2cm}
  {\LARGE\bf Moritz McGarrie}

  \vfill

  {\Large A Thesis presented for the degree of\\
         [1mm] Doctor of Philosophy}
  \vspace*{0.9cm}
  
   \begin{center}
   \includegraphics{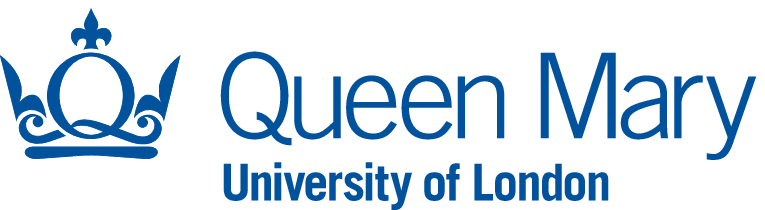}
   \end{center}

  {\large Centre for Research in String Theory\\
  		  [-3mm]Department of Physics\\
          [-3mm] Queen Mary\\
          [-3mm] 327 Mile End Road\\
          [-3mm] England\\
          [1mm]  August 2011}

\end{center}


\newpage
\thispagestyle{empty}
\addcontentsline{toc}{chapter}{\numberline{}Abstract}
\begin{center}
  \textbf{\Large Gauge Mediated Supersymmetry Breaking in Five Dimensions}

  \vspace*{1cm}
  \textbf{\large Moritz McGarrie}

  \vspace*{0.5cm}
  {\large Submitted for the degree of Doctor of Philosophy\\ August 2011}

  \vspace*{1cm}
  \textbf{\large Abstract}
\end{center}
In this thesis we focus on the construction of models in which a supersymmetry breaking hidden sector is located on one fixed point of an extra dimensional interval and the effects are gauge mediated across this interval to the other fixed point where the supersymmetric standard model is located. We use the formalism of current correlators to encode supersymmetry breaking effects and explore flat, warped and deconstructed extra dimensions.  We also apply these techniques to models of metastable supersymmetry breaking in $\mathcal{N}=1$ Supersymmetric Quantum Chromodynamics.
\chapter*{Declaration}
\addcontentsline{toc}{chapter}{\numberline{}Declaration}
The work in this thesis is based on research carried out at the
Centre for Research in String Theory, Queen Mary University of London. Except where specifically acknowledged in the text the work in this thesis is the original
work of the author. Chapter 2 is based on the publication \cite{McGarrie:2010kh} with Rodolfo Russo and similarly chapter 6 is based on the publication \cite{McGarrie:2010yk} with Daniel C. Thompson.  Additional results in this thesis are based on publications by the author \cite{McGarrie:2010qr,McGarrie:2011dc}.  I also acknowledge a complementary publication with Steven Thomas and Daniel Koschade \cite{Koschade:2009qu}, the results of which are not included in this thesis.

\chapter*{Acknowledgements}
\addcontentsline{toc}{chapter}{\numberline{}Acknowledgements}
I would like to thank Queen Mary, Centre for Research in String Theory and especially my supervisors Steven Thomas and Rodolfo Russo. I would like to thank John, Jane, Carolyn, Madeleine, Kirsten and Ian. I would like to thank Camilla and Lurpak for supporting my passion and finally to my friends and colleagues.
\tableofcontents
\clearpage


\pagenumbering{arabic}
\setcounter{page}{1}
\setcounter{equation}{0}
\chapter{Introduction}\label{chapter1}
The Standard Model (SM)  of particle physics is a widely accepted and accurately tested theory that describes the bosonic interactions of the strong, weak and electromagnetic forces between the fermionic quarks and leptons that make up the familiar and everyday world around us. It is a Lorentz invariant Quantum Field Theory (QFT) of gauge interactions. These standard model forces, are described by the exchange of excitations of bosonic gauge fields.  The quarks and leptons are excitations of fermionic fields. Additionally, the standard model includes a Higgs scalar field.   Unfortunately, as a theory it is sick and incomplete.  Firstly there is nothing within the theory that determines the electroweak scale from which most of the theory relies.  Secondly this scale, set by the Higgs mass, is extremely unstable under quantum corrections unless there is a fine tuning to the 38th decimal place \cite{Susskind:1978ms}: the Hierarchy problem.  Other criticisms may be mounted against it:  the standard model makes no attempt to unify with the force of gravity and seems rather poor at explaining  physics on large scales, such as the composition of Dark matter and Dark energy relative to the visible matter of the universe.

The Large Hadron Collider (LHC) at CERN is expected to test various extensions of the standard model of high energy physics.  One very well motivated extension is supersymmetry (SUSY).   Mathematically, the boson and fermion fields of the standard model are representations of the Lorentz symmetry group.  The generators of this group are represented only by bosonic generators $P^{\mu}$ and $M^{\mu \nu}$.  As there are bosonic and fermionic fields within the standard model, it is logical to suggest that the Lorentz symmetry group may be extended to include generators that are also fermionic: a supersymmetric quantum field theory.   From a mathematical viewpoint, this is a rather innocuous suggestion.  When taken as a serious description of nature it has some profound and far reaching  consequences.  Firstly, it predicts a supersymmetric standard model of particle physics, in which all known fields have superpartners, with differing spins.  The constraints of the observed standard model on these unobserved fields leads to the conclusion that this symmetry, if it exists, is broken.  Perhaps more profoundly, these constraints imply that it is broken in a local hidden sector that is not included within the standard model and that this breaking is then mediated to the supersymmetric standard model.  Further, for this symmetry to be broken dynamically, and importantly to supply a natural hierarchy of scales, implies a hidden sector that is described by a strongly coupled gauge theory. 

A softly broken supersymmetric standard model is a natural solution to many of the puzzles of the standard model such as the Hierarchy problem \cite{Martin:1997ns}.  Additionally, supersymmetry has useful dark matter candidates which the most accurate astrophysical observations and the Cosmological model \cite{Komatsu:2008hk}, Lambda CDM (the cosmological constant plus cold dark matter), predicts makes up $25\%$ of the universe, of which visible matter makes up only $4\%$.   Additional hints of  supersymmetry arise from its enhancement of gauge coupling unification of the standard model as required by Grand Unified Theories (GUT)  \cite{Georgi:1974sy}.  Finally, supersymmetry is a necessary component of superstring theory, which does supply a consistent framework from which the standard model and gravity may both arise.

This introductory chapter will motive the study of supersymmetry, its breaking and its mediation across and extra dimension via gauge interactions.  Of course there are many alternative proposals for extensions of the standard model with or without supersymmetry and with or without extra dimensions.  We will simply outline some of the guiding motivations for choosing this particular research direction.

\section{Why Supersymmetry?}
The (non supersymmetric) standard model, and in particular electroweak symmetry breaking, is unnatural \cite{PhysRevD.13.974,Weinberg:1979bn,Susskind:1978ms,Drees:1996ca,Terning:2006bq} as a field theory as the Higgs mass is extremely unstable under quantum corrections unless there are finely tuned counterterms in its Ultra Violet (UV) completion.  A straightforward calculation of the one loop correction to the physical Higgs field $h$, two point function,  from couplings with fermions, for instance top quarks \footnote{We are using four component notation here.}
\be
\mathcal{L}=-\lambda_{f}  \bar{f}_L f_R \phi - \lambda_{f}  \bar{f}_R f_L \phi^{*} .\label{equation1intro}
\ee  
The complex scalar may be written as\footnote{A different parameterisation is $\phi(x)=\frac{1}{\sqrt{2}}(v+\rho(x))e^{-i\zeta(x)/v}$}
\be
\phi(x)=\frac{1}{\sqrt{2}}[v+h(x)+i\chi(x)]
\ee
where $\text{Re} [\phi]= \frac{1}{\sqrt{2}}(h+v)$ in unitary gauge where $\chi=0$, gives a quadratically divergent integral, as well as logarithmically divergent one (to show this use $d^4p = 2\pi^2 p^2 dp^2$ and $x=p^2+m^2_f$:
\be
\Pi(0)^{f}_{hh}=(-1)\int \frac{d^4p}{(2\pi)^4}\text{Tr}(\frac{-i\lambda_f}{\sqrt{2}})\frac{i}{\slashed{p}-m_{f}} \text{Tr}(\frac{-i\lambda_f}{\sqrt{2}}) \frac{i}{\slashed{p}-m_{f}}\nonumber
\ee
\be
 =-2\lambda^2_{f}\int \frac{d^4p}{(2\pi)^4} [\frac{1}{p^2-m^2_f}+\frac{2m_f^2}{(p^2-m^2_f)^2} ]
\ee
The divergence may be absorbed into counter terms leaving a finite correction to the propagator of order $\frac{\lambda_{f}}{8\pi} m_{f}^{2}$, which one may argue is not so fine tuned, albeit rather arbitrary.  However, any UV GUT completion will inevitably introduce similar contributions whose mass scale is now $M_{\text{GUT}}\sim 10^{16}-10^{18} \text{GeV}$, such that even the usual finite terms result in a cancellation between bare mass and renormalised mass to within a few $100$ GeV.

The introduction of superpartners for fermions\footnote{see appendix \ref{conventions} for superspace notaton}, the complex scalars $\tilde{f}_L$, $\tilde{f}_{R}$, will generate couplings of the form
\be
\tilde{\lambda}_{f}|\phi|^2(|\tilde{f}_L|^2 +|\tilde{f}_R|^2)  \label{equation2intro}
\ee
These provide additional contributions to the two point function:
\be
\Pi(0)^{\tilde{f}}_{hh}=-\tilde{\lambda_f}\int \frac{d^4p}{(2\pi)^4}(\frac{1}{p^2-m^2_{f_{L}}} +\frac{1}{p^2-m^2_{f_{R}}})\nonumber
\ee
\be
+(\tilde{\lambda}_{f}v)^2 \int \frac{d^4p}{(2\pi)^4}[(\frac{1}{(p^2-m^2_{f_{L}})^2}+\frac{1}{(p^2-m^2_{f_{R}})^2}]\nonumber
\ee
\be
 =-2\lambda^2\int \frac{d^4p}{(2\pi)^4} [\frac{1}{p^2-m^2_f}+\frac{2m_f^2}{(p^2-m^2_f)^2} ]
\ee
Remarkably if the Yukawa coupling $\lambda^2_f=-\tilde{\lambda}_f$, the combined quadratic divergences from scalars and fermions cancel exactly, regardless of any of the mass scales introduced.  The presence of supersymmetry in the Lagrangian naturally acts to remove quadratic divergences and are therefore rather soft.   Remarkable also is that the remaining divergences depend on the mass splitting between fermion and sfermion masses $\delta^2=m^2_{\tilde{f}}-m^2_f$ such that if even the vacuum of the theory is supersymmetric, all divergences cancel exactly (with a little help from \ref{valofb}):
\be 
\Pi(0)^{f}_{hh}+\Pi(0)^{\tilde{f}}_{hh}\simeq -i\frac{\lambda^2_{f}}{16\pi^2}[4\delta^2 (1+\frac{1}{2}\log\frac{m^2_f}{\mu^2})]+O(\delta^4)
\ee
 In the superspace formalism (See appendix \ref{conventions})  which we adopt throughout this thesis both \refe{equation1intro} and \refe{equation2intro} compactly arise from the superpotential 
\be 
\int d^2\theta \lambda_f  H \Phi_L \Phi_R + \int d^2\bar{\theta} \lambda_f  H^{*} \Phi_L^{*} \Phi_R^{*}
\ee
when computing the Yukawas and F-term potential, respectively.  Extending the standard model with $\mathcal{N}=1$ supersymmetry with two Higgs superfields for anomaly cancellation and to keep the superpotential holomorphic, plus terms that break supersymmetry softly as demonstrated above, is known as the Minimal Supersymmetric Standard Model (MSSM).  The soft terms we may add to the MSSM are 
\be 
\mathcal{L}_{Soft}=-\phi^{*}_i(m^2_{\phi})_{ij}\phi_j + (\frac{1}{3!}A_{ijk}\phi_i\phi_j\phi_k -\frac{1}{2}B_{ij}\phi_i\phi_j+C_{i}\phi_i+h.c.)-\frac{1}{2}(M_\lambda \lambda \lambda+M_\lambda \lambda^{\dagger} \lambda^{\dagger})
\ee
This thesis will focus on the theoretical computation of the soft terms for scalars $m^2_{\phi}$ and the gauginos $M_\lambda$.

In summary,  (1) The introduction of superpartners for scalars, in particular the Higgs of the standard model,  protects  these scalar fields from quadratic divergences in its self energy. (2) As long as the superpartners of a scalar have a similar mass to the scalar, the scalar mass is shielded from loop corrections from other heavier particles that may arise in any Grand Unified Theory (GUT) or higher dimensional extension of the standard model which may act as a UV completion.  In summary, supersymmetry is a natural way to improve the standard model and protect electroweak physics from UV effects.  It does not yet explain the electroweak scale.

\section{Dynamical Supersymmetry Breaking}
Dynamical supersymmetry breaking is a natural way to explain the electroweak scale:
\begin{quote}
``As was first suggested by Witten \cite{Witten:1981nf}, we would like the mechanism which
spontaneously breaks supersymmetry to be dynamical. This means that it arises from an
exponentially small effect, and therefore it naturally leads to a scale of supersymmetry
breaking, $M_s$, which is much smaller than the high energy scales in the problem $M_{cutoff}$
(which can be the Planck scale or the grand unified scale):
\be 
M_s =  e^{- c/g(M_{cutoff} )^2}M_{cutoff} \ll M_{cutoff} . 
\ee
This can naturally lead to hierarchies. For example, the weak scale $m_{W}$ can be dynamically
generated, explaining why $m_{W}/m_{Pl}\sim   10^{−17}$.''(From \cite{Intriligator:2007cp})
\end{quote}

As the dynamics is strongly coupled, perturbation theory in the fundamental ``quarks'' and ``gluons'' is of little use here.  However much more may be learned about the low energy dynamics by writing an \emph{effective Lagrangian} which contains the low energy degrees of freedom and preserves the symmetries and interactions of the system. For instance the linear sigma model is used to describe mesons which encapsulate the collective behaviour of quarks and gluons at low energies. Corrections to the effective description are due to an expansion of operators in $k^2/\Lambda^2$, where $k^2$ is the the momomenta and $\Lambda^2$ is the scale at which the fundamental variables become strongly coupled.  The strongly coupled microscopic theory and the effective macroscopic theory are said to be within the same ``Universality Class'' as they both flow to the same infrared fixed point, within the space of dimensionless couplings.  Precisely at the fixed point the descriptions are no longer effective, they are exact.  This provides the notion that two (or more) descriptions may be \emph{dual}.  One description may offer a more useful perturbative description to examine its low energy properties. Equally one may start with dual descriptions and, by adding various operators or deformations, build useful effective descriptions. Further, one may on occasion make these flows arbitrarily close, such that the duality holds along an entire flow \cite{Strassler:2003qg,Strassler:2005qs}.

Much progress has been made in understanding dynamical supersymmetry breaking via the use of dualities \cite{Peskin:1997qi,Shifman:1999mv,Strassler:2003qg,Kitano:2010fa,Dine:2010cv} and in particular Seiberg duality \cite{Seiberg:1994pq,Affleck:1983mk}, in which exactly dual descriptions can be found at conformal fixed points, which then become effective descriptions after adding deformations such as small quark masses in the ``electric'' or fundamental microscopic description.  This progress has culminated in the viewpoint that supersymmetry is broken dynamically in a metastable minima \cite{intriligator2006dsb}, the ISS model, which is usually explored in its ``magnetic" or IR-free description.   This development in supersymmetry breaking is rather compelling and many questions seems to have a natural footing  within this construction.  For instance stability of the metastable broken vacuum is ensured by a parametrically small parameter 
\be \epsilon=\sqrt{\frac{m}{\Lambda}}\ll 1\ee
where $m$ is the quark mass scale and $\Lambda$ is the strong coupling scale. Stability also arises dynamically: the universe naturally cools towards a vacuum with more light degrees of freedom, and at lower temperatures the vacuum becomes trapped and parametrically long lived \cite{Abel:2006cr,Abel:2006my}.  GUT unification and its breaking may also be dynamically explained by the behaviour of the breaking of flavour symmetries of the hidden sector \cite{Koschade:2009qu}.  Additionally the potential between the two vacua of the ISS model \cite{intriligator2006dsb} is generically long and flat, providing a natural candidate for inflation.  Throughout this thesis we have in mind that supersymmetry is broken in this, or some related, way.  What we hope to contribute, in this thesis, is a rigorous theoretical framework for analysing different methods of mediating this type of supersymmetry breaking. We will argue that we are compelled to suggest gauge mediation and in particular some sort of effective extra dimension, in which massive vector like objects participate in this mediation.  We will discuss these issues further in chapters \ref{chapter3} and \ref{chapter6}.

A perhaps alternate viewpoint due to Randall and Sundrum \cite{Randall:1999ee}, as a solution to the Hierarchy problem, is when the geometry of spacetime generates a hierarchy of scales.  Starting from the $AdS_5$ metric given by 
\be
d^2s =e^{-2k y }\eta^{\mu \nu}dx_{\mu}dx_{\nu}+dy^2.
\ee
$1/k$ is the AdS curvature scale, we may look at intervals of this space with $0\le y\le \ell$  with $\ell=\pi R$.  Mass terms geometrically located at $y=\ell$ will be exponentially suppressed such that 
\be
M=e^{-k\ell} M_{0}
\ee
In general, one still needs supersymmetry to protect this hierarchy.  In fact it is rather enticing to suggest that if the local hidden sector that breaks supersymmetry dynamically is both approximately conformal and has a large gauge group, then one may construct a low energy effective Lagrangian via the AdS/CFT correspondence. One may then argue that the exponential warp factor is similarly capturing the dynamically generated hierachy of scales between supersymmetry breaking scale and that of the UV scale. The supersymmetric ``slice of AdS'' with supersymmetry breaking on the IR fixed point, mediated to the UV fixed point by gauge interactions may act as a powerful \emph{effective} description of the period of approximately conformal running of the hidden sector gauge group.   Importantly, fields that live in the full five dimensions will have a Kaluza-Klein (kk) mode expansion, with masses associated with the AdS warp factor, roughly $M_n\sim n\pi k e^{-k\ell}$ due to the interval of length $\ell$.

To be clear, we are not constructing an AdS description of the MSSM, but rather taking a conceptually different construction. For concreteness, we have in mind an $\mathcal{N}=1$, $SU(N_F)\times SU(N_{c})_{CFT}$ in the UV description.  The $SU(N_{c})_{CFT}$ is approximately conformal between the UV scale $M_{planck}$, down to some scale which will equate to $M_{kk}$ the mass of the lightest kk mode.  In the IR the $SU(N_{c})_{CFT}$ breaks supersymmetry which is, essentially captured on the IR brane of the AdS dual description.   This gauge group and matter content may admit an AdS dual low energy effective description, just as under Seiberg duality, a UV-free electric description admits an IR dual magnetic description.  The $SU(N_F)$ is a global flavour symmetry for which some subgroup is weakly gauged and associated with the MSSM gauge group or parent GUT $SU(5)$.  The global currents of this flavour symmetry would in principle require all order corrections from the hidden sector in terms of the electric description, when in the IR (the currents in the electric desciption are well defined in the UV), but since we have a low energy effective description, we can in principle determine those corrections in terms of the dual variables.  We also point out that $SU(3)_c$ QCD has well defined UV currents of the $SU(2)$ and $U(1)$ global groups.  In the IR, various tools are applied such as operator product expansions, sum rules, Pad\'e approximations and AdS/QCD to determine the structure of these currents.

In particular, we should expect that on the IR brane, should be some O'Raifeartaigh type model \cite{O'Raifeartaigh:1975pr}, perhaps rather similar to the magnetic description of the ISS construction \cite{intriligator2006dsb}.  This is simply parameterised by currents located on the IR brane. Under the correspondence, a global symmetry of the CFT is interpreted as a gauge symmetry of the bulk and resonances of the CFT act as Kaluza-Klein modes of the gauge symmetry.  On the CFT side, these fields should be thought of as part of the currents themselves, form factors, which on the AdS side appear as mediators of the IR brane localise currents. Most compelling in this construction is that as the gauge fields live in the bulk, Kaluza-Klein modes of the gauge fields will participate in the mediation of supersymmetry breaking.  A closely related phenomena arises both in QCD as ``Vector Meson Dominance" \cite{Bando:1984ej,Georgi:1989xy} and also rather naturally, although most often a subleading effect, within ISS models themselves, whenever one embeds the standard model gauge groups within a flavour group that exhibits ``magnetic colour-flavour locking'' \cite{Komargodski:2010mc}. It may help conceptually to think of chapter \ref{chapter3} as capturing a truncated version of the AdS model we are suggesting, which includes only the first Kaluza-Klein modes and the IR brane supersymmetry breaking sector.  We will return to these topics in chapters \ref{chapter3} and \ref{chapter6}.

\section{Why a Hidden Sector?}
So far we have argued that supersymmetry protects the hierarchy of scales between the Planck scale and the electoweak scale and that dynamically broken supersymmetry arising from strong dynamics naturally generates this hierarchy in the form of the supersymmetry breaking.  This equation is modified by the introduction of supergravity.  In fact the constraints of supersymmetry require that the standard model does not directly participate in the breaking of supersymmetry and the argument is as follows:

In superspace notation, the supersymmetric standard model (SSM) action $S$, will have a superpotential  $W$:
\be
S= \int d^8z [\mathcal{W}\delta^2 (\bar{\theta})+ W^{\dagger}\delta^2 (\theta)] ,\label{super1}
\ee
where $\int d^8z= \int d^4 x d^2\theta d^2 \bar{\theta}$. Where we have explicitly assumed a perturbative description with canonical K\"ahler potential. Through derivatives of the superpotential, one encodes fermion mass terms, scalar mass terms and Yukawa couplings.  As a result these quantities are constrained and related. The upshot are some supertrace mass sum rules \cite{Ferrara:1979wa}:
\be
\text{STr} \mathcal{M}^2= \sum_{J=0}^{1/2}(-1)^{2J}(2J+1)m^2_{J}=0 
\ee
with $J$ labelling the spin of the states. In fact this can be shown to be true over each chiral supermultiplet and including arbitrary gauge interactions \cite{Dimopoulos:1981zb}
\be
 \text{STr} \mathcal{M}^2=-2 T^a \braket{D^a}
\ee
These mass sum rules imply that if supersymmetry is broken within the SSM then for every standard model particle, there must be a superpartner with a mass lighter or equal to the standard model particle \emph{clearly ruled out by experiment} \cite{Nakamura:2010zzi}.  These sum rules are protected as a consequence of the Nonrenormalisation theorem which states that the superpotential \ref{super1}, which is holomorphic, does not receive any renormalisation at any order in perturbation theory.  This theorem may be motivated intuitively by realising that all local\footnote{in coordinate space} loops written in superspace must have a minimum of two internal propagators or equivalently $\int d^8 z D D\bar{D}\bar{D}$ must appear, which is not holomorphic. 

This leads to the requirement of a hidden sector where supersymmetry is broken.  The hidden sector can supply the necessary explicit breaking soft terms, such as the gaugino and sfermion masses, necessary for the SSM superpartner masses and to evade the constraints of the sum rules.  From the perspective of the visible sector, the soft terms explicitly break the sum rules however these terms may be generated by a supersymmetry breaking sector that necessarily obeys these sum rules.

\section{Interlude: Form factors}
The scattering part of the S-matrix between an ingoing state $A$ and an out going state $B$ is given by 
\be
\braket{B|\int dt H_{ext}|A} =  \int d^4x A^{\mu}(x)\braket{B|j_{\mu}(x)|A}.
\ee
If the external states are not fundamental, i.e. composite, we can therefore think of the matrix element as the definition of the current.  Translation invariance allows us to define
\be 
\braket{B|j_{\mu}(x)|A}=\braket{B|j_{\mu}(0)|A}e^{i(P_{A}-P_{B}).X}.
\ee
The simplest example \cite{Peskin:1995ev} is of the (4-component Dirac) electron.  Inclusion of radiative corrections will give a result of the form 
\be 
\braket{\bar{e}(p)|j_{\mu}(0)|e(p')} = g_{e}\left( F_{1}(q^2)\gamma^{\mu} +\frac{ i\Sigma^{\mu\nu} q_{\nu}  }{m}  F_{2}(q^2) \right)
\ee
One may choose to contract with spinor wavefunctions $\psi(x)=u(p)^{-ip.x}$ to obtain a spin index contracted form factor.  $F(q^2)$ is a momentum dependent form factor ($q=p'-p$), which to lowest order $F_1=1$  and $F_2=0$. In the Hadronic world, one may think of the electromagnetic and weak forces as ``external fields''.  The external probe fields therefore couple to Hadronic currents \cite{sakurai1969currents}. A simple example is that of the $\pi^{+}$ meson 
\be 
\braket{\pi(p)|j_{\mu}(0)|\pi (p')}= g_{e} F_{\pi}(q^2) (p+p')_{\mu}
\ee
What one finds experimentally is not $F_{\pi}(q^2)=1$, and additionally it is presumed that $F(\infty)=0$.  In fact it is argued that if an intermediate $\rho$ meson is exchanged then from general considerations the form factor is 
\be
F (p^2)= \frac{m^2_{\rho}}{p^2-m^2_{\rho}}.
\ee
This is vector meson dominance. What is perhaps less obvious is that one may think of the $\rho$ meson as a Kaluza-Klein mode of the photon, such that  the form factor is given by a bulk propagator. Schematically  with $m_0=0$ and $m_1=m_{\rho}$,
\be
F(p^2)=- p^2 \sum_{n=0}^{n=1}\frac{(-1)^n}{p^2-m^2_{n}} .
\ee
That hadronic form factors appear from weakly gauged global symmetries of the UV theory, and that their form may be modelled by a gauge symmetry in some higher dimensional space, has lead to Hidden Local Symmetry models and AdS/QCD as a very compelling description of low energy QCD  \cite{Bando:1984ej,O'Connell:1995wf,Erlich:2005qh,Son:2003et,RodriguezGomez:2008zp,Harada:2010cn}.  We are strongly compelled to argue that this generic feature of strongly coupled gauge theories will manifest itself in interactions with a supersymmetry breaking hidden sector, including in the generation of soft terms for the MSSM.
\section{Gauge Mediation}
So far we have a picture of supersymmetry being broken in a sector that is not part of the Minimal Supersymmetric Standard Model (MSSM) and additionally we have a sensible dynamical method of breaking supersymmetry.  The transmission of the supersymmetry breaking effects to the MSSM may arise from either gauge or gravitational interactions.  In this thesis we shall explore gauge mediation and not comment further on gravity mediation.

\subsection{General Gauge Mediation}
\begin{figure}[ht]
\centering
\includegraphics[scale=1]{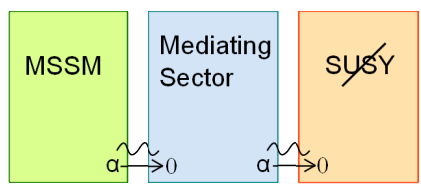}
\caption{A pictorial of gauge mediation.  The visible MSSM and susy breaking hidden sector completely decouple as the standard model gauge couplings ($\alpha$) go to zero.}
\label{ggm}
\end{figure}
General gauge mediation \cite{Meade:2008wd} can be defined as follows;  In the limit that the standard model gauge couplings $\alpha_{i}\rightarrow 0$, $i=1,2,3$ labelling $U(1), SU(2), SU(3)$ respectively, the theory decouples into \emph{three} sectors: a sector that breaks SUSY, the MSSM matter and the gauge mediating sector, as in figure \ref{ggm} . Strictly speaking, it is only necessary that the $\alpha$ coupling the standard model/mediating sector decouple. Computations may be made perturbatively in the MSSM gauge couplings $\alpha_i(\mu)$ between $M_{SUSY}\geq \mu \geq M_{z}$. Typically, this construction may be achieved by first building a hidden sector such as the ISS construction and then weakly gauging a flavour symmetry of the hidden sector and associating this with the MSSM gauge groups.  The currents and correlators of this weakly gauged global symmetry encode the supersymmetry breaking effects.  Within this framework, one may look at variations on this theme by changing either of the three sectors \cite{Buican:2008ws,deGouvea:1997tn,Intriligator:2008fr,Ooguri:2008ez,Distler:2008bt,Benini:2009mz,Buican:2009vv,Argurio:2009ge,Luo:2009kf,Kobayashi:2009rn,Lee:2010kb,Intriligator:2010be}.  For the most part four dimensional models have focused on comparing different SUSY breaking sectors, i.e. different currents, whilst holding the other two sectors fixed. In this thesis we will change the mediating sector, \emph{whilst staying within} the definition of general gauge mediation.  In particular we would like to explore models where the mediating sector has Kaluza-Klein modes and for that we will explore both five dimensional models and deconstructed four dimensional ones.

There is an important additional point on the framework of GGM is that it allows strongly coupled hidden sectors to be encoded into current correlators so that they may be extensively studied.  If the hidden sector is strongly coupled then in principle the current correlators must be known exactly with regard to $\alpha_{hidden}$.  In this case the current correlators simply parameterise our ignorance in terms of the current correlators and we compute soft masses perturbatively in the $\alpha_i$'s.

 \emph{A far more powerful} use of the current correlator framework arises from the use of dualities.  As the hidden sector and visible sector completely decouple, one may use a dual description of the currents of the weakly gauged global symmetry, in terms of the effective variables.  Within the ISS construction this naturally arises as one takes the two point functions of the magnetic degrees of freedom of the hidden sector even though the true fundamental current correlators are in terms of strongly coupled variables.  In this thesis we will show that \emph{even truly four dimensional models}, with regard to the microscopic description, may have an apparent extra dimensional structure, within the macroscopic description. This arises both in the sense of (De)constructed extra dimensions \cite{McGarrie:2010qr} and through the AdS/CFT correspondence  \cite{McGarrie:2010yk}.   Heuristically, in principle all-order current correlators on the left hand side may be factorisable:
\be
 \braket{J(p) J(-p)}\rightarrow [F(\frac{p^2}{m_{v}^2})]^2 \braket{\tilde{J}(p)\tilde{J}(-p)}+...
\ee
such that it may be described by a current correlator of the effective degrees of freedom, multiplied by a momentum dependent form factor $F(p^2/m_{v}^2)$.  The ellipses denote that the right hand side is a perturbative expansion in $g_{weak}$ of the all-order current correlators on the right hand side.     From the perspective of the weak side of the duality the form factor arises due to Kaluza-Klein modes of the vector superfield, with mass $m_{v}$, participating in the mediation of supersymmetry. From the perspective of field theory these are vector-like resonances of the strongly coupled description.   The form factor is associated with a bulk propagator
\be
F(p^2/m_{v}^2)\sim \braket{\phi(x,0)\phi(x,\ell)}.
\ee
In this sense the GGM construction truly captures the essence that a four dimensional calculation on one side is equivalent to a 5d calculation on the other side of a duality transformation.  For the particular case of a two site lattice model, we will show that this form factor is completely dominated by the first kk mode mass, and we recover exact vector meson dominance.  In fact it is quite reasonable that this type of factorisable structure should occur under the duality, as both the four dimensional general gauge mediation \cite{Meade:2008wd} equations for the scalar masses should be valid and so should the five dimensional scalar mass equations of the dual five dimensional models. 

This motivation for exploring extra dimensions, which may be understood by comparing with QCD physics. The strong force binding quarks as fundamental objects at high energies, is a gauge theory which at low energies has an effective description of baryons and mesons.  In particular the vector like spin-$1$ mesons for example the octet containing $\rho$ mesons, appear to arise from a purely \emph{emergent} local symmetry due to the effects of strong coupling \cite{Georgi:1989xy,Bando:1984ej}.  This symmetry is not apparent in the high energy quark description.  In fact these resonances behave like the lightest Kaluza-Klein particles of an extra dimension.  Not only are vector mesons detectable in experiment, they also influence many physical processes of the standard model through an effect known as ``vector meson dominance" \cite{Sakurai:1960ju}, whereby the form factor of $\pi^{+}$ interacting with electromagnetism, is dominated by the $\rho$ meson.   If the hidden sector is strongly coupled, it seems likely that the analogue of vector mesons should arise in the low energy effective description and play an important role in phenomenology.

Gauge mediation in five dimensions captures the analogous effect.  In the process of this thesis we will demonstrate how purely four dimensional models of SQCD may develop effective Kaluz-Klein modes analogous to $\rho$ mesons and that they act to screen scalar masses analogous to vector meson dominance.  We will then show that by interpreting the hidden sector as having an effective AdS dual description, one may further capture similar features.  All these developments will become more apparent in the later chapters: chapter \ref{chapter2} will simply focus on a rigorous five dimensional extension of general gauge mediation. 

\subsection{Gaugino Mediation and GGM5D}
\begin{figure}[ht]
\centering
\includegraphics[scale=0.4]{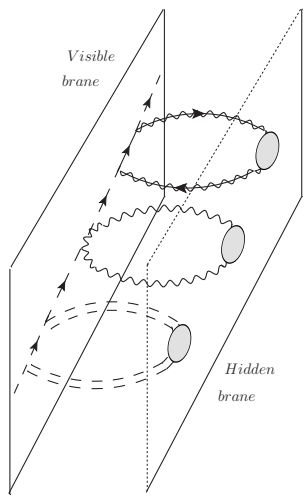}
\caption{A pictorial of the leading order ($\alpha^2$) sfermion mass contributions due to general gauge mediation across an interval.  Sfermion mass contributions on the visible brane are generated by supersymmetry effects encoded in current correlators (blobs) located on the hidden brane.}
\label{flatfigure}
\end{figure}

\begin{figure}[htcb]
\centering
\includegraphics[scale=0.6]{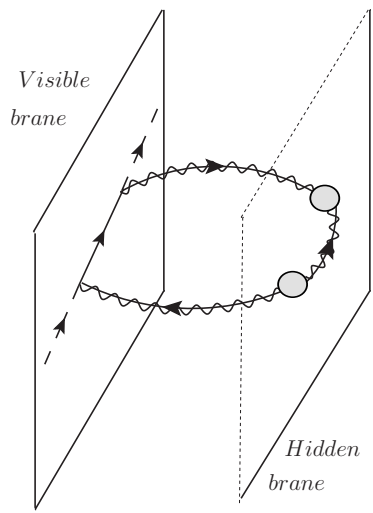}
\caption{A diagram representing a subleading (order $\alpha^3$) contribution to sfermion masses ($p_{ext}=0$) through a gaugino mediated double mass insertion of the Majorana soft mass on the hidden brane. For non-zero external scalar momenta, this diagram is a correction to the scalar kinetic term, leading to the soft mass contribution to one loop renormalisation group equations (RGE's) for sfermions.}
\label{gauginomed}
\end{figure}

Typically in extra dimensional models of gauge mediation, the Kaluza-Klein modes of the vector multiplet participate in the mediation of supersymmetry.  This acts to screen scalar masses of the MSSM, as was first shown rather rigorously by Mirabelli \& Peskin \cite{Mirabelli:1997aj} and is pictured in figure \ref{flatfigure}.
\begin{flushleft}
\end{flushleft}
As a result, for a minimal hidden sector, typically the gaugino mass is the only non-zero soft term and additionally, gaugino mass contributions in the renormalisation group equations (RGE's) \cite{Drees:2004jm,Martin:1993zk,Yamada:1994id} pictured in figure \ref{gauginomed}, will generate scalar masses for the MSSM. This was explored phenomenologically in models of Gaugino mediation \cite{Kaplan:1999ac,Chacko:1999mi,Schmaltz:2000gy,Schmaltz:2000ei}. As the whole theory depends on only one soft term, it is rather predictive. 
\begin{flushleft}
\end{flushleft}
We would like to empasise that this is simply a particular, and minimal case, of a far broader space of possible five dimensional constructions.  General gauge mediation in five dimensions may be shown to interpolate between the screened and unscreened (or four dimensional) limit of scalars masses, as is shown in chapter \ref{hybridchapter}: Hybrid mediation.  Further, as in general gauge mediation, both four and five dimensions, the gaugino and sfermion current correlators are not related, so it is rather simple to construct a model in which for example, the gaugino Majorana mass is vanishingly small through an approximate R-symmetry, the scalar masses are only partially screened in the Hybrid regime, and the erstwhile massless gaugino gets a Dirac mass instead \cite{Benakli:2008pg}.  This is just one example of the rich set of phenomenologically unexplored possibilities that we hope general gauge mediation in five dimensions may be applied to, which seem no more or less likely than the benchmark construction of Gaugino mediation. Additionally, we emphasise that general gauge mediation in five dimensions gives a natural geometric interpretation of the decoupling limit of GGM, as the hidden and visible sectors are spatially separated on different fixed points it makes direct couplings impossible by construction.   Further, the screening effect on the scalar masses may alleviate many problems found in models of ISS \cite{intriligator2006dsb} of anomalously light gauginos relative to scalars.

\section{Chapter Summary}
This chapter has summarised the motivations for a supersymmetric standard model in which supersymmetry breaking is described by a hidden sector and gauge mediated to the MSSM.  We have also given an outline for this thesis in which we will focus on gauge mediated supersymmetry breaking across an extra dimension within the program of current correlators which allows for a model independent construction which is also applicable for analysing strongly coupled hidden sectors.  The remaining of this thesis will work as follows:  in chapter \ref{chapter2} we will outline the framework of using current correlators to encode brane to brane mediation of supersymmetry on a $R^{1,3}\times S^1/\mathbb{Z}_{2}$ background.  In chapter \ref{chapter3} we will demonstrate the utility of this framework by applying it to models of metastable supersymmetry breaking.  In the same chapter, we will show how ISS-like models may generate an effect extra dimension through ``magnetic colour-flavour locking'' which will motive chapter \ref{chapter4} in which we focus on deconstructing an extra dimension from purely four dimensional lattice constructions.  In chapter \ref{hybridchapter} will focus on a simplified model of an extra dimension in which there is only one massive kk mode and show that this allows for analytic analysis of a \emph{new} hybrid regime where $m_v\sim M$.  In chapter \ref{chapter6} we will apply the framework of current correlators to an $AdS_5$ background which goes some way to achieving the aim of using general gauge mediation to explore the supersymmetry breaking of strongly coupled hidden sectors.  In chapter \ref{chapter7} we conclude and discuss future directions.

It may be helpful for the readers to note the following publications that each chapter is broadly based upon:
\begin{itemize}
\item Chapter2: ``General Gauge Mediation in Five Dimensions" \cite{McGarrie:2010kh}, M.McGarrie and R.Russo.
\item Chapter3: ``General Gauge Mediation in Five Dimensions" \cite{McGarrie:2010kh}, M.McGarrie and R.Russo. \\ \emph{and}
``Direct Gaugino Mediation,”" \cite{Green:2010ww}, D. Green, A. Katz, and Z. Komargodski.
\item Chapter4: ``General Gauge Mediation and Deconstruction" \cite{McGarrie:2010qr}, M.McGarrie.
\item Chapter5:  ``Hybrid Gauge Mediation'' \cite{McGarrie:2011dc}, M.McGarrie.
\item Chapter6: ``Warped General Gauge Mediation''  \cite{McGarrie:2010yk}, M.McGarrie and D.C.Thompson.
\end{itemize}

\chapter{General Gauge Mediation in Five Dimensions}\label{chapter2}
In this chapter we use the ``General Gauge Mediation'' formalism to describe a 5D setup with an $S^{1}\!/\mathbb{Z}_{2}$ orbifold.  The original publication is \cite{McGarrie:2010kh}.  We first consider a model independent SUSY breaking hidden sector on one boundary and generic chiral matter on another.  In our analysis the susy breaking dynamics is confined to a 4d brane and thus the current correlators are exactly those of the usual GGM formalism and are independent of momentum in the 5th dimension.   In the ``bulk'' or five dimensional space we place five dimensional Super Yang Mills whose conventions can be found in appendix \ref{NON}; this sector plays the role of the mediating sector.  Throughout this chapter we apply the analysis of~\cite{Mirabelli:1997aj} in the context of GGM and rederive the scalar and gaugino masses in terms of the current correlators.   We find the gaugino, sfermion and hyperscalar mass formulas for minimal and generalised messengers in different regimes of a large, small and intermediate extra dimension of length $\ell$.  We also briefly discuss Semi-Direct gauge mediation.

Realistic models of hidden sectors that break supersymmetry are commonly described by $\mathcal{N}=1$ SQCD messengers coupled to a spurion.  The use of Seiberg duality often allows for an otherwise strongly coupled hidden sector to be perturbatively described using its IR free dual magnetic description in which at lowest order in $\alpha_{mag}$ the sector reduces to the form of messengers coupled to a spurion.  For these reasons we will demonstrate in section \ref{sec:general}  the explicit encoding of a perturbative hidden sector into current correlators and give more explicit formula for these types of models.  These results will build the foundations which will then allow us to explore a fully realistic ISS scenario in the next chapter.
\setcounter{equation}{0}
\section{Framework}\label{sec:Frame}
In this section we recall the main features of $\mathcal{N}=1$ super Yang-Mills and hypermultiplet matter in $5d$. Once compactified on an orbifold of $S^{1}/\mathbb{Z}_{2}$, a positive parity vector multiplet couples to the boundaries of the orbifold and we associate this with the standard model gauge groups in the bulk.  The remaining fields fill a negative parity chiral multiplet which we do not couple to the boundaries. Similarly we will outline features of the orbifold compactified hypermultiplet.  A complete description is found~\cite{Hebecker:2001ke}, see also Appendix \ref{NON}. 

We first focus on the pure super Yang-Mills theory. The action written in components is
\be
S_{5D}^{SYM}= \int d^{5} x
~\text{Tr}\left[-\frac{1}{2}(F_{MN})^2-(D_{M}\Sigma)^2-i\bar{\gl}_{i}\gamma^M
D_{M}\gl^{i}+(X^a)^2+g_{5}\, \bar{\gl}_i[\Sigma,\gl^i]\right].
\ee
The coupling $1/g^2_{5}$ has been rescaled inside the covariant derivative, $D_{M}=
\partial_{M}+ig_{5} A_{M}$.  The other fields are a real scalar $\Sigma$, an $SU(2)_{R}$ triplet of
real auxiliary fields $X^{a}$, $a=1,2,3$ and a symplectic Majorana
spinor $\gl_{i}$ with $i=1,2$ which form an $SU(2)_R$ doublet. The reality condition is $\gl^i= \epsilon^{ij} C\bar{\gl}_{j}^{T} $.
Next, using an orbifold $S^{1}\!/\mathbb{Z}_{2}$ the boundaries will
preserve only half of the $\mathcal{N}=2$ symmetries. We choose to
preserve $\epsilon_{L}$ and set $\epsilon_{R}=0$.  We have a parity
operator $P$ of full action $\mathbb{P}\phi(y)=P\phi(-y)$ and define
$P\psi_{L}=+\psi_{L}$ $P\psi_{R}=-\psi_{R}$ for all fermionic fields
and susy parameters. One may then group the
susy variations under the positive parity assignments and they fill an off-shell $4d$ vector
multiplet $V(x^{\mu},x_{5})$.  Similarly the susy variations of odd parity
form a chiral superfield $\Phi(x^{\mu},x_{5})$. We may therefore write a $5d$
$\mathcal{N}=1$ vector multiplet as a $4d$ vector and
chiral superfield:
\begin{alignat}{1}
V=&- \theta\sigma^{\mu}\bar{\theta}A_{\mu}+i\bar{\theta}^{2}\theta\gl-
i\theta^{2}\bar{\theta}\bar{\gl}+\frac{1}{2}\bar{\theta}^{2}\theta^{2}D\\
\Phi=& \frac{1}{\sqrt{2}}(\Sigma + i A_{5})+
\sqrt{2}\theta \chi + \theta^{2}F\,,
\label{fields}
\end{alignat}
where the identifications between $5d$ and $4d$ fields are
\begin{equation}
D=(X^{3}-D_{5}\Sigma) \quad F=(X^{1}+iX^{2})\,,
\end{equation}
and we used $\lambda$ and $\chi$ to indicate $\lambda_{L}$ and
$-i\sqrt{2}\lambda_R$ respectively. 
The vector and chiral superfield have a Kaluza-Klein (kk) mode expansion given by 
\begin{alignat}{1}
V(x,y)= &\frac{1}{\sqrt{\ell}}V^0 (x)+\frac{\sqrt{2}}{\ell}\sum^{\infty}_{n=1}V^n (x) \cos \frac{n\pi y}{\ell}\\
\Phi(x,y)= &\frac{\sqrt{2}}{\ell}\sum^{\infty}_{n=1}\Phi^n (x) \sin \frac{n\pi y}{\ell}.
\end{alignat}
The bulk hypermultiplet action \bea
S_{5D}^{H} &=& \int d^5 x[-(D_MH)^\dagger_i(D^MH^i)-
i\bar{\psi}\gamma^MD_M\psi+ F^{\dagger i}F_i-g_5
\bar{\psi}\Sigma\psi\\
&&+g_5 H^\dagger_i(\sigma^aX^a)^i_jH^j
+g_5^2 H^\dagger_i\Sigma^2H^i+ig_5\sqrt{2}\bar{\psi}
\lambda^i\epsilon_{ij} H^j-i\sqrt{2}g_5H^{\dagger}_{i}
\epsilon^{ij}\bar{\gl}_{j}\psi\,]\nonumber \label{hyperaction2}
\eea 
decomposes into a positive and negative parity chiral superfield, $PH=+H$ and $PH^c=-H^c$:
\bea
H &=& H_1+\sqrt{2}\theta\psi_L+\theta^2(F_1+D_5H_2-g_5\Sigma H_2)\\
H^c &=& H^\dagger_2+\sqrt{2}\theta\psi_R+\theta^2(-F^{\dagger }_{2}-D_5
H^\dagger_1- g_5 H^\dagger_1\Sigma)\,.
\eea
With our conventions,the dimensions of ($H_{i},\psi,F_{i}$) are ($\frac{3}{2},2,\frac{5}{2}$).  The hypermultiplets are intriguing, as in the simplest case they also only couple to the branes via the gauge coupling $g_5$ so they satisfy the framework of general gauge mediation.

In the next section we will locate a susy breaking hidden sector on one boundary of the orbifold.  We will encode the hidden sector into a set of current correlators, and use the positive parity vector multiplet to generate, a gaugino mass, construct loops across the bulk to generate sfermion masses on the other orbifold boundary and finally construct loops to generate a mass for the zero mode of the bulk hypermultiplets. 

\section{General Gauge Mediation for bulk and boundaries}\label{sec:3}
In this section we follow~\cite{Meade:2008wd} and use the formalism of
current correlators in a 5D orbifold $R^{1,3}\times S^1 \!
/\mathbb{Z}_{2}$ where supersymmetry is broken only on one of the two
planes at the end of the interval. The gaugino and sfermion mass are
written in terms of current correlators on the supersymmetry breaking
plane. Additionally we explore the hypermultiplet scalar and fermion
masses via the same set of current correlators.

In a supersymmetric gauge theory, global current superfields
$\mathcal{J^A}$ have the component form
\begin{equation}
\mathcal{J^A}=J^\mathcal{A}+i \theta j^\mathcal{A} -
i\bar{\theta}\bar{j}^\mathcal{A}-
\theta \sigma^{\mu} \bar{\theta} j_{\mu}^\mathcal{A} + 
\frac 1 2 \theta^2 \bar{\theta} \bar{\sigma}^{\mu} 
\partial_{\mu} j^\mathcal{A} - \frac 1 2 \bar{\theta}^2 \theta 
\sigma^{\mu} \partial_{\mu} \bar{j}^\mathcal{A} - 
\frac 1 4 \theta^2 \bar{\theta}^2 \square J^\mathcal{A}\;,
\end{equation}
which by definition satisfies the conditions
\begin{equation}
\bar{D}^2 \mathcal{J}^\mathcal{A}=D^2 \mathcal{J}^\mathcal{A}=0\;.
\end{equation}
This implies the usual current conservation on $j_{\mu}$:
$\partial^{\mu} j^\mathcal{A}_{\mu}=0$.  
We now gauge the global symmetry and couple the current to the vector
superfield with
\begin{equation}
\mathcal{S}_{int}=2g_{5}\!\int\! d^5x d^{4}\theta \mathcal{J} 
\mathcal{V}\delta(x_{5})= \!\int\! d^5x g_{5}(JD- \gl j \!-
 \!\bar{\gl} \bar{j}-j^{\mu}A_{\mu})\delta(x_{5})
\end{equation}
We may relate 4d brane localised currents as 5d currents by
$J_{5d}=J_{4d}\delta(x_{5})$.  The vector multiplet is five
dimensional but is written in 4 dimensions as $V(x^{\mu},x^{5})$ and
has been coupled to the boundary fields. The 5d coupling, $g_{5}$, has
mass dimension, $\text{Dim}[g_{5}]= \frac{(4-D)}{2}$.  In this
normalisation, following \refe{fullaction}, the mass dimensions of
$(A_{\mu},\Sigma,\gl_{i},X_{a})$ are $(3/2,3/2,2,5/2)$.  It follows
that 5d currents that couple to these fields,
$(J_{\mu},J_{\gl^{i}},J_{X^{a}})$ have mass dimension
$(4,7/2,3)$. $\delta(x_{5})$ carries a mass dimension $1$. We
explicitly insert the relation for the $D$ term and keep the auxiliary
fields $X^{3}$. The change of the effective Lagrangian to
$O(g_{5}^{2})$ is
\begin{alignat}{1}\label{effectiveL}
\delta \mathcal{L}_{eff}=-& g^{2}_{5}\tilde{C}_{1/2}(0) i \lambda \sigma^{\mu} \partial_{\mu} \bar{\lambda} - g^{2}_{5}\frac {1} {4} \tilde{C}_1(0) F_{\mu\nu} F^{\mu\nu} \\ \nonumber &-g^{2}_{5}\frac {1}{ 2}(M \tilde{B}_{1/2}(0) \lambda \lambda + M \tilde{B}_{1/2}(0)\bar{\gl}\bar{\gl})\\ \nonumber & +\frac {1}{ 2 }g^{2}_{5}\tilde{C}_0 (0)(X^{3})^{2}+\frac {1}{ 2 }g^{2}_{5}\tilde{C}_0 (0)(D_{5}\Sigma)^{2}-g^{2}_{5}\tilde{C}_0 (0)(D_{5}\Sigma)X^{3} \\&+g_{5}^{2}\braket{J j^{\mu}}((D_{5}\Sigma)A_{\mu}-X^{3}A_{\mu})+\cdots\;\nonumber.
\end{alignat}
These are evaluated in the IR ($p^{\mu}_{\text{ext}}=0$).  When using these components to construct the diagrams in Figure 1, one must include the full momentum dependence. The $\tilde{B}$ and $\tilde{C}$ functions are related to momentum space current correlators, found below.  The last 4 terms require comment:  the first three of these replace the $D^{2}$ term, in the last line the current correlator is found to be zero \cite{Buican:2009vv}.  In position space, the current correlators can be expressed in terms of their mass dimension\footnote{Renormalised operators of conserved currents receive no rescalings $Z_{J}=1$ and no anomalous dimension $\gamma_{J}=0$} and some functions $C_{s}$ and $B_{\frac{1}{2}}$,
\begin{alignat}{1}
\braket{J(x,x_{5})J(0,x'_{5})}=&\frac{1}{x^{4}}C_0(x^{2}M^{2})\delta(x_{5})\delta(x'_{5}) \\
\braket{j_\alpha(x,x_{5})\bar j_{\dot\alpha}(0,x'_{5})}=&-i\sigma_{\alpha\dot\alpha}^\mu \partial_\mu(\frac{1}{x^{4}}C_{1/2}(x^{2}M^{2}))\delta(x_{5})\delta(x'_{5})\\
\braket{j_\mu(x,x_{5})j_\nu(0,x'_{5})}=&(\partial^2\eta_{\mu\nu}-\partial_\mu \partial_\nu)(\frac{1}{x^{4}}C_1(x^{2}M^{2}))\delta(x_{5})\delta(x'_{5})\\
\braket{j_\alpha(x,x_{5})j_\beta(0,x'_{5})}=&\epsilon_{\alpha\beta}\frac{1}{x^{5}}B_{1/2}(x^{2}M^{2})\delta(x_{5})\delta(x'_{5}) \\
\braket{j_\mu(x,x_{5})J(0,x'_{5})}=& cM^{2}\partial_{\mu}(\frac{1}{x^{2}}) \delta(x_{5})\delta(x'_{5}) \label{scale}
\end{alignat}
$M$ is a characteristic mass scale of the theory (e.g. the fermion mass of the SUSY breaking messenger multiplet).   $B_{1/2}$ is a complex function, $C_{s}$, $s=0,1/2, 1$, is real.
When supersymmetry is unbroken
\begin{equation}
C_0=C_{1/2}=C_1\;,\qquad \text{and} \qquad B_{1/2}=0\;.
\end{equation}
Supersymmetry is restored in the UV such that
\begin{equation} \label{BCUV}
\lim_{x \rightarrow 0} C_0(x^2 M^2)=\lim_{x \rightarrow 0} C_{1/2}(x^2 M^2) =\lim_{x \rightarrow 0} C_1(x^2 M^2)\;, \text{and}  \lim_{x \rightarrow 0} B_{1/2}(x^2 M^2)=0\;.
\end{equation}
 $\tilde{C}_s$ and $\tilde B$ are Fourier transforms of $C_s$ and $B$,
\begin{equation}
\begin{split}
\tilde{C}_s\left(\frac{p^2}{M^2};\frac{M}{\Lambda}\right)&=\int d^4 x e^{ipx} \frac 1 {x^4} C_s(x^2 M^2)\\
M\tilde{B}_{1/2}\left(\frac{p^2}{M^2}\right)&=\int d^4 x e^{ipx} \frac 1 {x^5} B_{1/2}(x^2 M^2)\;.
\end{split}
\end{equation}
The $\tilde{C}_s$ and $\tilde B$ terms are the nonzero current correlator functions of the components of the current superfield.  The correlators  have positive parity ($P=+1$) as they live on the wall. Using the full action of $\mathbb{P}$, the Fourier transforms over $x_{5}$ and $x'_{5}$ removes the delta functions. In this off-shell formalism $\delta(0)$ \emph{does not} enter explicitly in the calculation (compare with \cite{Mirabelli:1997aj}). In momentum space we have,
\begin{alignat}{1}
\braket{J(p,p_{5})J(-p,p'_{5})} =&\tilde{C}_0(p^2/M^2)  \label{eq:c0}   \\
\braket{j_\alpha(p,p_{5})\bar j_{\dot\alpha}(-p,p'_{5})} =&-\sigma_{\alpha\dot\alpha}^\mu p_\mu\tilde{C}_{1/2}(p^2/M^2) \label{eq:c1/2}		\\ 
\braket{j_\mu(p,p_{5})j_\nu(-p,p'_{5})} =&-(p^2\eta_{\mu\nu}-p_\mu p_\nu)\tilde{C}_1(p^2/M^2)	\label{eq:c1}	\\ 
\braket{j_\alpha(p,p_{5})j_\beta(-p,p'_{5})} =&\epsilon_{\alpha\beta}M\tilde{B}_{1/2}(p^2/M^2)     \label{eq:b1/2}	\\ 
\braket{j_\mu(p,p_{5})J(-p,p'_{5})} =& c M^{2}\frac{2{\pi}^2 i p_{\mu}}{p^2}\label{eq:fsdf}
\end{alignat}
We see that the current correlators are completely independent of the momentum in the fifth dimension. The analysis of \cite{Buican:2009vv} demonstrates that $c=0$ in the last equation.

We shall frequently express our results in terms of the ``supertraced'' set of these current correlators 
\be
 [3\tilde{C}_1(p^2/M^2)-4\tilde{C}_{1/2}(p^2/M^2)+\tilde{C}_{0}(p^2/\hat{M}^2)]=\Omega \left(\frac{p^2}{M^2} \right)\label{Omega1}\, .
\ee
The numerical coefficient in front of the $\tilde{C}_{s}$ terms in \refe{Omega1} is associated with the off-shell degrees of freedom of the bulk propagating vector multiplet and arise from taking an index contraction of the current correlators.


\subsection{Gaugino masses}
At $g^2$ order the susy breaking contribution to the gaugino mass can be read directly from the
Lagrangian \eqref{effectiveL} after rescaling $\lambda$ so as to
canonically normalise the bulk action~\eqref{fullaction}:
\begin{equation}\label{E:Gaugino}
M^{nm}_{\lambda } = g^2_{4} M \tilde{B}_{1/2} (0)\;.
\end{equation}
These terms are of Majorana type and couple every Kaluza-Klein mode
with every other mode with the same coefficient. In addition we have
the usual Kaluza-Klein tower of masses ($p_{5}=\frac{n\pi}{\ell}$)
which are of Dirac type and mix $\lambda^L_n$ and $\lambda^R_n$.  The mass
eigenstates will be in general a linear combination involving
different Kaluza-Klein modes. This is similar in vein to the ``see-saw'' mechanism and for large $\ell$ the lowest mass eigenstate can become very light. This highlights that for bulk mediation, the scale $\Lambda_{G}$ is not a good scale and must be replaced by the lightest gaugino mass eigenvalue. We comment on three regimes: 
\begin{description}
\item[Small $\ell$] 
When the scale of the extra dimension $1/\ell$ is much bigger than the scales $\sqrt{F}$ and $M$ then we return to an effective 4d theory and the zero mode mass is given by~\eqref{E:Gaugino}.
\item[Intermediate $\ell$]
When $F\leq 1/\ell^2\ll M^2$ the susy breaking mass $M^{mn}_\lambda$ is of order $F/M$ and the K.K. mass is much bigger because $F/M \leq (1/M) (1/\ell^2) \ll 1/\ell$. In this case the gaugino mass is still given by~\eqref{E:Gaugino} and $\Lambda_{G}$ is a good scale.  
\item[Large $\ell$]
When  $1/\ell^2 \ll F$, then one must be careful and see how $F$ and $M$ scale.   For instance if $F \sim M/\ell$, which is possible in this regime, then there is a sizeble mixing between the various K.K. modes and the first mass eigenstate is lighter than $M^{nm}_\lambda$. If $M^{nm}_\lambda \gg 1/\ell$ the lightest gaugino eigenstate can have a \emph{much} lower mass than $M^{nm}_\lambda$ due to mixing with the tower of K.K. modes.  In this case $\Lambda_{G}$ is \emph{not} a good scale.
\end{description}

\subsection{Sfermion masses}
\begin{figure}[ht]
\centering
\includegraphics[scale=0.7]{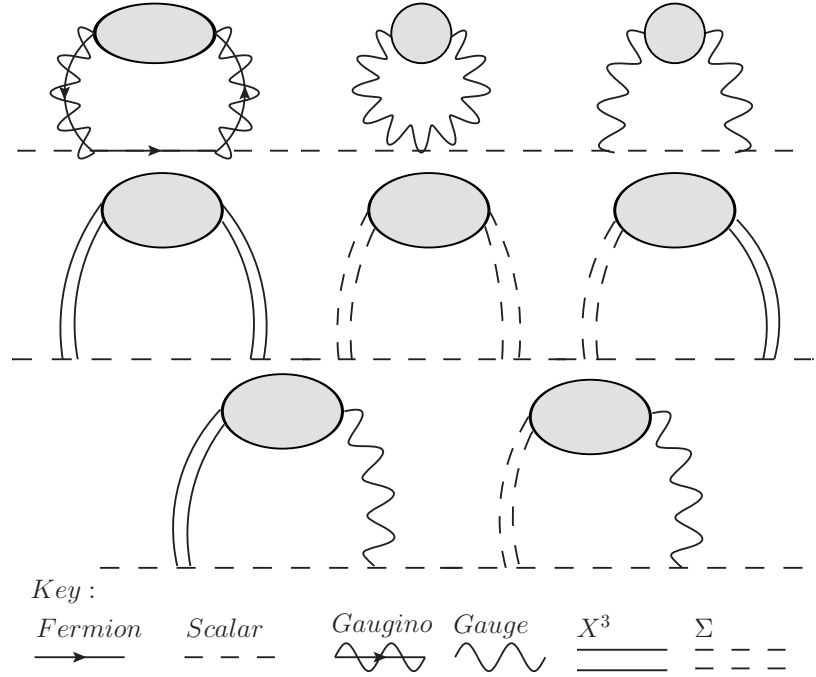}
\caption{The graphical description of the two point functions to the soft sfermion masses.  In the top row, the first diagram is from $\braket{j_{\alpha} \bar{j}_{\dot{\alpha}}}$ and the second and third are from $\braket{j_{\mu} j_{\nu}}$.  The middle row are all separately related to $\braket{J J}$, only the middle diagram survives propagation across the bulk. The final row is constructed from couplings to $\braket{J j_{\mu}}$ and are exactly equal to zero \cite{Buican:2009vv}.}
\label{od3}
\end{figure}

At leading order in $\alpha$, the sfermion masses can be determined in terms of the $\tilde{C}_{s}$ current correlator functions and propagation of the vector multiplet in the bulk.  This corresponds to the 8 diagrams in figure \ref{od3}.  The ``blobs'' are current correlators located on the hidden brane.  The scalar lines are located on the visible brane.  The intermediate propagators are the bulk fields and are components of the vector multiplet in the bulk.  The full momentum dependence of the current correlators must be taken into account as they form a part of a loop on the scalar propagator. The full set of diagrams are accounted for in \cite{Mirabelli:1997aj} including the two vanishing diagrams associated with $\braket{j_{\mu} J}$  (see also \cite{Buican:2009vv}). The top right most diagram contributes nothing to the mass due to transversality, when taking the external momentum to zero. The middle row has an auxiliary field $X^3$ which cannot propagate across the bulk so its diagrams vanish and only the middle one of that row survives.  In conclusion, when computing the soft mass terms, only the first two diagrams and the middle diagram of the middle row survive. They are the final ``supertraced'' combination with the same structure as in the 4D case.

To compute the three diagrams we need the propagator of a free massless bulk field 
\begin{equation}
  \left\langle{ a(x,x^5) a(y,y^5)} \right\rangle = 
	\int_{p5} \frac{i }{p^2 - (p_5)^2}\,
 e^{-ip\cdot (x-y)} (e^{ip_5(x^5-y^5)}
     + P e^{ip_5(x^5+y^5)}) \ ,
\label{fiveprop}\end{equation}
where 
\begin{equation}
   \int_{p5} = \int \frac{d^4p}{ (2\pi)^4} \frac{1}{ 2\ell}\sum_{p_5} \ ,
\label{intdef}\end{equation}
with $p_5$ summed over the values $ \pi n /\ell$, $n =$ integer. We propagate from $x^{5}=0$ to $y^{5}=\ell$.  The propagator was found by use of the full action of $\mathbb{P}$ on the Fourier transformation from position to momentum space, where $P$ is the parity eigenvalue $\pm 1$. The mass eigenstates arise as a Higgs mechanism and is naturally found from a geometric sum of mass insertions.  The exponents of the 5th dimension, in brackets,  for two propagators will reduce to 
\begin{equation}
4(-1)^{n+\hat{n}}.
\end{equation}
In particular, this factor encodes the finite separation of the branes and will allow for a convergent finite answer for the soft mass.  It should also be noted neither brane (current correlator) conserves the incoming to outgoing $p_{5}$ momenta.  All vertex couplings can be determined by expanding out a canonical K\"ahler potential for a chiral superfield, which can be seen by example in Appendix \ref{generalised}.   
We would like to factor out all the extra dimensional contributions to the sfermions so that it leaves the GGM result multiplied by higher
dimensional contributions. We find
\begin{equation}
m_{\tilde{f}}^2= \sum _r g_{r(5d)}^4  c_2(f;r)E_r
\end{equation}
where
\begin{equation}
E_r= -\! \!\int\! \frac{d^4p}{ (2\pi)^4}\frac {1}{\ell^{2}}\sum_{n, \hat{n}} \!  \frac{(-1)^{n+\hat{n}}}{p^2-(p_{5})^{2}}\frac{p^{2}}{p^2-(\hat{p}_{5})^{2}} \Omega^{r}(\frac{p^2}{M^2}), \label{Primeresult}
\end{equation}
where we used the regularisation of the K.K. sum described in Eq.(29) of~\cite{Mirabelli:1997aj}. $r=1,2,3.$ refer to the gauge groups $U(1),SU(2), SU(3)$. $c_2(f;r)$ is the quadratic Casimir for the representation of $f$ under the gauge group $r$. We have followed the convention of \cite{Intriligator:2010be} by using $E$, reserving $A$ for A-terms. The numerical coefficient in front of the $\tilde{C}_{s}$ terms in \refe{Primeresult} are essentially set by taking an index contraction of the current correlators \refe{eq:c0} to \refe{eq:c1}. 

To make further progress we must identify different expansion limits of this result. In particular for large and small $\ell$ regimes one may obtain either four dimensional general gauge mediation, or the screened five dimensional limit.

We use Matsubara frequency summation to identify
\begin{equation}
  \frac{1}{ \ell}\sum_{n}  (-1)^n \frac{1}{ k^2 + (k_5)^2}    
    = \oint \frac{dk^5}{ 2\pi} \frac{2 e^{ik_5 \ell}}{ e^{2ik^5\ell}-1} 
 \frac{1}{ k^2 + (k_5)^2}  = \frac{1}{ k}\frac{1}{ \sinh k\ell} .
\label{firstcontour}\end{equation} 
\begin{figure}[htcb]
\centering
\includegraphics[scale=0.6]{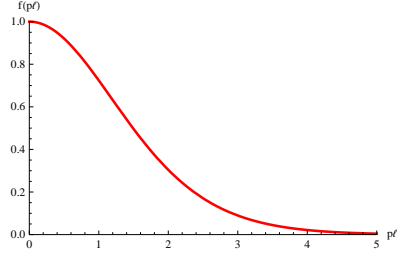}
\caption{A plot of the momentum dependent form factor that suppresses leading order sfermion masses due to bulk mediation of supersymmetry breaking, after a Matsubara summation of all kk modes.  When $p\ell\rightarrow 0 $ one recovers four dimensional general gauge mediation.  When $p\ell \ge 0 $ one obtains screening of the leading order scalar soft mass.}
\label{plot1}
\end{figure}
We obtain
\begin{equation}
E_r= \! -\!\int\! \frac{d^4p}{ (2\pi)^4}(\frac{1}{p\sinh p\ell})^{2} p^{2} \Omega^{r}(\frac{p^2}{M^2})
\label{mainresult}
\end{equation}
The full answer is a the 4d GGM  \cite{Meade:2008wd} answer multiplied by a momentum dependent form factor 
\be
f(p\ell)=   (\frac{p\ell}{\sinh p\ell})^{2}.
\ee
plotted in figure \ref{plot1}.  We may recover the four dimensional limit $\ell \to 0$, where
\begin{equation}
\frac{1}{k}\frac{1}{ \sinh k\ell} \rightarrow \frac{1}{\ell k^{2}}.
\end{equation}

\subsubsection{Subleading contributions}
In this section we would like to discuss the subleading contribution to sfermion masses ($p_{ext}=0$) pictured in figure \ref{gauginomed}  (See also \cite{Chacko:1999mi,Kaplan:1999ac,Csaki:2001em}).  In the all order kk model \cite{McGarrie:2010kh}  its contribution is given by
\be
\delta m_{\tilde{f}}^2= \sum_r c_2(f;r) \frac{g^6_{5}}{2\ell^3}\int \frac{d^4p }{(2\pi)^4}p^2\sum_{n,\hat{n},\hat{\hat{n}}}\frac{(-1)^{n+\hat{n}}}{p^2+p^2_{5}}\frac{M\tilde{B}_{1/2}(p^2/M^2)}{p^2+\hat{p}^2_{5}}\frac{M\tilde{B}_{1/2}(p^2/M^2)}{p^2+\hat{\hat{p}}^2_{5}}.
\ee
In general, this integral is divergent due to the brane to same brane propagator on the hidden brane that connects the double mass insertions.   The four dimensional limit, when $M\ll \frac{1}{\ell}$ i.e. $\ell$ is small one finds 
\be
 \delta m_{\tilde{f}}^2=\sum_r c_2(f;r) \frac{g^6_{4d}}{2} \int \frac{d^4p }{(2\pi)^4}\frac{1}{p^4}(M\tilde{B}_{1/2}(p^2/M^2))^2.
\ee
In the limit $\frac{1}{\ell}\ll M $ one may carry out a Matsubara summation and finds 
\be
\delta m_{\tilde{f}}^2= \sum_r c_2(f;r)\frac{g^6_{5}}{2}\int \frac{d^4p }{(2\pi)^4}\frac{\left(M\tilde{B}_{1/2}(0)\right)^2}{p \sinh^2 (p\ell)\tanh (p\ell) }
\ee
where in this limit 
\be 
\lim_{\frac{p^2}{M^2}\rightarrow 0}M\tilde{B}_{1/2}(p^2/M^2)=M\tilde{B}_{1/2}(0).
\ee
The integral is IR divergent and must be regulated. We choose  
\be
\delta m_{\tilde{f}}^2=\sum_r c_2(f;r) \frac{g^6_{4}}{2}\left(M\tilde{B}_{1/2}(0)\right)^2 \int_{0}^{\infty} \frac{2 \pi^2 dy }{(2\pi)^4}[\frac{y^2}{ \sinh^2 (y)\tanh (y) }-\frac{e^{-\Lambda y}}{y}]
\ee
\be
\delta m_{\tilde{f}}^2= \sum_r c_2(f;r)\frac{g^6_{4}}{16\pi^2}\left(M\tilde{B}_{1/2}(0)\right)^2[3/2 + \gamma +\log \Lambda/2 ],
\ee
Schematically, for a messenger sector, one can compare the leading order and subleading contribution
\be
 m^2_{\tilde{f}}\sim  (\frac{\alpha}{4\pi})^2 \Lambda^2_{S}\frac{1}{(M \ell)^2}+ (\frac{\alpha}{4\pi})^3 \Lambda^{2}_{G}
\ee
such that it is not always clear which term is truly leading order. $\Lambda_S$ is the four dimensional scalar mass scale and $\Lambda_G$ is the gaugino mass scale.  When $\Lambda_{S}\sim \Lambda_{G}$ comparing $\alpha/4\pi$ versus $1/(M\ell)^2$ may be sufficient, however models with an approximate R-symmetry may also suppress $\Lambda_{G}$.  It is worth emphasising that for the generalised messenger sector discussed in this paper and for typical values of $M$, $\ell$ and $\alpha$ when $\frac{1}{\ell}\ll M $, the double mass insertion in figure \ref{gauginomed} is most likely to be the largest contribution to sfermion masses, however for ISS-like models \cite{McGarrie:2010kh,Green:2010ww} where the gaugino mass is suppressed due to an R-symmetry, figure \ref{flatfigure} is the leading contribution.   We also \emph{stress} that it is completely reasonable to build models whereby $\frac{1}{\ell}\sim M $, in which case this model is in a Hybrid regime where both digrams will play significant roles in the contribution of soft masses to scalars, for example in chapter \ref{hybridchapter}.  It is also useful to note that the subleading diagram may act as a bound, at the high scale $M$, on the ratio of masses:
\be
\frac{m^2_{\lambda} }{m^2_{\tilde{f}}}\lesssim (\frac{4\pi}{\alpha})\sim 300 
\ee
where in the last line we took $\alpha(M_{GUT})$=0.04.

\subsection{Hypermultiplet scalar masses}
The supersymmetry breaking masses of the bulk hypermultiplet scalars and hypermultiplet fermions can also be computed in the gauge mediation setup and couple to the hidden brane exclusively via $g_5$, when using the action \refe{hyperaction2}. The diagrams for the scalars are similar to those of  Figure 1, but include an additional contribution with a bulk propagator, indicated with  the $\otimes$ symbol, coupling the positive parity gaugino to the negative parity bulk fermion 
\be
\braket{\gl_{L\alpha} \gl_{R\beta}}= \int \frac{d^4 k }{(2\pi)^4}\frac{1}{2\ell}\sum_{k_{5}}\frac{ i k_{5}\epsilon_{\alpha \beta}}{k^2-k^2_{5}}e^{-ik.(x-y)}(e^{ik_{5}.(x_{5}-y_{5})}+Pe^{-ik_{5}.(x_{5}+y_{5})}) 
\ee
A similiarly constructed propagator can be written in the $\sin (k_{5}x_{5})$ and $\cos (k_{5}x_{5})$ basis for the 5D wavefunction. 
\begin{figure}[ht]
\centering
\includegraphics[scale=0.8]{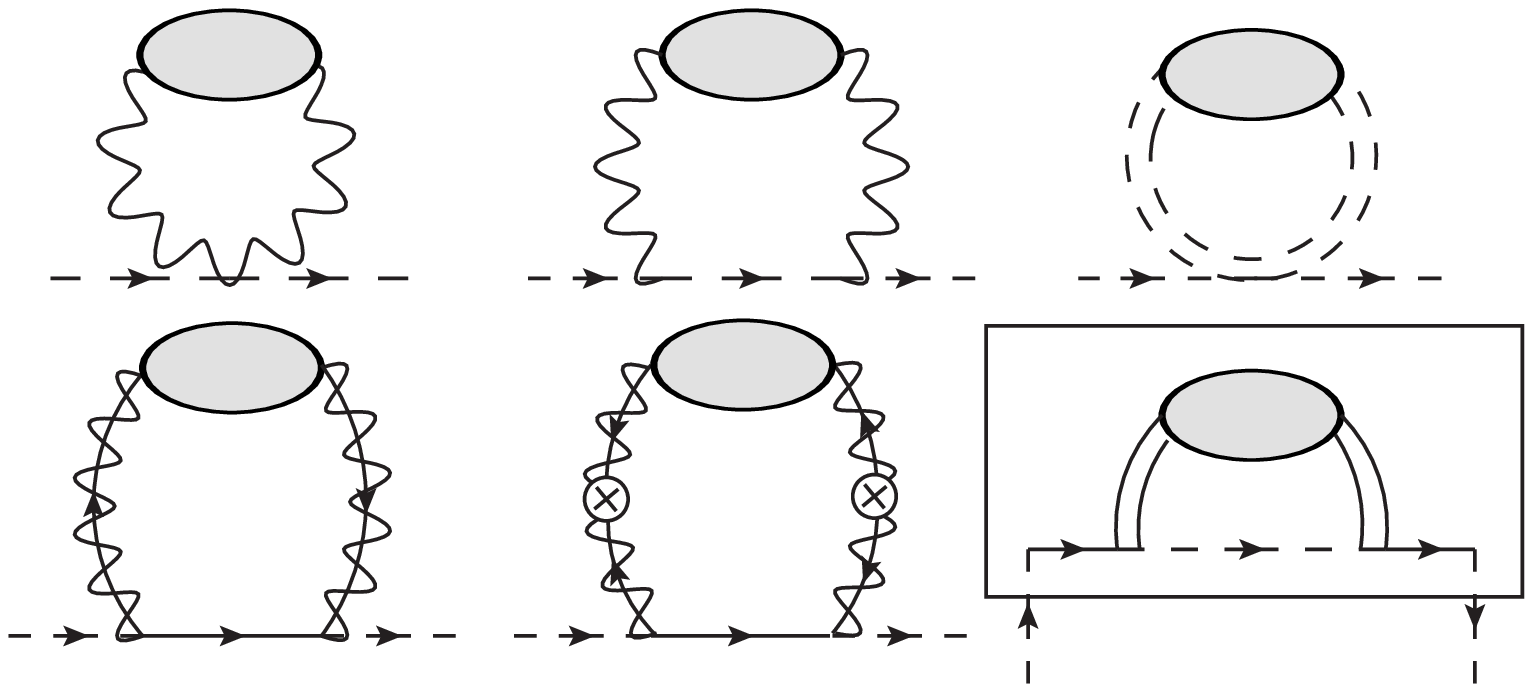}
\caption{The graphical description of the two point functions to the hypermultiplet scalar masses. Unlike the sfermion diagrams, the scalar field propagator does \emph{not} lie on a brane; the position of the vertex point must be integrated over for \emph{both} external sewing points when computing the diagrams. The parity of the external bulk scalar legs must also be specified.}
\end{figure}
To compute the diagrams: The current correlators (blobs) are brane localised, however the vertices joining the vector fields to the hyperscalar must be integrated over all of $y_{5}$. One must also specify the 5d wavefunction ($p_{5}$ momenta) of the external hyperscalar legs using
\be
\frac{1}{\sqrt{2\ell}}(e^{i\frac{n\pi}{\ell}y_{5}}+Pe^{-i\frac{n\pi}{\ell}y_{5}})\;\; (\text{for } n\neq 0), \;\;\; \;\frac{1}{\sqrt{\ell}} \;\; (\text{for } n=0)
\ee  
at the sewing point $y_{5}$, which is then integrated over. The second diagram does not contribute to the mass due to transversality. The rectangle in the final diagram represents that the diagram is completely localised on the hidden sector brane, including the vertices that couple to the external hyperscalar legs.

We will focus on the zero mode mass ($m_{0}\bar{H}_{0}H_{0}$)
\begin{equation}
m_{H_{0}}^2= \sum _r g_{r(5d)}^4 c_2(f;r)D_r
\end{equation}
where
\begin{equation}
D_r= -\! \!\int\! \frac{d^4p}{ (2\pi)^4}\frac {1}{2\ell^{2}}\sum_{n} \!  (\frac{p}{p^2+(p_{5})^{2}})^2 \Omega^r(\frac{p^2}{M^2}),\label{secondresult2}
\end{equation}

\begin{equation}
D_r= \! -\!\int\! \frac{d^4p}{ (2\pi)^4}\frac{\coth (p\ell)+ p \ell \text{csch}^2(p \ell)  }{2 p \ell} \Omega^r(\frac{p^2}{M^2}).
\label{secondresult}
\end{equation}
The momentum integral is UV divergent. Physically this is to be expected as the hypermultiplet is not brane localised and so unlike the sfermion masses there is no brane separation to suppress large momenta contributions.  We can extract the $\ell$ dependent susy breaking mass  $\slashed{D}_{r}$ \cite{Buchbinder:2003qu}. We take $D_{r}= \slashed{D}_{r}+ \text{independent of $\ell$}$  where 
\be 
\slashed{D}_r= \! -\!\int\! \frac{d^4p}{ (2\pi)^4}\frac{\coth (p\ell)+ p \ell \text{csch}^2(p \ell) -1 }{2 p \ell} \Omega^r(\frac{p^2}{M^2}) \label{secondresult3}
\ee

Both in the case of bulk scalars and in the previous section on brane-localised scalars, we focused just on the case of external states at zero momentum. It would be interesting relax this condition and study, in a GGM setup, higher derivative operators by following the analysis of~\cite{Ghilencea:2005hm,Ghilencea:2006qm}.

\subsection{Vacuum Energy}
The propagation of supersymmetry breaking effects in the bulk produces a non zero vacuum energy.   The computation of the vacuum diagrams at the order $O(g_5^2)$ is similar to the scalar masses in the previous sections. Similarly to the computation of the hypermultiplet scalar masses, the diagram with $X^3$ and that with $D_5 \Sigma$ combine and yield a contribution proportional to $p^2 \tilde{C}_0$. By including all other diagrams we obtain
\begin{equation}
E_{Vac}/V_{4d}= \frac{1}{4} g^{2}_{5}  d_{G} \int \frac{d^{4}p}{(2\pi)^{4}} \frac{p}{\tanh (p\ell)}\Omega^r(\frac{p^2}{M^2}). 
\end{equation}
$d_{G}$ is the dimension of the adjoint representation of the gauge group $r$. The vacuum energy is also UV divergent. The Casimir energy is the component of the vacuum energy where the bulk propagation winds $x^{5}$. This, in general, will contribute to the determination of the physical value of $\ell$, along with other supergravity corrections.  The Casimir energy is given by
\begin{equation}
\frac{E_{\text{Cas}}}{V_{4}}= \frac{1}{2}g^{2}_{5} d_{G}\int \frac{d^{4}p}{(2\pi)^{4}}\frac{p}{ e^{2p \ell}-1} \Omega^r(\frac{p^2}{M^2}).\label{casimirenergy}
\end{equation}
\setcounter{equation}{0}
\subsection{Semi-Direct Gauge Mediation via the bulk}
In \cite{Argurio:2009ge} semi-direct gauge mediation in a $4d$ setup is explored using current correlators. In this section we comment on the semi-direct case in our $5d$ setup with two $4d$ Branes and a $5d$ bulk.  One brane is the MSSM brane, described by some generic chiral matter, charged under the visible gauge group $G_{v}$, which lives in the $5d$ bulk.  The other brane is a SUSY breaking brane.  The messenger fields are located on this brane and are charged under both $G_{v}$ and a brane localised gauge group $G_{h}$.  The messengers do not participate directly in the susy breaking dynamics, however they couple to the brane localised susy breaking sector via gauge interactions with gauge group $G_{h}$ and by construction the messengers and susy breaking sector decouple as $g_{h}\rightarrow 0$.   

The Gaugino masses vanish at leading order (three loops) precisely because of the argument of \cite{Argurio:2009ge}. 
The sfermion masses in a flat bulk are found to be
\begin{equation}
m_{\tilde{f}}^2= \sum _r g_{r(5d)}^{(v)4} g_{r(4d)}^{(h)4} c_2(f;r)E_r
\label{useme}
\end{equation}
where
\begin{equation}
E_r= +\! \!\int\! \frac{d^4p}{ (2\pi)^8}\frac {1}{\ell^{2}}\sum_{n, \hat{n}} \!  \frac{(-1)^{n+\hat{n}}}{p^2-(p_{5})^{2}}\frac{p^{2}K(p^{2}/m^{2})}{p^2-(\hat{p}_{5})^{2}} \Omega^r(\frac{p^2}{M^2}).
\end{equation}
 $m$ is the mass of the messengers. $K_{s}(p^{2}/m^{2})$ are the kernels, which in principle could be different for each of $s=0,1/2,1$. In \cite{Argurio:2009ge}, it was checked that $K_{0}=K_{1/2}=K_{1}$, so we will ignore this supscript index.  
 
As a final comment, one motivation for GGM5D is that it makes the partitioning of the hidden and visible sector a geometric feature.  One may be motivated to make semi-direct mediation a geometric feature too by placing the SUSY breaking sector $X$, the messengers $\phi,\phi^\dagger$ and the MSSM on three distinct branes. It would be interesting to study explicitly if it is possible to realise such a possibility in a concrete model.  
\setcounter{equation}{0}
\section{Generalised Messenger sector}\label{sec:general}
In this section we give a concrete description of the
4-dimensional susy breaking brane and consider two sets of chiral
messenger $\phi_i,\ti\phi_{i}$ coupled to a spurion field $X$. We
follow~\cite{Marques:2009yu} and extend the usual setup of a
generalised messenger sector to the case where the gauge multiplet
propagates in a 5d orbifold.

The superpotential describing the coupling of the messengers and the
spurion is identical to that considered in~\cite{Marques:2009yu} and
is localised in the fifth dimension on the susy breaking brane
\be W_{\phi} = {\cal M}(X)_{ij}\ \phi_i \ti \phi_j = (m + X
\lambda)_{ij}\ \phi_i \ti \phi_j \label{superpotential}\ee
where $m$ and $\gl$ are generic matrices.  We assume that all chiral
fields have canonical kinetic term and so, after a field redefinition,
we can take ${\cal M}=m+\langle X \rangle \lambda$ to be diagonal with
real eigenvalues $m_{0k}$, where, as usual, $\langle X \rangle$ is the
vev of the scalar component of the spurion superfield $X =\langle X
\rangle+  \theta^2 F  $. Further, we take $F\lambda$ to be hermitian and
by using unitary matrices one may diagonalise the bosonic mass-squared
matrix
\be
\mathcal{M}^{2}_{\pm}=U^{\dagger}_{\pm}(\mathcal{ M}^{2} \pm
F\lambda)U_{\pm} \ee Such that $\mathcal{M}^{2}_{\pm}$ has real
eigenvalues $m^{2}_{\pm k}$.  We define two mixing matrices: \be
A^{\pm}_{kn}=(U^{\dagger}_{\pm} )_{kn}(U_{\pm} )_{nk} \quad \quad
B^{\pm}_{kn}=(U^{\dagger}_{\pm}U_{\mp})_{kn}
(U^{\dagger}_{\mp}U_{\pm} )_{nk}
\ee
The calculations are carried out explicitly in section \ref{generalised}; in this section we simply display the results.
\subsection{Gaugino masses}

As we have seen in the previous section, the Majorana gaugino mass
matrix of the 5d model couples every Kaluza-Klein mode to every other
with the same coefficient and this contribution is
captured in~\eqref{E:Gaugino}.
In this case we can compute explicitly the correlator determining
$M \ti {B}_{1/2}$ by using~\eqref{supercurrent}

\be M_{r} = g^2 M \tilde{B}_{1/2}(0) =\frac{\alpha_r}{4 \pi}
\Lambda_{G} \ , \ \  \Lambda_{G} = 2 \sum_{k,n = 1}^N \sum_\pm \pm\ d_{kn}\ A_{kn}^\pm\ m^0_n 
 \frac{(m_k^\pm)^2 \log ((m_k^\pm)^2/(m_n^0)^2)}{(m_k^\pm)^2 - (m_n^0)^2}. \label{44}\ee
$k,n$ are messenger indices running from $1$ to $N$, the number of messengers, while $d_{kn}$ is nonzero and equal to $d_{k}$ or $d_{n}$ only if $\phi_{n}$ and $\tilde{\phi}_{k}$ are in the same representation.  In the full mass matrix one must take into account the Dirac masses of
the Kaluza-Klein tower itself. However, the susy breaking
contribution~\refe{44} is identical to the purely 4-dimensional case
and it is possible to follow~\cite{Marques:2009yu} for various case by
case simplifications. For instance, when $F\ll M^2$, to lowest order in $F/M^{2}$ and for
$SU(N)$ fundamentals one finds
\be
\Lambda_{G}=\sum^{N}_{k=1}\frac{F\gl_{kk}}{m^0_{k}}=F \partial_{X}\log \text{det}
\mathcal{M}
\ee
which is a familiar $4d$ result.
\subsection{Sfermion masses}\label{chap:sfermionmasses}
The sfermion masses are sensitive to the extra dimension $\ell$.  In the small $\ell$ limit the 4d results are recovered and this is fully explored in \cite{Marques:2009yu}. 
\subsubsection{Large $\ell$}
When $1/\ell^2$ is smaller than the scales $F$ and $X^{2}$  the sfermions masses can be written as 
\be m_{\ti f}^2 = 2 \sum_{r =1}^3 C_{\ti f}^r  \ \left(\frac{\alpha_r}{4\pi}\right)^2 \ \Lambda_{S}^2 
\label{alpha}
\ee
$C_{\ti f}^r$ are the quadratic Casimirs of $\ti f$ in the gauge group $r$. The sfermion scale $\Lambda_S^2$ is\footnote{See Appendix \ref{generalised} for its derivation.}
\bea 
\Lambda_S^2  =\sum_{k,n} \frac{\zeta (3)}{\ell^2} \!\!\! &&\!\!\! \sum_{\pm}d_{kn}[B^{\pm}_{kn}[\frac{2m^{2}_{\pm k}}{m^{2}_{\pm k}-m^{2}_{\mp n}}\log m^2_{\pm k}-1] + \delta_{kn}[\log m^{2}_{\pm k}m^4_{0k}] \nn\\
\!\!\! &&\!\!\!-\frac{4A^{\pm}_{kn}}{(m^{2}_{\pm k}-m^{2}_{0 n})}[m^2_{\pm k}\log m^2_{\pm k}-m^2_{0 n}\log m^2_{0 n}-1]     \label{LambdaS2}\\
\!\!\! &&\!\!\!-\frac{2A^{\pm}_{kn}}{(m^{2}_{\pm k}-m^{2}_{0 n})^2}[ (m^2_{\pm k}-m^2_{0 n})(m^2_{\pm k}+m^2_{0 n})-2 m^2_{\pm k}m^{2}_{0 n}\log \frac{m^2_{\pm k}}{m^2_{0 n}}]] \nn \eea
We may reduce to minimal gauge mediation \cite{Martin:1996zb} by setting $m = 0$ in equation (\ref{superpotential})
\be 
\Lambda_S^2 =  \left(\frac{F}{X}\right)^2 \left(\frac{1}{\lambda_{k} X \ell}\right)^2\sum_{k=1}^N    \zeta(3)
d_{kk} h(x_k)\label{LSMGM}\ee
\be
x_{k}=\frac{F}{\lambda_{k}X^2}
\ee
\be h(x) =\frac{3}{2}[\frac{4+x-2x^2}{x^4}\log(1+x)+\frac{1}{x^2}] + (x\rightarrow -x)
\ee
$h(x)$ for $x<0.8$ can be reasonably approximated by $h(x)=1$ and $\gl_{k}$ are the eigenvalues of $\gl$ in \eqref{superpotential}.
\begin{figure}
\begin{center}
\includegraphics[scale=1]{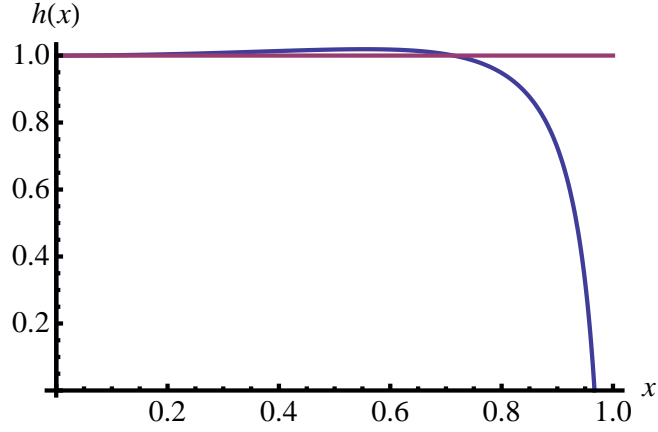}
\end{center}
\caption{A plot of the function $h(x)$ between  $x=0$ and $x=1$.}
\end{figure}
The limit of small multi-messenger mixing effects  gives 
\be \Lambda_S^2 =   \sum_{k = 1}^N  \zeta(3) d_{kk}\ \frac{F^2 \lambda_{k}^2}{\ell^2(m_k^0)^4} \ h\left(\frac{F \lambda_{k}}{(m_k^0)^2}\right) \label{ExpansionFSfermions}\ee
In the $h(x)=1$ limit, and small multi messenger mixing, we would like to derive the analog sfermion formula found in \cite{Cheung:2007es}.  The additional $|\ell^{2} M_{i}|^{2}$ factors cannot be taken inside such that we find
\be
\Lambda_{S}^{2} =\frac{\zeta(3)}{2}\sum_{i=1}^{N}\frac{|F|^{2}}{|\ell \mathcal{M}_{i}|^{2}}\frac{\partial^{2}}{\partial X \partial X^{*}}(\ln |\mathcal{M}_{i}|^{2})^{2}
\ee
Where $\mathcal{M}_{i}$ are in this case the complex eigenvalues of $\mathcal{M}$.  
\subsubsection{Intermediate $\ell$}
In the intermediate limit that $F\leq 1/\ell^2 \ll M^2$, the large $\ell$ results are still valid and $h(x)=1$. 
Reducing to minimal gauge mediation ($W=X\Phi\bar{\Phi}$) we find
 \be
\Lambda_S^2 = \frac{1}{\ell^2} 
\frac{2 F^2}{m^4}   \zeta(3)\sum_{k=1}^{N}d_{kk}
\ee
The limit of small multi-messenger mixing effects gives
 \be
\Lambda_S^2 = \frac{1}{\ell^2}\sum_{k=1}^{N}d_{kk} 
\frac{2 F^2\gl^{2}_{k}}{ (m^{0}_{k})^4} \zeta(3)
\ee

Finally we comment on the ratio $\frac{\Lambda^2_{G}}{\Lambda^{2}_{S}}$. In ``(Extra) Ordinary Gauge Mediation'' \cite{Cheung:2007es}, this quantity is defined as $N_{eff}$ and it may vary continously between $0$ and $N$, the number of messengers.  This definition is peculiar to the 4d models; we may easily  have $\Lambda_{S}^{2}\rightarrow 0$ and $N_{eff}\rightarrow \infty$ in this 5D construction.  To avoid confusion we will refer to it as a ratio and not as $N_{eff}$.

\subsection{Hyperscalar zero mode mass}
The positive parity scalar of the hypermultiplet have a susy breaking mass term. In this paper we compute only the zero mode scalar.  We may use the same expansions of the function in square brackets (Appendix \ref{sec:cterms}) we used for the sfermion masses.  In general we can define 
\be m_{H_{0}}^2 = 2 \sum_{r =1}^3 C_{H}^r  \ \left(\frac{\alpha_r}{4\pi}\right)^2  \Lambda^2_{H_{0}}.
\label{beta}
\ee
$C_{H}^r $ is the Casimir of the representation $r$ of the positive parity hypermultiplet $H$.  We now look at the various limits. 
\subsubsection{Small $\ell$}
In the small $\ell$ limit we start with \refe{secondresult2} and truncating the tower we find 
\begin{equation}
D_r= -\! \!\int\! \frac{d^4p}{ (2\pi)^4}\frac{1}{p^2} \Omega^r(\frac{p^2}{M^2})   .\label{secondresult4}
\end{equation}
This  is exactly the 4d result for sfermion masses found in \cite{Meade:2008wd}.  We may use all the generalised messenger results of \cite{Marques:2009yu}, with the identification
\be
\Lambda^{2}_{H_{0}} = \Lambda^2_{S (4d)}
\ee
where $\Lambda^2_{S (4d)}$ is the $4d$ result found in \cite{Marques:2009yu}.
\subsubsection{Large $\ell$}
When $1/\ell^2$ is smaller than the scales $F$ and $X^{2}$ we may start from \refe{secondresult3}. We know that the function in square brackets is independent of $p$ and can be found in appendix \ref{sec:cterms},  so we may evaluate the $p$ dependent integral independently. The result is 
\be
\Lambda^2_{H_{0}} =\frac{2}{3}\Lambda^2_{S}
\ee
$\Lambda^2_{S}$ is written explicitly in the previous subsection. The intermediate $\ell$ limit can be found by setting $h(x)=1$.  The zero mode fermions of the hypermultiplet receive no susy breaking mass corrections at order $g^4_{5d}$.
\subsection{Casmir Energy}
The Casimir energy can be computed for the generalised messenger sector just as for the sfermion and hyperscalar zero mode mass by using the results of appendix \ref{generalised}. In particular using \refe{casimirenergy} and restricting to the case of minimal gauge mediation and for a single set of fundamental messengers we find
\be
 E_{\text{Cas}}/V_{4d}=- g^2_{4} \frac{d d_{g} \zeta(5)}{512\pi^4} \frac{F^2_{X}}{(X \ell)^4}h(x).\label{vacua}
\ee
This result reproduces exactly the result of \cite{Mirabelli:1997aj}.  

\setcounter{equation}{0}
\section{Chapter Summary}\label{chapter2summary}
We have computed bulk gauge mediated supersymmetry breaking of $\mathcal{N}=1$ supersymmetry from a hidden to a visible brane, using the current correlator techniques of \cite{Meade:2008wd}. We obtained analytic results for both semi-direct and direct coupling of the messengers and susy breaking sector, located on a hidden brane.  

To summarise, we have found that there are two limits, one in which the leading order scalar masses are four dimensional and one in which kk modes are lighter than the scale $M$, in which case we have found the leading order scalar masses are screened.  Overall we have shown that the lightest gaugino mass is unaffected by the extra dimension.  We would like to emphasise one point, the RGE equation for scalar masses in the five dimensional limit will differ from the four dimensional RG equation there will be many more gaugino masses.

In the next chapter will demonstrate how the framework so far constructed, may be applied to model building using Seiberg duality and in particular metastable supersymmetry breaking models of the ISS type.

\setcounter{equation}{0}
\chapter{Dynamical Supersymmetry breaking: ISS on the brane}\label{chapter3}
A particularly popular and natural form of gauge mediated supersymmetry breaking is the construction of ISS \cite{intriligator2006dsb}.  In ISS, supersymmetry is broken non-perturbatively in the electric $SU(N_c)$ description and is metastable.  It is a simple $\mathcal{N}=1$ SQCD model and as a result one may apply Seiberg duality to obtain an effective magnetic description in which supersymmetry breaking can be explored perturbatively.  It is well known that this theory has a signature of light gauginos relative to heavier sfermions and this is seen as an unfortunate drawback. 

In this section we turn to an explicit application of 5D general gauge mediation which will alleviate the above mentioned problem  by screening the leading order scalars masses.  We construct a scenario in which the ISS model   \cite{intriligator2006dsb,Intriligator:2007py} is located on the hidden brane.  We choose to explore supersymmetry perturbatively in the macroscopic (magnetic) variables in the window $N_c+2 \leq N_{f}\leq \frac{3}{2} N_c$.  We have an $\mathcal{N}=1$ SQCD with magnetic gauge group $SU(N)$ and $N_{f}$ flavours where $N=N_f-N_c$.  The superpotential is
\begin{equation}
W_{\text{ISS}}= h\text{Tr}\tilde{\varphi}M\varphi-h \text{Tr}[\mu^{2}M]+ [\text{Deformations}]
\end{equation}
The magnetic meson $M$ is a gauge singlet and an adjoint of the flavour group. The magnetic quarks $\varphi$ and $\tilde{\varphi}$ are fundamental (antifundamental) of the gauge group and antifundamental (fundamental) of the flavour group, respectively.  Supersymmetry is broken by rank condition when $N_{f}>N$, as only the first $N$  F terms of $M$ can be set to zero. Following the model explored in \cite{Kitano:2006xg,Zur:2008zg,Anguelova:2007at,Koschade:2009qu}, the matrix $\mu$ is explicitly broken
\begin{equation}
\mu^2_{AB} = \left(
\begin{array}{cc}
m^2 \mathbb{I}_{N} & 0  \\
0 & \mu^2 \mathbb{I}_{N_F-N} 
\end{array} \right)_{A B}
\label{t}
\end{equation}
with $\mu<m$, and  additionally we include the deformation
\be 
\delta W= h^2 m_{z} \text{Tr} \tilde{Z}Z.
\ee 
As mentioned mentioned more fully in \cite{Kitano:2006xg,Zur:2008zg,Anguelova:2007at,Koschade:2009qu}, this deformation explicitly breaks R-symmetry thus allowing gaugino masses.   The final unbroken vacuum symmetry groups and matter content is 
\begin{center}
\begin{tabular}{|ccc|}
\hline
Field & $SU(N)_{D} $& $SU(N_{f}-N)_{f} $\\
\hline
$M = \left(
\begin{array}{cc}
Y_{\text{{\tiny $N$x$N$}}} & Z_{\text{{\tiny $N$x($N_{f}$-$N$)}}}  \\
\tilde{Z}_{\text{{\tiny ($N_{f}$-$N$)x$N$}}} & X_{\text{{\tiny ($N_{f}$-$N$)x($N_{f}$-$N$)}}}
\end{array} \right)_{\text{{\tiny $N_{f}$x$N_{f}$}}} $ &

$\left(
\begin{array}{cc}
\text{$\mathbf{Adj}+1$} & \bar{\square} \\
\square & 1
\end{array} \right) $ &

$\left(
\begin{array}{cc}
1 & \square \\
\bar{\square} & \text{$\mathbf{Adj}+1$}
\end{array} \right)$\\

$\varphi=\left ( \begin{array}{c}
\lambda_{N \times N} \\
\rho_{N_{f}-N\times N}
\end{array} \right)_{\text{\tiny{NfxN}}}$

&$\left (\begin{array}{c}
\text{$\mathbf{Adj}+1$}\\ 
\square \end{array} \right)$ 

& $\left(\begin{array}{c}
1 \\
 \bar{\square}
\end{array} \right)$\\

$\tilde{\varphi}=\left ( \begin{array}{c}
\tilde{\lambda}_{N \times N} \\
\tilde{\rho}_{N\times N_{f}-N }
\end{array} \right)_{\text{\tiny{NxNf}}}$

&$\left (\begin{array}{c}
\text{$\mathbf{Adj}+1$}\\ 
\bar{\square} \end{array} \right)$ 

& $\left(\begin{array}{c}
1 \\
 \square
\end{array} \right)$ \\
\hline 
\end{tabular}
\end{center} 
Where $SU(N)_D$ arose from the magnetic ``colour flavour locking" of $SU(N)_{mag}\times SU(N)_f$.  The superpotential is
\be
W= h\text{Tr}(\tilde{\lambda} Y \lambda + \tilde{\rho}Z\lambda + \tilde{\lambda}\tilde{Z}\rho+\tilde{\rho}X \rho)-h^2 m^2 \text{Tr}Y -h^2 \mu^2 \text{Tr}X+h^2m_{z}\text{Tr}\tilde{Z}Z.\label{super2}
\ee
We have two choices of embeddings: we choose to weakly gauge either of the flavour symmetry groups $SU(N)_{f}$ or $SU(N_{f}-N)_{f}$ and associate it with the gauge group in the bulk. One may choose for simplicity an $SU(5)$ standard model ``parent'' gauge group. The messengers of this model are the $\rho$'s and $Z$'s in either embedding.  Additionally for the embedding of $SU(N_f-N)_f$ the degrees of freedom of $X$ may also contribute as messengers as has been explored in \cite{Koschade:2009qu}. The other fields do not contribute as messengers.  We take these fields to have a canonical K\"ahler potential and all the matter on the MSSM brane to have a canonical K\"ahler potential coupled to this bulk gauge superfield. Classically the potential is 
\be
V_{ISS}=(N_{f}-N)|h^2\mu^4| .
\ee
$\text{Tr}X$ singlet is the Goldstino superfield and the scalar component is a classical modulus and its vacuum expectation value, $X_{0}$ is found by minimising
\begin{equation}
V_{Total}= V_{ISS}+ V_{CW} + V_{Casimir}
\end{equation}
where $V_{CW}$ is the corresponding Coleman-Weinberg potential
\begin{equation}
V_{CW} = \frac{1}{64\pi^{2}}\text{STr}\mathbf{M}^4\text{Log}\frac{\mathbf{M}^{2}}{\Lambda^{2}}=
\frac{1}{64\pi^{2}}(\text{Tr}\mathbf{m}^{4}_{B}\text{Log}\frac{\mathbf{m}_{B}^{2}}{\Lambda^{2}}-
\text{Tr}\mathbf{m}^{4}_{F}\text{Log}\frac{\mathbf{m}_{F}^{2}}{\Lambda^{2}}).
\end{equation}
The Casimir energy is sufficiently suppressed to be ignored at this stage.  We find
\begin{equation}
X_{0}=\braket{X} = \frac{1}{2} h m_z, \; \; \; \; M^{2}_X = \frac{h^{4}\hat{\mu}^{2}}{12\mu^{2}\pi^{2}}\left(
\begin{array}{cc}
\hat{\mu}^{2}&  -\frac{9}{40}{X_0}^{2}  \\
-\frac{9}{40}{X_0}^{2} &\hat{\mu}^{2}
\end{array} \right)
\end{equation}
where we have expanded to first order in $h, m_z$ and in $\hat{\mu} / \mu$ up to first non-vanishing order \cite{Koschade:2009qu}. We have suppressed factors of $N(N_{f}-N)$
in the expression for $M^{2}_X $ coming from tracing over degenerate mass eigenavalues.

\section{Large $\ell$ mediation} 
The susy breaking contribution to the gaugino masses are
\be
m^{n \hat{n}}_{\gl} =\frac{\alpha}{4\pi}\Lambda_{G}=\frac{\alpha}{4\pi}F_{X}\sum_{i}\frac{\partial_{X}M_{i}}{M_{i}} \label{nice}
\ee
where $M_{i}$ are the eigenvalues of the fermion mass matrix of messengers derivable from \refe{super2} and computed in \cite{Zur:2008zg,Koschade:2009qu}.  The messenger sector is dominantly $(\rho,Z)$ in both embeddings of the standard model and we find the susy breaking mass to be
\be 
\Lambda_{G}= \frac{N h^2 \mu^2 m_{z}}{(m^2-hX_{0}m_{z})}.
\ee 
As we have highlighted throughout, the mass eigenstates must be found after inclusion of the kk masses, to the mass matrix.  The sfermion masses can be found, using the results of the previous section, from
\be
m^{2}_{\tilde{f}} =2 C_{\tilde{f}}(\frac{\alpha}{4\pi})^{2}\Lambda^{2}_{S}
\ee
with
\be
\Lambda_{S}^{2}= \frac{N \zeta(3)}{\ell^2} |F_{X}|^{2} \sum_{i}|\frac{\partial_{X}M_{i}}{M^{2}_{i}}|^{2}\label{moremasses}
\ee
We find
\bea
\Lambda_{s}^{2}=& \frac{\zeta(3)N \mu^4}{\ell^{2}} \times \\&
\frac{[2m^6 \!+\! 2h^3 m^2 m_{z}^{3}(3hm_{z}-2X)\!+\! h^4 m^{4}_{z}(X_{0}-hm_{z})^{2}+m^4(9h^2m^2_{z}+2hm_{z} X_{0}\!+\! X^2_{0})]}{(m^2-hm_{z}X_{0})^4(4m^2+(X_{0}-hm_{z})^2)} \nonumber
\eea
Which is non-zero but highly suppressed. 

\section{Intermediate $\ell$ mediation} 
In the intermediate range that $F\leq 1/\ell^2\ll M^2$ the vacuum energy is given by \refe{vacua}, with $h(x)=1$ This time, however, the zero mode gaugino mass \emph{can} be approximated by \refe{nice}: 
\be 
m^{0}_{\gl}= \frac{N \alpha h^2 \mu^2 m_{z}}{4\pi(m^2-hX_{0}m_{z})}.
\ee
and there is no problem of kk mixing.  The sfermion masses are still given by \refe{moremasses}. As a first approximation, one may ignore this contribution to sfermion masses entirely.  One then lets the renormalisation group flow at the high scale down to the low scale generate sufficiently heavy sfermions to avoid current bounds.
\section{The Emergent Lattice}
In the previous section we placed the ISS model on the fixed point of the extra dimension and the theory is truly five dimensional.  In this section we will demonstrate how an effective extra dimension may emerge in the magnetic description of certain classes of SQCD theories and will motivate a lattice (de)construction of the general gauge mediation formalism.    

First a crucial comment is necessary. Under Seiberg duality global symmetries are preserved. This must be so as global symmetries are associated with observable currents and so must be the same under the duality.  The gauge symmetries on the other hand, are a redundancy in the description.  One may ask how it is possible that two theories may have differing number of gauge bosons and gauginos.  Importantly the two theories are at a scale invariant fixed point, where there are no asympotically free states and no well defined particle interpretation.  We can of course deform the model by adding quark masses to the electric description $W=m\text{Tr}[Q.\tilde{Q}]$ and now particle states are well defined and the duality becomes an \emph{effective} one.  In fact it is curious to ask what becomes of the gauge bosons of the $SU(N_f-N_c)$ magnetic group?  The answer lies in that the dual of the quark mass matrix completely Higgs' the magnetic gauge group and those degrees of freedom become vector mesons.  As a result in terms of the low energy effective description the gauge fields are now observable as vector mesons.  

A further insight is necessary. If we weakly gauge a global symmetry, then additional massless vector fields arise which are not related to the Seiberg duality.  One can think of these fields as external ``probes'' of the SQCD theory.  However, if the magnetic $SU(N_f-N_c)$ group, or some subgroup of this, undergoes a ``magnetic colour-flavour locking" with the weakly gauged global symmetry, then the vector fields will appear to have massless modes and Kaluza-Klein modes from the vector mesons:  a deconstructed extra dimension emerges.  From the perspective of the electric description, resonances have filled the first level of a Kaluza-Klein tower on top of the massless external fields.  From the perspective of the magnetic description, this can be seen to take place dynamically as the system undergoes the locking. 

The remainder of this chapter is essentially a review of \cite{Green:2010ww}, which demonstrates how the locking arises dynamically, which we feel is crucial in developing many of the concepts in the remainder of this thesis and as a building block in the AdS/CFT construction of chapter \ref{chapter6}.  

In \cite{Green:2010ww}  An electric SQCD hidden sector was chosen 
\begin{center}
\begin{tabular}[c]{|ccccc|} \hline
Field & $SU(N_c) $&$SU(k)_f$&$SU(N_{f}-k)_f$& $U(1)'$ \\
\hline
$Q=\binom{Q^a}{Q^i} $ &  $\square $&$\binom{1}{\square} $&$\binom{\square}{1}$& 1\\
$\tilde{Q}=(\tilde{Q}_a,\tilde{Q}_i)$ &$\overline{\square } $&$(\overline{\square },1) $& $(1,\overline{\square })$ &-1\\
$S^a_i$ &1&$\overline{\square } $& $\square $ &-1\\
$\tilde{S}_a^i$ &1&$\square  $& $\overline{\square }$ &1
\\ \hline
\end{tabular}
\end{center}
where the indices $i,j= 1,...,k$ and $a,b =k,...,N_f$, with the superpotential 
\be
W_{electric}= m^J_I Q^I \tilde{Q}_{J} + S^a_i Q^i \tilde{Q}_a+ \tilde{S}_a^i Q^a \tilde{Q}_i
\ee
The factor $h=\Lambda_e/\lambda$ of the previous section has been set to one for simplicity.  The flavour symmetires may be thought of as arising from  $SU(N_f)_L\times SU(N_f)_R\hookrightarrow SU(N_f)_{diag}$ after the introduction of the quark mass term, which is then further broken to $SU(N_f)_{diag} \hookrightarrow SU(k)_f \times SU(N_{f}-k)_f$ after the introduction of the $S^a_i Q^i \tilde{Q}_a$ operators.   The flavour symmetry preserving masses are $m_{I}^J =m_1 \delta^j_i \oplus  m_{2} \delta^b_a$. 

In the window $N_2+2\leq N_f \leq \frac{3}{2}N_c$, this theory flows to an IR free magnetric Seiberg dual description with gauge group $SU(N_f-N_c)=SU(N)$ and UV cutoff $\Lambda$. The superpotential is given by 
\be
W_{mag}=q \Phi \tilde{q}+ \Lambda S\tilde{T} + \Lambda \tilde{S}T+ \Lambda \text{Tr}(m\Phi)
\ee
The representations of the magnetic description are given by 
\begin{center}
\begin{tabular}[c]{|cccc|} \hline
Field & $SU(N_f-N_c)_{mag} $&$SU(k)_{\chi}$&$SU(N_{f}-k)$ \\
\hline
$\Phi = \left(
\begin{array}{cc}
N & \tilde{T}^i_a \\
T_i^a & M
\end{array} \right)_{\text{{\tiny $N_{f}$x$N_{f}$}}} $ & $1$&
$\left(
\begin{array}{cc}
\text{$\mathbf{Adj}+1$} & \square \\
\bar{\square} & 1
\end{array} \right) $&$\left(
\begin{array}{cc}
1 & \bar{\square} \\
\square & \text{$\mathbf{Adj}+1$}
\end{array} \right) $
\\
$q=\binom{\chi_i}{\psi_a}$ &$\square  $&$\binom{\overline{\square }}{1} $& $\binom{1}{\overline{\square }}$ \\
$\tilde{q}=(\tilde{\chi}^i,\tilde{\psi}^a)$ &$\overline{\square} $&$(\square ,1) $& $(1,\square )$ \\
$S^a_i$ &1&$\overline{\square } $& $\square $ \\
$\tilde{S}_a^i$ &1&$\square  $& $\overline{\square }$ 
\\ \hline
\end{tabular}
\end{center}
where we have chosen subscript labellings of the groups to make clearer the patterns of breaking.  After integrating out the heavy states using the equations of motion of the $S$'s, one obtains
\be
W= \chi N \tilde{\chi}+ \psi M\tilde{\psi} -\mu^2_{1}N^i_i- \mu^2_{2}M^a_a
\ee
where $\mu^2_1,2=-m_1,2\Lambda$.  The theory now has two sectors which in principle decouple when $g_{mag} \rightarrow 0$ but more importantly the MSSM completely decouples when $g_{MSSM}\rightarrow 0$

Next, we proceed to explore the vacuum symmetries of the theory.  If we choose for simplicity $k=N_f-N_c=N$ then we see that all $\chi$'s are will be Higgsed as that F terms of $N^i_i$ can all be set to zero:
\be 
\braket{\chi \tilde{\chi}}=\mu_1^2 \mathbb{I}_N
\ee
Using the remaining symmetries, one may further diagonalise the $\chi$'s to the vacuum 
\be
\braket{\chi}=\mu_1  \mathbb{I}_N \ \ \ \braket{\tilde{\chi}}=\mu_1  \mathbb{I}_N 
\ee
Crucially, these $\chi$ fields will play the role of linking chiral superfields in a deconstructed model and they link the $SU(N)_{mag}$ lattice site with the $SU(k)_\chi$ flavour site.  

 In the second sector $M$ has maximum rank $N_f-k=N_c$ however the matrix $(\psi .\tilde{\psi})$ has rank $N$.   For $N_c > N$  supersymmetry is broken by rank condition as only the first $N$ F terms may be set to zero.  Hence the global symmetries will break further from $SU(N_f-k)\hookrightarrow SU(N_f-k-N)\times SU(N)_\lambda $ and one finds
\begin{center}
\begin{tabular}[c]{|cccc|} \hline
Field & $SU(N_f-N_c)_{mag} $&$SU(N_{f}-k-N)_{\rho}$&$SU(N)_{\lambda}$ \\
\hline
$M = \left(
\begin{array}{cc}
Y & Z \\
\tilde{Z} & X
\end{array} \right) $ & $1$&
$\left(
\begin{array}{cc}
\text{$\mathbf{Adj}+1$} & \square \\
\bar{\square} & 1
\end{array} \right) $&$\left(
\begin{array}{cc}
1 & \bar{\square} \\
\square & \text{$\mathbf{Adj}+1$}
\end{array} \right) $
\\
$\psi=\binom{\lambda}{\rho}$ &$\square  $& $\binom{1}{\overline{\square }}$&$\binom{\overline{\square }}{1} $ \\
$\tilde{\psi}=(\tilde{\lambda},\tilde{\rho})$ &$\overline{\square} $& $(1,\square )$ &$(\square ,1) $
\\ \hline
\end{tabular}
\end{center}
where $\braket{\lambda}=\mu_2 \mathbb{I}_N$ , $\braket{\tilde{\lambda}}=\mu_2 \mathbb{I}_N$, $\braket{Y}=0$.  As the flavour symmetries are broken we may choose different masses identify $\mu_3 \mathbb{I}_{N_{f}-k-N}$ as the unhiggsed part of the original $\mu_2 \mathbb{I}_{N_f-k}$ masses.  The classical vacuum is now given by 
\be
V= (N_f-k-N)|\mu_3^4| 
\ee
The $\text{Tr}X$ singlet is the Goldstino superfield of spontaneous supersymmetry breaking. As the fermionic component is a massless Goldstino, the scalar is a pseudomodulus.  The supersymmetry breaking sector has the symmetry groups $SU(N_f-N_c)_{mag}\times SU(N)_{\lambda}\hookrightarrow SU(N)_{D}$ times the unaltered flavour group $SU(N_{f}-k-N)_{\rho}$. The supersymmetry breaking sector has an approximate R-symmetry and one may choose different deformations such as the $m_{z}\text{Tr}\tilde{Z}Z$ demonstrated in the previous section.  The sector of messengers and supersymmetry breaking field has a similar structure to the one analysed in the previous section, althought the rank conditions arise differently.  In this model the  $SU(k)_\chi$ flavour group must be weakly gauged and associated with the standard model.  This weakly gauged group then shares properties with the visible sector fixed point of the previous model. Similarly the hidden brane is equivalent to the $SU(N_F-N_c)_{mag}$ lattice site.  As discussed before, this model is crucial for developing a dynamical description of how a system undergoes the  ``magnetic colour-flavour locking'' and vector mesons fill up as kk modes above the massless external ``probe'' gauge fields of the weakly gauged global symmetry.  This strongly motivates the AdS/CFT construction of chapter \ref{chapter6} whereby a weakly gauged global symmetry of the UV CFT description arises as a gauge symmetry in the bulk AdS description.

Many questions still remain.  It would be very interesting to see if one can populate more levels of the Kaluza-Klein tower dynamically using Seiberg duality and finally arise at a contintum limit of AdS space, perhaps starting from the conformal window with large $N_c$.  This also strongly motivates a study of observing these kk modes at the LHC and furthermore by interpreting them as resonances of a strongly coupled hidden sector, extract from them properties of the hidden sector, such as their internal degrees of freedom as UV quarks, and their quantum numbers under the $SU(N_c)$ UV gauge group.  Some work in this direction has already been completed \cite{Contino:2006nn,Kilic:2008pm,Kilic:2009mi}.


\newcommand{\drawsquare}[2]{\hbox{%
\rule{#2pt}{#1pt}\hskip-#2pt
\rule{#1pt}{#2pt}\hskip-#1pt
\rule[#1pt]{#1pt}{#2pt}}\rule[#1pt]{#2pt}{#2pt}\hskip-#2pt
\rule{#2pt}{#1pt}}

\newcommand{\sfrac}[2]{{\textstyle\frac{#1}{#2}}}
\def\tv#1{\vrule height #1pt depth 5pt width 0pt}

\newcommand{\Yfund}{\raisebox{-.5pt}{\drawsquare{6.5}{0.4}}}
\newcommand{\Ysymm}{\raisebox{-.5pt}{\drawsquare{6.5}{0.4}}\hskip-0.4pt%
        \raisebox{-.5pt}{\drawsquare{6.5}{0.4}}}
\newcommand{\Ythrees}{\raisebox{-.5pt}{\drawsquare{6.5}{0.4}}\hskip-0.4pt%
          \raisebox{-.5pt}{\drawsquare{6.5}{0.4}}\hskip-0.4pt%
          \raisebox{-.5pt}{\drawsquare{6.5}{0.4}}}
\newcommand{\Yfours}{\raisebox{-.5pt}{\drawsquare{6.5}{0.4}}\hskip-0.4pt%
          \raisebox{-.5pt}{\drawsquare{6.5}{0.4}}\hskip-0.4pt%
          \raisebox{-.5pt}{\drawsquare{6.5}{0.4}}\hskip-0.4pt%
          \raisebox{-.5pt}{\drawsquare{6.5}{0.4}}}
\newcommand{\Yasymm}{\raisebox{-3.5pt}{\drawsquare{6.5}{0.4}}\hskip-6.9pt%
        \raisebox{3pt}{\drawsquare{6.5}{0.4}}}
\newcommand{\Ythreea}{\raisebox{-3.5pt}{\drawsquare{6.5}{0.4}}\hskip-6.9pt%
        \raisebox{3pt}{\drawsquare{6.5}{0.4}}\hskip-6.9pt
        \raisebox{9.5pt}{\drawsquare{6.5}{0.4}}}
\newcommand{\Yfoura}{\raisebox{-3.5pt}{\drawsquare{6.5}{0.4}}\hskip-6.9pt%
        \raisebox{3pt}{\drawsquare{6.5}{0.4}}\hskip-6.9pt
        \raisebox{9.5pt}{\drawsquare{6.5}{0.4}}\hskip-6.9pt
        \raisebox{16pt}{\drawsquare{6.5}{0.4}}}
\newcommand{\Yadjoint}{\raisebox{-3.5pt}{\drawsquare{6.5}{0.4}}\hskip-6.9pt%
        \raisebox{3pt}{\drawsquare{6.5}{0.4}}\hskip-0.4pt
        \raisebox{3pt}{\drawsquare{6.5}{0.4}}}
\newcommand{\Ysquare}{\raisebox{-3.5pt}{\drawsquare{6.5}{0.4}}\hskip-0.4pt%
        \raisebox{-3.5pt}{\drawsquare{6.5}{0.4}}\hskip-13.4pt%
        \raisebox{3pt}{\drawsquare{6.5}{0.4}}\hskip-0.4pt%
        \raisebox{3pt}{\drawsquare{6.5}{0.4}}}
\newcommand{\Yflavor}{\Yfund + \overline{\Yfund}} 
\newcommand{\Yoneoone}{\raisebox{-3.5pt}{\drawsquare{6.5}{0.4}}\hskip-6.9pt%
        \raisebox{3pt}{\drawsquare{6.5}{0.4}}\hskip-6.9pt%
        \raisebox{9.5pt}{\drawsquare{6.5}{0.4}}\hskip-0.4pt%
        \raisebox{9.5pt}{\drawsquare{6.5}{0.4}}}%

\setcounter{equation}{0}
\chapter{General Gauge Mediation and Deconstruction}\label{chapter4}
In this chapter we locate a supersymmetry breaking hidden sector and supersymmetric standard model on different lattice points of a quiver gauge diagram \cite{Hill:2000mu,ArkaniHamed:2001ca,ArkaniHamed:2001nc,Cheng:2001vd,Cheng:2001nh,Kunszt:2004ps,DiNapoli:2006kc}.  The hidden sector is encoded in a set of current correlators and the effects of the current correlators are mediated by the lattice site gauge groups with ``lattice hopping'' functions and through the bifundamental matter that links the lattice sites together.  We show how the gaugino mass, scalar mass and Casimir energy of the lattice can be computed for a general set of current correlators and then give specific formulas when the hidden sector is specified to be a generalised messenger sector. The results reproduce the effect of five dimensional gauge mediation of chapter \ref{chapter2} from a purely four dimensional construction  \cite{Csaki:2001em}.

In this construction each lattice site is a copy of a four dimensional supersymmetric standard model parent gauge group $SU(5)$. Each lattice site is connected using bifundamental chiral superfields that link the lattice together.  The combination of super Yang-Mills with specified ``lattice hopping" vectors and the bifundamental linking matter, will mediate the supersymmetry breaking effects from the hidden sector lattice site to the standard model lattice site, to generate sfermion masses \cite{Csaki:2001em,Cheng:2001an}.  In this model, the lattice site spacing $a$ will be the scale that suppresses the momentum in the loops for sfermion masses. We shall see that effective five dimensional suppression effects arise when $\frac{1}{(Na)^2}\ll M^2 $ and that we recover four dimensional effects when $M^2 \ll \frac{1}{(Na)^2}$. At low energies this lattice construction is equivalent to the five dimensional orbifold model mentioned above.  The scale $a$  arises in the masses of the propagating gauge fields and bifundamental matter which resemble KK states at low energies.  The masses of the KK spectrum are generated by the vacuum expectation value of the scalars of the bifundamental chiral superfields that link the lattice together and so we see that these suppression methods are all related.
\setcounter{equation}{0}
\section{Framework}\label{section:Frameworkdecon}
This section concisely reviews the construction of the orbifold moose (quiver gauge diagram) following the work of \cite{Csaki:2001em}. We start with a lattice of four dimensional, $SU(5)_{i}$ super Yang-Mills gauge groups all identified with the gauge groups of the supersymmetric standard model.  The matter content of the lattice is:
\begin{equation}
 \begin{array}{c|cccccccc}
        & SU(5)_0 & SU(5)_1 & \cdots & SU(5)_{N-2} & SU(5)_{N-1} \\ \hline
\tv{15} \tilde{P}_1, \ldots, \tilde{P}_5
        & \overline{\Yfund} & 1 & \cdots & 1 & 1 \\
           Q_1 & \Yfund  & \overline{\Yfund} & \cdots & 1 & 1 \\
        \vdots & \vdots & \vdots & \ddots & \vdots & \vdots \\
         Q_{N-1} & 1 & 1 & \cdots & \Yfund & \overline{\Yfund} \\
P_1, \ldots, P_5 & 1 & 1 & \cdots & 1 & \Yfund \\
\overline{\bf 5}_{1,2,3} & 1 & 1 & \cdots & 1 & \overline{\Yfund} \\
{\bf 10}_{1,2,3} & 1 & 1 & \cdots & 1 & \Yasymm \\
  H_d & 1 & 1 & \cdots & 1 & \overline{\Yfund} \\
  H_u & 1 & 1 & \cdots & 1 & \Yfund \\
\Phi & \Yfund & 1 & \cdots & 1 &  1\\ 
 \tilde{\Phi} & \overline{\Yfund} & 1 & \cdots & 1 &  1\\
  X & 1 & 1 & \cdots & 1 & 1 \\
 \end{array} \nonumber 
\end{equation}
The $Q_{i \alpha}^{\phantom{\beta}\beta}$ are bifundamental chiral superfields that link the $SU(M)^N$ lattice sites together. The $\alpha,\beta$ are gauge indices labelling the fundamental or antifundamental of the gauge group $SU(M)$.  The bifundamental scalars all obtain a vacuum expectation value $v$, which may be generated by some dynamical superpotential \cite{ArkaniHamed:2001ca,Csaki:2001em}.  The indices $i,j,k$ label the lattice, running from $i=0$ to $N-1$ (mod $N$) with lattice spacing $a=1/(\sqrt{2}g v)$, such that $\ell=N a$ is the length of the lattice. The five chiral multiplets $\tilde{P}$ and five $P$ are localised on the endpoint lattice sites and are required to cancel anomalies due to the breaking of ``hopping" symmetry at the lattice end points.  These are inessential to the discussion of gauge mediation.  The $5,10$ and $H_{u},H_{d}$ play the role of the standard model matter and Higgs superfields. Additionally, we may add the fields $\Phi,\tilde{\Phi}$ as messenger superfields coupled to a spurion $X=\braket{X}+\theta^2 F$. These fields will enter the discussion when we specify a generalised hidden sector in a later section. 

The bifundamental scalars are given a vacuum expectation value and fluctuations about that value, $Q_{i \alpha}^{\phantom{\beta}\beta}=v \delta_{\alpha}^{\phantom{\beta}\beta}+ \phi_{i \alpha}^{\phantom{\beta}\beta}$. The scalar kinetic terms of the bifundamental matter are used to generate a mass matrix for the gauge bosons, via the Higgs mechanism. The mass spectrum is computed in \cite{Csaki:2001em} and in \cite{ArkaniHamed:2001ca,Hill:2000mu,Cheng:2001vd}.   For the gauge bosons, the masses are 
\be
m^2_{k}=8g^2 v^2 \sin^2 (\frac{k\pi}{2N}) \phantom{AAAAA} k=0, ..., N-1.
\ee
A key attribute of this setup is that lattice eigenstates are not mass eigenstates of the system.  
The mass eigenstates are given by 
\be
\tilde{A}_{k}=\sqrt{\frac{2}{2^{\delta_{k0}}N}}\sum_{j=0}^{N-1} \cos \frac{(2j+1)k\pi}{2N}A_{j} \phantom{AAAAA} k=0, ..., N-1. 
\ee
These are even parity modes.    The fermion mass matrix of bifundamental fermions $q_{i}$ and gauginos $\gl_{i}$ must be diagonalised as in \cite{Csaki:2001em}. The even states  $\gl_{i}$ have masses
\be
 m^2_{k}=8g^2 v^2 \sin^2 (\frac{k\pi}{2N}) \phantom{AAAAA} k=0, ..., N-1
\ee
and eigenvectors
\be
\tilde{\gl}^{+}_{k}=\sqrt{\frac{2}{2^{\delta_{k0}}N}}\sum_{j=0}^{N-1} \cos \frac{(2j+1)k\pi}{2N} \gl_{j} \phantom{AAAAA} k=0, ..., N-1. 
\ee 
The odd parity fermions $q_{i}$ have masses
\be
 m^2_{k}=8g^2 v^2 \sin^2 (\frac{k\pi}{2N}) \phantom{AAAAA} k=1, ..., N-1
\ee
with eigenstates 
\be
\tilde{\gl}^{-}_{k}=\sqrt{\frac{2}{N}}\sum_{j=1}^{N-1} \sin \frac{j k\pi}{2N} q_{j} \phantom{AAAAA} k=1, ..., N-1. 
\ee
In the continuum limit, the fermions $q_{i}$ should coincide with the adjoint fermion $\chi$  of a negative parity chiral superfield.  To construct this adjoint fermion one identifies the bifundamental scalar with the unitary link variable: $Q_{\alpha i}^{\phantom{\beta}\beta}=v U_{ \alpha i}^{\phantom{\beta}\beta}$ where $U$ is a unitary matrix. Keeping track of indices  we may relabel $(U_{\alpha}^{\phantom{\beta} \beta})_{i}^{\dagger}q_{\gamma}^{\phantom{\beta}\beta}=\chi_{\gamma i}^{\phantom{\beta}\alpha}$, where both indices $\alpha,\gamma$ are valued at the $i$'th lattice site. A negative parity adjoint scalar $\Sigma$ is similarly defined and as the multiplet is supersymmetric, the mass spectrum and eigenstates are equivalent to that of the fermion.  

Using the above eigenfunctions, the mixed space scalar propagator can readily be determined by insertion of a closure relation for the mass eigenstates.   The result is
\be
\braket{p^2;k,l} =\frac{2}{N}\sum_{j=0}^{N-1}\frac{1}{2^{\delta_{j0}}}\cos (\frac{(2k+1)j\pi}{2N})\cos (\frac{(2l+1)j\pi}{2N})\frac{1}{p^2+(\frac{2}{a})^2\sin^2 (\frac{j\pi}{2N})}.\label{propagator1}
\ee
In summary, the resulting low energy degrees of freedom and field content for large $N$ is that of an $\mathcal{N}=1$ positive parity vector multiplet and negative parity chiral superfield of $\mathcal{N}=1$ super Yang-Mills in five dimensions compactified on $R^{1,3}\times S^1/\mathbb{Z}_{2}$ \cite{Hebecker:2001ke,McGarrie:2010kh}.

\subsection{The periodic lattice}
It is useful to compare this construction with that of a periodic lattice corresponding to a $5d$ theory compactified on a circle \cite{Csaki:2001em}. 
The anomaly cancelling $P$ and $\tilde{P}$ fields are unwanted in the periodic construction. All the bulk fields have a mass spectrum given by
\be
m^2_{k}=8g^2 v^2 \sin^2 (\frac{k\pi}{N}) \phantom{AAAAA} k=0, ..., N-1.
\ee
The vector superfield mass eigenstates are also related to the lattice eigenstates through
\be
\tilde{V}_{k}=\frac{1}{\sqrt{N}}\sum_{j=0}^{N-1} e^{i(2\pi k j)/N}V_{j} \phantom{AAAAA} k=0, ..., N-1.
\ee
The mixed space scalar propagator for the circle may also be written as
\be
\braket{p^2;k,l} =\frac{1}{N}\sum_{j=0}^{N-1}e^{-i(2\pi k j)/N}e^{i(2\pi \ell j)/N}\frac{1}{p^2+m^2_{j}}.\label{propagator2}
\ee
The low energy degrees of freedom are of a full $4d$ $\mathcal{N}=2$ model, including restoring the zero modes that had been projected out by negative parity, in the interval case. 
\setcounter{equation}{0}
\section{Lattice localised currents} \label{section:latticecurrents}
This section will encode a SUSY breaking sector, localised on the lattice site i=0 in terms of current correlators \cite{Meade:2008wd}.  The lattice has a set of vector superfields $\{V_{i}\}$ and a set of current multiplets $\{\mathcal{J}_{i}\}$ associated to the bifundamental matter linking the lattice\footnote{In this discussion we will ignore the currents of the fields $P_{i}$ and $\tilde{P}_{i}$.}.  The supersymmetric standard model matter (visible sector) will form a current multiplet $\mathcal{J}^v_{N-1}$. Additionally, we may locate a hidden sector at lattice point $i=0$, $\mathcal{J}^{h}_{0}$. 

We may couple the hidden sector current multiplet to the lattice gauge fields
\be 
S_{int}=2g\!\int\! d^4 x  d^{4}\theta \mathcal{J}^h_{0}
\mathcal{V}_{0}= g \int d^4 x (JD_{0}- \gl_{0} j \!-
 \!\bar{\gl}_{0} \bar{j}-j_{\mu}A^{\mu}_{0})\ee
As this whole discussion refers to the hidden sector current multiplet $\mathcal{J}^{h}_{0}$ at $i=0$, we will drop the hidden sector index.  The change of the effective Lagrangian on the i=0 lattice site to
$O(g^{2})$ is
\begin{align}  
\delta \mathcal{L}_{eff}= &-g^{2} \tilde{C}_{1/2}(0) i \lambda_{0} \sigma^{\mu} \partial_{\mu} \bar{\lambda}_{0}
- g^{2}\frac {1} {4} \tilde{C}_1(0) F_{0 \mu\nu} F^{\mu\nu}_{0} \nonumber + \frac{1}{2}g^2 \tilde{C}_{0}(0)D^2_{0}\\
&-g^{2}\frac {1}{ 2}(M \tilde{B}_{1/2}(0) \lambda_{0} \lambda_{0} + M \tilde{B}_{1/2}(0)\bar{\gl}_{0}\bar{\gl}_{0}).\nonumber
\end{align}
In the next subsections we will use this effective action  to determine soft mass formulas and the Casimir energy of the lattice.
\subsection{Gaugino masses}
A soft supersymmetry breaking gaugino mass for the $i=0$ lattice sites arises at tree level. It is given by 
\be 
m_{\gl, i=0}= g^2 M\tilde{B}_{1/2}(0) \label{gauginomass1}
\ee
In the continuum orbifold scenario \cite{McGarrie:2010kh}, the orbifold fixed points break Lorentz invariance in the fifth dimension and this is signified by the ``brane''\footnote{The word ``brane'' being used to denote the ends of an interval, where matter is located in five dimensional orbifold constructions.} localised current correlators not preserving incoming and outgoing $p_{5}$ momenta.  These current correlators therefore couple, equally, to all states of the K.K. tower of vector superfields.  In the lattice picture the currents only coupled to fields at a single lattice site.  However it is precisely because the lattice fields are a sum of mass eigenstates, that the current correlator still generates a correction to all mass eigenstates in the lattice picture. The soft term mass must be included in the full mass matrix of all fermions ($\gl_{i}, q_{i}$) in the lattice and if we assume that $m_{\gl, i=0}$ is small, we may treat this as a perturbation of the full mass matrix.  This process is outlined in \cite{Csaki:2001em} and one finds the zero mode mass is
\be
 m_{0}= \frac{g^2}{N} M\tilde{B}_{1/2}(0).
\ee
The four dimensional coupling is determined from $g^2_{4d}=g^2/N$.  The mass splittings are also similarly obtained
\be
m^2_{k} =4 g^2 v^2 (\sin \frac{k\pi}{2N}\pm \frac{g}{\sqrt{2}vN}M\tilde{B}_{1/2}(0)\cos^2 \frac{k\pi}{2N})\sin\frac{k\pi}{2N}, \phantom{AAAA} k=1,...,N-1.
\ee
Heuristically, we see that it is the process of moving from lattice states to mass eigenstates that reproduces the orbifold effect of a soft mass coupling to all K.K. modes.
\subsection{Sfermion masses}\label{latticesfermionmasses}
\begin{figure}[ht]
\centering
\includegraphics[scale=0.8]{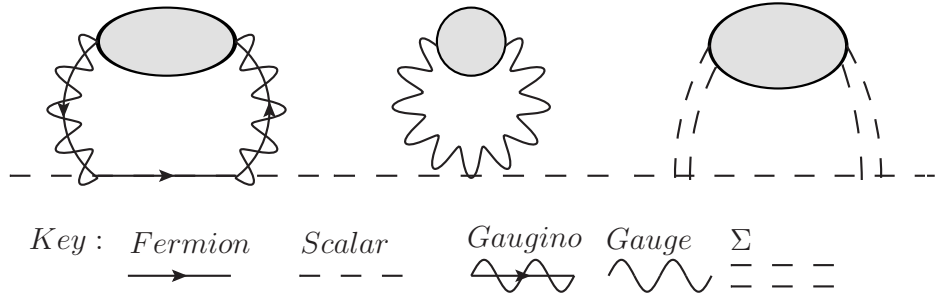}
\caption{The graphical description of the contributions of the two point functions
to the sfermion soft masses.  The ``blobs'' represent hidden sector current correlators on the lattice site $i=0$.  From left to right the diagrams represent the correlator $\braket{j_\alpha \bar j_{\dot\alpha}}$ mediated by the ``lattice hopping" gauge bosons, $\braket{j_\mu j_\nu }$ by the gaugino and  $\braket{JJ}$ by the adjoint scalar built from the bifundamental linking scalars fields.  The scalar and fermion lines at the bottom are located at the lattice site $i=N-1$.}
\label{sfermion3}
\end{figure}
We would like to propagate the effects of supersymmetry breaking from the $i=0$ lattice site to the $i=N-1$ site to generate scalar masses for the MSSM located at that site.  This is a loop diagram with intermediate gauge boson, gaugino and $D_{i}$ term as can be seen in figure \ref{sfermion3}.   The gauge boson and gaugino are dynamical and they may propagate around the lattice using the ``lattice hopping'' wavefunction.  The lattice scalar propagator \refe{propagator1} from the hidden sector lattice site to visible lattice site is 
\be 
\braket{q; 0,N-1}=\frac{2}{N}\sum_{j=0}^{N-1}\frac{1}{2^{\delta_{j0}}}\frac{(-1)^{j}\cos^{2}{\frac{j\pi}{2N}}}{q^{2}+(\frac{2}{a})^{2}{\sin^2{\frac{j\pi}{2N}}}} =a^2\prod_{j=0}^{N-1}\frac{1}{(aq)^{2}+4\sin^{2}{\frac{j\pi}{2N}}} .
\ee 
For the periodic lattic case the same final equation holds, after adjusting for the mass eigenstates by $2N\rightarrow N$ in the denominator.  We highlight that from interval to interval fixed point propagators, the factor $(-1)^{j} $ arises.  This factor is crucial in the cancellation of alternate states of the K.K. tower which generates the suppression of sfermion masses at large momenta and keeps ``brane to brane'' diagrams UV finite.  This factor does not arise in ``brane'' to same ``brane'' diagrams, which is why the vacuum energy diagram is divergent.  Equally, diagrams with a double insertion of the gaugino mass will also be UV divergent.  To generate a gaugino or gauge boson propagator we supplement this scalar propagator with the correct Lorentz structure of a fermion, $\sigma_{\alpha\dot\alpha}^\mu \partial_\mu$ or transverse gauge projector $P^{\mu\nu}= (\partial^2\eta^{\mu\nu}-\partial^\mu \partial^\nu)/\partial^2$ as necessary.  The remaining diagram propagated by $\Sigma$, is not so simple. In orbifold models \cite{McGarrie:2010kh}, the $D$ term is given by $D= (\partial_{5}\Sigma+iX^3)$, where $\Sigma$ is a dynamical negative parity scalar field and $\partial_{5}$ is a derivative in the fifth direction. This field contributes to the propagation of supersymmetry across the interval.  To make the $D_{i=0}$ dynamical on the lattice, one may integrate out this auxiliary field in terms of the lattice currents and find 
\be 
D_{i}=gJ_{i}= g \text{tr}((Q_{\alpha}^{\beta})^{\dagger}_{i}T^{a \gamma}_{\alpha} Q_{\gamma i}^{\phantom{\beta}\beta}-Q^{\phantom{\beta}\beta}_{\alpha i-1}T_{\beta}^{a\gamma} (Q_{\alpha}^{\phantom{\gamma}\gamma })^{\dagger}_{i-1}).
\ee
The index $a$ is a generator index running from $1$ to $N^2-1$.  Next, using $Q_{\alpha i}^{\phantom{\beta}\beta}=v U_{ \alpha i}^{\phantom{\beta}\beta}$ where $U$ is a unitary matrix and keeping track of the indices of each gauge group one finds 
\be 
D_{i}= \frac{1}{\sqrt{2}a}(T^a \Sigma_{i}-\Sigma_{i-1} T^a )+...
\ee
where we have relabelled $(U_{\alpha}^{\phantom{\beta} \beta})_{i}^{\dagger}Q_{\gamma}^{\phantom{\beta}\beta}=\Sigma_{\gamma i}^{\phantom{\beta}\alpha}$.  The contraction of indices has resulted in an adjoint scalar $\Sigma$ of the $i$'th lattice site and both indices $\alpha,\gamma$ are valued at the same lattice site. In the continuum limit, the $D$ term is a lattice derivative  of $\Sigma$ \cite{Bauer:2003mh}:
\be
\frac{1}{\sqrt{2}}\partial_{5}\Sigma = \frac{1}{\sqrt{2}} \lim_{a\rightarrow 0}\frac{\Sigma(y+a)-\Sigma(y)}{a}.
\ee
We may now associate $\Sigma$ with the negative parity scalar of the $5D$ $\mathcal{N}=1$ super Yang-Mills action.  An additional manipulation used to calculate this diagram in orbifold models, is to use 
\be 
\delta(0)= \frac{1}{2\ell} \sum_{n}\frac{p^2-m^2_{n}}{p^2-m^2_{n}}
\ee
to exchange $m^2_{n}$ terms, generated by the derivative $\partial_{5}$, for $p^2$. Whilst this manipulation is rather less precise on the lattice than in the continuum limit, it is \emph{necessary} to ensure that corrections to the beta function due to a supersymmetric hidden sector cannot generate sfermion masses. Collecting the contributions from all three diagrams, the sfermion formula is 
\begin{equation}
m_{\tilde{f}}^2= \sum _r g_{r}^4 c_2(f;r)E_r
\end{equation}
where
\begin{equation}
E_r\!\!=\!\! -\! \!\int\!\! \frac{d^4p}{ (2\pi)^4}p^{2}\!\braket{p^2;0,N\!-\!1}\! \!\braket{p^2;0,N\!-\!1}\!
[3\tilde{C}_1^{(r)}(p^2\!/M^2)-4\tilde{C}_{1/2}^{(r)}(p^2\!/M^2)+\tilde{C}_{0}^{(r)}(p^2\!/M^2)]. 
\end{equation}
For the interval model, using \refe{propagator1} gives 
\begin{equation}
E_r\!\!=\!\! -\! \!\int\!\! \frac{d^4p}{ (2\pi)^4}a^4p^{2}\!\left[\prod_{j=0}^{N-1}\!\frac{1}{(ap)^{2}+4\sin^2{\frac{j\pi}{2N}}}\right]^2\!\!\!\!
[3\tilde{C}_1^{(r)}(p^2\!/M^2)-4\tilde{C}_{1/2}^{(r)}(p^2\!/M^2)+\tilde{C}_{0}^{(r)}(p^2\!/M^2)]. \label{sfermionmasses}
\end{equation}
The same equation holds in the periodic case, using the periodic mass eigenstates $4\sin^2{\frac{j\pi}{N}}$. In this equation, we have used a standard model lattice with gauge coupling $g_{r}$ where $r=1,2,3$ refering to the group $U(1),SU(2),SU(3)$ respectively. 
$c_2(f;r)$ is the quadratic Casimir of the representation $f$ of the scalar mass in question, under the gauge group $r$. The integral is UV and IR finite. As discussed in the introduction, one can see that the momentum integral in this equation will be suppressed by the product of KK propagators with length scale $a$ entering from the mass of the KK modes.  The limit in which there is a single lattice site ($N=1$) is the corresponding four dimensional limit.  We find 
\begin{equation}
E_r= -\! \!\int\! \frac{d^4p}{ (2\pi)^4}\frac{1}{p^2}
[3\tilde{C}_1^{(r)}(p^2/M^2)-4\tilde{C}_{1/2}^{(r)}(p^2/M^2)+\tilde{C}_{0}^{(r)}(p^2/M^2)]. \label{4dlimit}
\end{equation}
This equation reproduces exactly the result of ``General gauge mediation'' \cite{Meade:2008wd}.  For $N=2$ lattice sites, we may take the two gauge eigenmasses to be $m_{0}=0$ and $m_{1}=m_{v}$.  For the sfermion mass formula, one obtains
\begin{equation}
E_r= -\! \!\int\! \frac{d^4p}{ (2\pi)^4}\frac{1}{4p^2}[\frac{m^2_{v}}{p^2+m^2_{v}}]^2
[3\tilde{C}_1^{(r)}(p^2/M^2)-4\tilde{C}_{1/2}^{(r)}(p^2/M^2)+\tilde{C}_{0}^{(r)}(p^2/M^2)]\label{n=2limit}
\end{equation}
where the unwanted factor of $1/4$ is absorbed into the gauge coupling $g^2_{4d}=g^2/N$.
\subsection{The Casimir energy}
In a globally supersymmetric theory the vacuum energy is zero. Supersymmetry breaking effects of the hidden sector will generate a vacuum energy and the lattice dependent part of this vacuum energy will correspond, in the continuum limit, to the Casimir energy of a higher dimensional theory \cite{Bauer:2003mh,Mirabelli:1997aj,McGarrie:2010kh}.  To calculate the Casimir energy we must compute the vacuum diagrams that appear in figure 2 of \cite{Intriligator:2008fr}. Each vacuum diagram may be generated by simply forming a closed loop with the field that propagates each current correlator in the effective action. The propagation is from the zeroth lattice site back to the zeroth lattice site in the loop. The zero to zero lattice site propagator is given by
\be
 \braket{q^2; 0,0}=\frac{1}{N q^2}  [1+\sum_{k=1}^{N-1} \frac{2(aq)^2 \cos^2 \frac{k\pi}{2N} }{(aq)^2+4\sin^2 \frac{k \pi}{2N} }].
\ee
This time there is no product form for the propagator as the $(-1)^j$ is absent.  The vacuum energy is given by
\be 
\frac{E_{vac}}{V_{4d}}\! =\!\sum_{r}g^2_{r}d_{g}\!\!\int\!\! \frac{d^4p}{ (2\pi)^4} p^2\braket{p^2; 0,0}
[3\tilde{C}_1^{(r)}(\frac{p^2}{M^2})-4\tilde{C}_{1/2}^{(r)}(\frac{p^2}{M^2})+\tilde{C}_{0}^{(r)}(\frac{p^2}{M^2})]. \label{VAC}
\ee
$d_{g}$ is the dimension of the adjoint representation of the gauge group $r$. This integral is UV divergent. To extract from this the finite Casimir energy, one must extract the continuum limit of this sum. The prescription for this is found in \cite{Bauer:2003mh}.  Additionally, the appendix includes relevant steps which are applied in the next section, where we focus on a generalised messenger sector for which the $\tilde{C}_{s}$ terms may be determined.
\setcounter{equation}{0}
\section{Generalised messenger sector}\label{section:general}
In this section we give a concrete description of matter content of the SUSY breaking sector located at the zeroth lattice site, following the construction of \cite{Martin:1996zb}. We consider sets of $N$ chiral superfield messengers\footnote{We hope that this index $i$ which runs $1$ to $N$, the number of messengers, does not get confused with the lattice site index of the previous section.}  $\Phi_{i},\tilde{\Phi}_{i}$ in the vector like representation of the lattice gauge group, coupled to a SUSY breaking spurion $X= M + \theta^2 F $ with $F \ll M^2$. Generalisations to arbitrary hidden sectors are a straightforward application of the results of \cite{Marques:2009yu,McGarrie:2010kh}.  The superpotential is 
\be W  =X \eta_{i}\
\Phi_i \ti \Phi_i \label{superpotential2}\ee
In principle $\eta_{ij}$ is a generic matrix which may be diagonalised to its eigenvalues $\eta_{i}$ \cite{Martin:1996zb}.  The messengers will couple to the bulk vector superfield as
\be \delta {\cal L} = \int d^2\theta d^2\bar\theta \left(\Phi^\dag_i
e^{2 g V^a T^a} \Phi_i + \ti\Phi^\dag_i e^{-2 g V^a T^a}
\ti\Phi_i\right) + \left(\int d^2\theta\  W +
c.c.\right) \label{hiddensector} \ee
We can extract the multiplet of currents from the kinetic terms in the above Lagrangian. The current correlators can then be computed and their results can be found in \cite{Meade:2008wd,McGarrie:2010kh,Marques:2009yu}.  We will use the result of these current correlators to determine the gaugino massses, sfermion masses and Casimir energy.
\subsection{Gaugino masses}
The SUSY breaking zero mode Majorana gaugino mass is found by first evaluating the current correlator in \refe{gauginomass1}, which may be found in \cite{Martin:1996zb,Meade:2008wd,McGarrie:2010kh}, and then diagonalising the full fermion mass matrix and extracting the zero mode as discussed in the previous section.  For the zero mode this simply fixes $g^2/N=g^2_{4d}$.  The zero mode gaugino mass is found to be 
\be 
m^r_{\gl_{0} }= \frac{\alpha_{r}}{4\pi} \Lambda_{G}\ , \ \ \ \  \Lambda_{G}=\sum_{i=1}^{N}[\frac{d_{r}(i) F}{M}g(x_{i})]
\ee
The label $r=1,2,3$ refers to the gauge groups $U(1),SU(2),SU(3)$, $d_{r}(i)$ is the Dynkin index of the representation of $\Phi_{i},\tilde{\Phi}_{i}$ and
\be 
g(x)= \frac{(1-x)\log(1-x)+(1+x)\log(1+x)}{ x^2}
 \ee
where  $x_{i}=\frac{F}{\eta_{i}M^2}$.  $g(x)\sim 1$ for small $x$ \cite{Martin:1996zb}.

\subsection{Sfermion masses}
The entirely four dimensional limit of the sfermion mass formula when both the hidden and visible sector are located on the same single lattice site, is displayed in \refe{4dlimit}.  For the generalised messenger sector, this four dimensional result can be found in \cite{Martin:1996zb}. To obtain an effective five dimensional behaviour from the lattice, one must require sufficient lattice sites to suppress large contributions to loop momenta in the diagrams contributing to sfermion masses.  We start with \refe{sfermionmasses} and when $\frac{1}{(Na)^2}\ll M^2 $, one may then expand the current correlators in the limit $\frac{p^2}{M^2} \rightarrow 0$ and find \cite{McGarrie:2010kh}
\be
 [3\tilde{C}_1(p^2/M^2)-4\tilde{C}_{1/2}(p^2/M^2)+\tilde{C}_0(p^2/M^2)]\approx -\frac{1}{(4\pi)^2}\frac{2 d}{3}x^2 h(x)+O(p^2) \label{cterms}
\ee  
which is independent of $p^2$ at this order.  The function $h(x)$ is given by
\be h(x) =\frac{3}{2}[\frac{4+x-2x^2}{x^4}\log(1+x)+\frac{1}{x^2}] + (x\rightarrow -x),
\ee
where $h(x)$ for $x<0.8$ can be reasonably approximated by $h(x)=1$.  We find
\begin{equation}
m_{\tilde{f}}^2= \sum _r g_{r}^4 c_2(f;r)E_r
\end{equation}
where 
\begin{equation}
E_r= \sum^{N}_{i=1}[\frac{d_{r}(i)}{128 \pi^4 a^2}]|\frac{F}{ \eta_{i}M^{2}}|^2\frac{2}{3}h(x_{i}) \mathcal{I}
\end{equation}
and $\mathcal{I}$ is an integral that depends on the number of lattice sites 
\begin{equation}
\mathcal{I}=\int^{\infty}_{0} d(ap) (ap) \prod_{j=1}^{N-1}\frac{1}{(ap)^{2}+4\sin^2{\frac{j\pi}{2N}}}
\prod_{i=1}^{N-1}\frac{1}{(ap)^{2}+4\sin^2{\frac{i\pi}{2N}}}.
\end{equation}
The function $\mathcal{I}$ behaves like $\mathcal{I} \sim c/N^4$ 
\begin{center}
\begin{tabular}{c r @{.} l c  c}
$N$ &
\multicolumn{2}{c}{$\mathcal{I}$} & $1/N^4$ & c\\
\hline
$2$ & 0& 25 & 0.0625 & 4\\
$3$ & 0&029 & 0.012 &   2.4      \\ 
$3$ & 0&0082   &   0.0039 &    2.1     \\
$5$ & 0 &0032     &0.0016&      2    \\
$6$ & 0 &0015      &0.00077&     2   \\
$7$ & 0 &00079    &0.00042 &   1.7      \\
\end{tabular}
\end{center}
Rescaling $g^2/N= g^2_{4d}$  and taking $\ell=Na$ we find that the sfermion mass scales as
\be 
m^2_{\tilde{f}}\sim  \frac{g^4_{4d}}{(M\ell)^2}\frac{F^2}{M^2}
\ee
which reproduces the results of the Mirabelli-Peskin model \cite{Mirabelli:1997aj,McGarrie:2010kh}.

\subsection{Casimir energy}
The Casimir energy of the lattice can be extracted from the vacuum energy by taking the difference between the lattice propagator and its continuum counterpart.  This will cancel the divergent parts of the momentum integral  in the vacuum energy.  The final answer will be an approximate result that approaches the continuum Casimir energy when the number of lattice sites is infinite.  We start with \refe{VAC}.  The sum of $C$ terms is still given by \refe{cterms}.  We then must solve the UV divergent momentum integral in \refe{VAC}:
\be 
\sum_{k=0}^{N-1} f(\frac{k}{N})= \! \!\int\! \frac{d^4p}{ (2\pi)^4} [\sum_{k=0}^{N-1} \frac{2(ap)^2 \cos^2 \frac{k\pi}{2N} }{2^{\delta_{k0}}(ap)^2+4\sin^2 \frac{k \pi}{2N} }].
\ee
Next we take the difference between the lattice and continuum momentum integral as 
\be
\sum_{k=0}^{N-1} f(\frac{k}{N})-N  \int_{0}^{\infty} ds f(s)=  \frac{2}{(4\pi)^2\ell^4} \mathcal{S}(N).
\ee
The divergent parts of the lattice and continuum limit of this function will cancel out, leaving the finite mass dependent parts.  Additional steps may be found in the appendix which follow the procedure of \cite{Bauer:2003mh}.  The renormalized function $\mathcal{S}(N)$ is given by 
\begin{align}
\mathcal{S}(N)=-[N^4 & \sum^{N-1}_{k=1}\cos^2 (\frac{k\pi}{N})(\Delta(k/N))^2\log (\Delta(k/N)) \\&-N^5\!\!\int_{0}^{\infty}\! \!\!ds \cos^2(\frac{s\pi}{2})(\Delta(s))^2\log (\Delta(s))]
\end{align}
where 
\be
\Delta(\frac{k}{N})=(a m(\frac{k}{N}))^2= 4\sin^2 \frac{k\pi}{2N}
\ee
In the continuum limit $\mathcal{S}(N)$ is 
\be
\lim_{N\rightarrow \infty} S(N)\rightarrow 3\zeta(5).
\ee  For the Casimir energy, we obtain
\be 
\mathcal{E}_{\text{Casimir}} =- \sum_{r}\sum^{N}_{i=1} 
\frac{g^2_{ r}}{N} \frac{d_{g} d_{r}(i)}{(4\pi)^4}\frac{2}{3\ell^4}|\frac{F}{ \eta_{i}M^2}|^2h(x_{i}) \mathcal{S}(N).
\ee
This result agrees with the Casimir energy found in the Mirabelli-Peskin model \cite{Mirabelli:1997aj}  when $\lim_{N\rightarrow \infty} S(N)$. The Casimir energy for the periodic case is similarly obtained
\setcounter{equation}{0}
\section{Summary and conclusion}\label{conclusion4}
In this chapter we combined the framework of encoding generic hidden sector in terms of current correlators \cite{Meade:2008wd}, with the four dimensional construction of supersymmetric extra dimensions on a lattice \cite{Csaki:2001em}.  This extends previous lattice constructions of supersymmetry breaking so that different hidden sectors may be explored.  We have demonstrated that the low energy description of this model matches that of extra dimensional supersymmetry breaking on an interval \cite{McGarrie:2010kh}.  In particular we have shown that when the scale of the lattice is much smaller than the characteristic scale of the hidden sector $\frac{1}{(Na)^2}<<M^2$, then for a perturbative messenger sector sfermion masses are suppressed by an additional factor $\frac{1}{(Na M)^2}$ relative to pure four dimensional gauge mediation. This suppression arises due to the suppression of momenta in the outer loop of the two loop diagrams generating sfermion masses.  

It is useful to note that the running gauge coupling for a ``flat'' lattice has been explored in \cite{ArkaniHamed:2001vr,Chankowski:2001hz,Csaki:2001zx}.  Also various related such as an open moose \cite{Son:2003et} have been shown to have similar features to that of the AdS/CFT correspondence which is a topic we will return to in chapter \ref{chapter6}. 

In the next chapter we will look at a special case of five dimensional gauge mediation in which there is only 1 massive kk mode of the vector superfield assisting in the mediation of the breaking effects.  This special case is equivalent to the two lattice site model mentioned earlier in this chapter.  We will find that due to special properties of two loop feynman diagrams, the leading order scalar soft masses of the two site lattice model may be analytically determined for all ratios of the lattice spacing $a$ versus $M$ and a new hybrid regime is uncovered.

\setcounter{equation}{0}
\chapter{Hybrid Gauge Mediation}\label{hybridchapter}
In this chapter we would like to examine a particular effective model in which the vector superfield that mediates supersymmetry breaking has only a massless zero mode and one Kaluza-Klein mode with mass $m_v$.   This model is particularly interesting for two principal reasons.  First the model arises quite naturally as the minimal case of a deconstructed lattice of chapter \ref{chapter4} and as an effective model of a flat extra dimension as in chapter \ref{chapter2}.  Secondly this model not only captures all the essential features of both the gauge and gaugino mediated limits, it also allows one to interpolate between these limits analytically and explore a new hybrid limit.  This hybrid regime is rather new and we hope it receives further attention.  

To understand this model we will recapitulate some of the key features of general gauge mediation in five dimensions:   In gauge mediated supersymmetry breaking, one may broadly split models into two classes, those that are gauge mediated and those that are RG gaugino mediated. Let us clarify further: using the construction of general gauge mediation found in chapter \ref{chapter2},  the soft term for scalars masses at lowest order in $\alpha$ is dependent on a super-traced set of current correlators 
\be 
[3\tilde{C}_1(p^2/M^2)-4\tilde{C}_{1/2}(p^2/M^2)+\tilde{C}_{0}(p^2/M^2)] 
\ee
as pictured in figure \ref{flatfigure}. In general, given a perturbative hidden sector, we may expand this set of current correlators in either of two limits to obtain analytic expressions:  the four dimensional gauge mediated limit $M^2/p^2\leq 1$ or the screened (five dimensional) limit $p^2/M^2< 1$ \cite{Mirabelli:1997aj,McGarrie:2010kh,McGarrie:2010qr,McGarrie:2010yk}. In this second limit the scalar soft masses are rather suppressed at the high characteristic scale $M$ of the hidden sector. However, due to the gaugino mass contributions in the renormalisation group equations (RGE's)\footnote{See \cite{Drees:2004jm,Martin:1993zk,Yamada:1994id} for the four dimensional RG equations.} \cite{Bhattacharyya:2010rm} as in figure \ref{gauginomed}, scalars develop soft masses at low scales through RG gaugino mediation \cite{Schmaltz:2000gy,Schmaltz:2000ei}.  We should of course point out the mechanism by which one discriminates between the two expansion limits: when a mass scale $m_v$ enters into the outer loop (see figure \ref{flatfigure}) of the leading order sfermion mass diagrams with $m_v \ll M$,  this mass scale suppresses loop momenta and warrants expanding the current correlators in the screened limit.  As higher dimensional models naturally introduce this mass scale through Kaluza-Klein (kk) modes,  it is customary that the screened limit become synonymous with higher dimensional mediation of supersymmetry breaking.

It is natural to ask if there is some intermediate type of mediation whereby the leading order scalar soft masses are still somewhat suppressed at the high scale but not as drastically as in gaugino mediation limit, such that both the leading order contributions and subleading RG contributions play a significant role: hybrid gauge mediation.  The hybrid regime, when $m_{v}\sim M$, cannot be accessed by expanding the current correlators in $M^2/p^2$ and one must find a new way to evaluate the sfermion diagrams.  However for the case of a minimally truncated kk tower with just one massive kk mode, the leading orders sfermion diagrams are analytically solvable for all ratios of $F$, $M$ and $m_{v}$ and one can not only interpolate between the four dimensional and screened five dimensional limit, but one can also access the hybrid regime.

\emph{The key message} of this chapter is that whereas gaugino mediated models typically give a leading order sfermion mass of 
\be
m^2_{\tilde{f}(5d)} \sim  m^2_{\tilde{f} (4d)}\frac{1}{(M\ell)^2},
\ee
more generally by changing the ratio of the first kk mass $m_{v}=\frac{\pi}{\ell}$ with $M$ one can obtain analytic formulas whereby
\be
m^2_{\tilde{f} (5d)} \sim m^2_{\tilde{f} (4d)} \frac{1}{(M\ell)^\rho}
\ee
where $\rho$ takes real values between $0$ and $2$. 

The hybrid regime has been explored for a (de)construction model \cite{Auzzi:2010xc,Auzzi:2010mb}.  It also naturally arises in ISS-like models that exhibit magnetic colour-flavour locking that generate linking fields which we focused on in chapter \ref{chapter3}. Phenomenological scans which implement hybrid gauge mediation have also already been explored to some degree \cite{Abel:2010vba,Abel:2009ve,Rajaraman:2009ga,Thalapillil:2010ek} by virtue of having a nonzero scalar soft mass at a high scale and effective four dimensional RG equations when $m_v\sim M$.  In this chapter we will obtain completely analytic solutions for the leading two loop scalar soft masses and highlight the hybrid regime.  We will do this predominantly for the five dimensional model of chapter \ref{chapter2} although we also highlight the usefulness of these results to the (de)construction model as first pointed out in \cite{Auzzi:2010xc,Auzzi:2010mb}.

\section{The minimally truncated model}\label{section:truncated}
\begin{figure}[ht]
\centering
\includegraphics[scale=0.6]{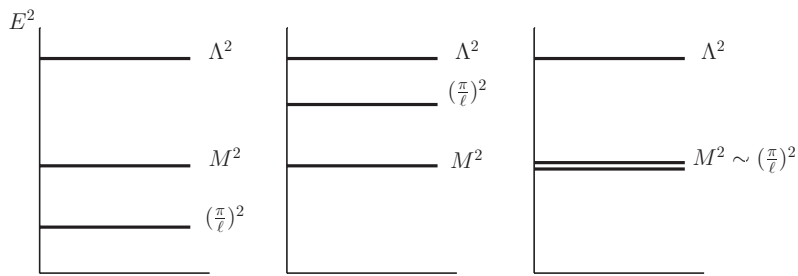}
\caption{Pictorial of the relative mass scales for the two state Kaluza Klein tower for a) gaugino mediation, b) gauge mediation and c) hybrid mediation.  $\Lambda$ is the cutoff of the effective model, $M$ is the characteristic mass scale of the hidden sector or vev of the spurion and the first Kaluza Klein mode is $m_{v}=\frac{\pi}{\ell}$ where $\ell$ is the length of the interval.  We emphasise that when the sfermion masses are screened in the gaugino mediated limit, there are always kk modes below the scale $M$.}
\label{energyscales}
\end{figure}

We saw in chapter \ref{chapter2} that all the kk modes of the vector superfield, in principle, propagate the supersymmetry breaking effects from the hidden to the visible brane.  If the mass scales of the model are sufficiently separated, we are at liberty to analyse an effective model in which only the zero mode $m_{0}=0$ and first mode of the kk tower with mass $m_{1}=m_{v}=\frac{\pi}{\ell}$ of vector superfields are part of the spectrum by placing a cutoff $\Lambda$ above these scales, as depicted in figure \ref{energyscales}. As any finite truncation of kk modes in the vector superfield may be related in the IR to a finite lattice (de)construction model \cite{McGarrie:2010qr,Csaki:2001em}, the results of this minimal truncation can be related to the minimal (de)construction model explored in \cite{Auzzi:2010xc,Auzzi:2010mb,Sudano:2010vt}. 
\begin{flushleft}
\end{flushleft}
\subsection{Gaugino soft masses}
We start by highlighting the 4 leading soft gaugino mass contributions which are given by
\be 
\mathcal{L}_{\text{soft}}\supset \frac{g^2_{5}}{2}MB_{1/2}(0)(\gl^0 \gl^0 + \gl^1 \gl^0 +\gl^0 \gl^1+ \gl^1 \gl^1) +  \text{c.c.} \label{gauginosoftmass}
\ee
Due to these soft masses, when computing the RG gaugino mediated contributions (as in figure \ref{gauginomed}) above the mass scale $m_{v}$ this will require evaluating the RG contributions from 6 diagrams.  When running RG equations at energy scales below $m_v$, only the diagram built from $m_{\gl}\gl^0 \gl^0$ contribute and the four dimensional RGE's are sufficient.

\subsection{Sfermion masses}
\begin{figure}[htcb]
\centering
\includegraphics[scale=0.6]{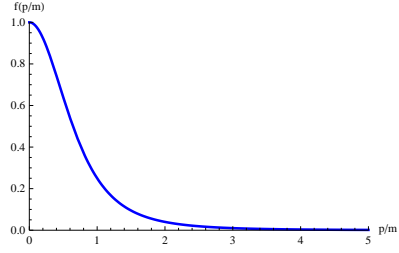}
\caption{A plot of the momentum dependent form factor, for the minimal model.  When $p/m_{v}\rightarrow 0 $ one recovers four dimensional general gauge mediation.  When $p/m_{v} \gg 0 $ one obtains screening of the leading order scalar soft mass.  This model captures the essential features of the all order kk model.}
\label{plot2}
\end{figure}

Supersymmetry breaking effects encoded on the hidden brane are mediated to the visible brane by both modes $\sum_{n=0}^{1}V_{n}(x,y)$. One may write the brane localised sfermion mass summations of \refe{Primeresult} as a product. After a Wick rotation one finds\footnote{We have rescaled the factors of $\ell$ into the coupling}
\begin{equation}
E= -\! \!\int\! \frac{d^4p}{ (2\pi)^4}\frac{1}{p^2}(\frac{m^2_{v}}{p^2+m^2_{v}})^2[3\tilde{C}_1(p^2/M^2)-4\tilde{C}_{1/2}(p^2/M^2)+\tilde{C}_{0}(p^2/M^2)] \label{product}
\end{equation}
with 
\be 
f(p/m_{1})=\left(1/(\frac{p^2}{m^2_{v}}+1)\right)^2.
\ee
The form factor is plotted in figure \ref{plot2} and captures the essential screening behaviour of the all order model plotted in figure \ref{plot1}.  This result has also been obtained for the (de)constructed version of this model \cite{Sudano:2010vt}. 
\begin{flushleft}
\end{flushleft}
\subsection{Bulk scalar masses}
In \cite{McGarrie:2010kh} the positive parity bulk hyperscalar zero mode masses $m^2_{H_{0}}$ were computed.  For the two state model it is given by  
\begin{equation}
m^2_{H_{0}}=  g_{4}^4   D
\end{equation}
where
\begin{equation}
D= -\! \!\int\! \frac{d^4p}{ (2\pi)^4}\sum_{n}(\frac{p}{p^2-m^2_{n}})^2 [3\tilde{C}_1(p^2/M^2)-4\tilde{C}_{1/2}(p^2/M^2)+\tilde{C}_{0}(p^2/M^2)]. \label{hyperscalar}
\end{equation}
This can be rewritten as (we Wick rotated and then manipulated)
\begin{equation}
D= -\! \!\int\! \frac{d^4p}{ (2\pi)^4}(\frac{1}{p^2} +\frac{1}{p^2+m^2_{v}}- \frac{m^2_{v}}{[p^2+m^2_{v}]^2}) [3\tilde{C}_1(p^2/M^2)-4\tilde{C}_{1/2}(p^2/M^2)+\tilde{C}_{0}(p^2/M^2)]. \label{Thekeyequation}
\end{equation}
We will show in section  \ref{section:hybrid} how this soft mass may be analytically determined for a specific hidden sector and specified currents.  In the next section we will show that this mass formula will determine the soft masses of linking fields in the minimal gaugino mediation model.
\subsection{Minimal gaugino mediation: Linking scalar soft masses}
\begin{figure}[ht]
\centering
\includegraphics[scale=0.6]{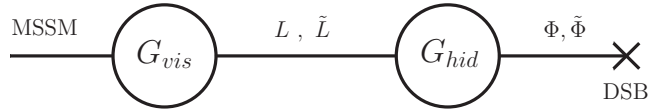}
\caption{Minimal gaugino mediation. MSSM matter located at the lattice site $G_{vis}$ corresponds to matter on the visible brane in figure 1. Similarly the fields $\Phi,\tilde{\Phi}$ located at $G_{hid}$ represents messengers located on the hidden brane.  The lattice linking fields $L$ and $\tilde{L}$ are bifundamental and antibifundamental respectively and corresponds to bulk hypermultiplets of figure 1. }
\label{model}
\end{figure}
The minimally truncated five dimensional model we have been describing so far can be related, in the IR, to the  minimal gaugino mediation model pictured in figure \ref{model}.  This means that the soft masses written above may be also be used for this lattice model.  This has been shown to be the case in \cite{Auzzi:2010xc,Sudano:2010vt,Auzzi:2010mb}.   In this section we would like to briefly comment on this model by showing how the linking scalar soft masses also match that of the positive parity hyperscalar mass formula \refe{hyperscalar}.

The kinetic terms for the bifundamental chiral superfields of the minimal model are given by
\be \delta {\cal L} = \int d^2\theta d^2\bar\theta \left(L^\dag
e^{2 g V_{0} -2 g V_{1} } L + \ti L^\dag e^{-2 g V_0+2 g V_{1} }
\ti L \right)  \label{linkingfields} \ee
where for simplicity we have taken $g_0=g_1=g$.  We see that for the minimal model of figure \ref{model} the linking fields will each pick up two leading order soft masses, being bifundamental.  The resulting soft mass for $m^2_{L} L^\dagger L$  is given by $m^2_{L}=\sum_{k=0}^{1}m^2_{k,L}$ where
\begin{equation}
m^2_{0,L}\!\!=\!\! -g^4\! \!\int\! \frac{d^4p}{ (2\pi)^4}p^{2}(\braket{p^2;0,1})^2
[3\tilde{C}_1(p^2/M^2)-4\tilde{C}_{1/2}(p^2/M^2)+\tilde{C}_{0}(p^2/M^2)],
\end{equation}
\begin{equation}
m^2_{1,L}\!\!=\!\! -g^4\! \!\int\! \frac{d^4p}{ (2\pi)^4}p^{2}(\braket{p^2;1,1})^2
[3\tilde{C}_1(p^2/M^2)-4\tilde{C}_{1/2}(p^2/M^2)+\tilde{C}_{0}(p^2/M^2)].
\end{equation}
Adding these contributions, one obtains the hyperscalar mass result \refe{hyperscalar} and emphasises the similarity between lattice and continuum models.  The mass eigenstates are then given by the full mass matrix including the kk masses and all linking fields of the same representation.

\section{Hybrid gauge mediation}\label{section:hybrid}
\begin{figure}[ht]
\centering
\includegraphics[scale=0.6]{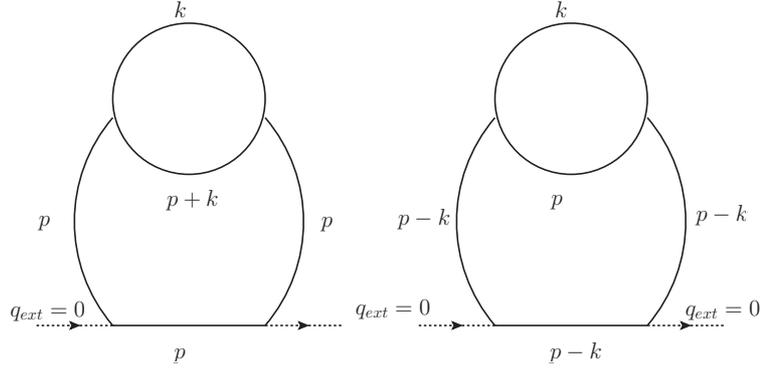}
\caption{A figure representing the labelling of momenta in the typical two-loop diagrams contributing to sfermion masses at leading order.  The inner loop has a characteristic mass scale $M$ and is typically a loop inside the current correlator, when the hidden sector has a perturbative description. The outer loop has a characteristic mass scale $m_{v}$.  The first case of labelling momenta is typical for GGM, however the mass scales $M$ and $m_{v}$ do not mix within each integral on $k$ or $p$ momenta and one must expand current correlators in a ratio of $M^2$ and $p^2$ to obtain an analytic limit. In the second case, the mass scales mix in either integral on momenta $k$ and $p$ and one may expand in a ratio of $m_{v}/M$.}
\label{labelmomentum}
\end{figure}

So far, we have given quite general soft mass expressions for gauginos, visible brane localised scalars and bulk positive parity scalars for the minimal model.  In this section we will specify the supersymmetry breaking hidden sector to be a generalised messenger sector coupled to a supersymmetry breaking spurion.  Specifying the hidden sector specifies the currents of the hidden sector (see appendix \ref{generalised}) and we may then use known expressions for these currents.  The relevant diagrams are typically two loop diagrams whose momentum integrals can be found in  \cite{Meade:2008wd,McGarrie:2010kh,Martin:1996zb,Marques:2009yu} and whose momenta is typically labelled as in figure \ref{labelmomentum}.  It is quite straightforward to shift momenta of these two loop diagrams and then apply the general expressions for massive two-loop Feynman diagrams, which are analytically solvable when the Mandelstam variables vanish \cite{Ghinculov:1994sd,vanderBij:1983bw,Martin:1996zb,Marques:2009yu}. 
\begin{flushleft}
\end{flushleft}
\subsection{Generalised messenger sector}\label{section:general2}
In this section we will quickly remind the reader of the conventions of a general messenger sector coupled to a spurion  \cite{Martin:1996zb}, whose conventions we first mentioned in chapter \ref{chapter1}.  We will then give useful mass formulas for this sector and mediation type.  We consider sets of $N$ chiral superfield messengers  $\Phi_{i},\tilde{\Phi}_{i}$ in the vector like representation of the lattice gauge group, coupled to a SUSY breaking spurion $X= M + \theta^2 F $. The superpotential is 
\be W  =X \eta_{i}\
\Phi_i \ti \Phi_i \ee
In principle $\eta_{ij}$ is a generic matrix which may be diagonalised to its eigenvalues $\eta_{i}$ \cite{Martin:1996zb}.  The messengers will couple to the bulk vector superfield as
\be \delta {\cal L} = \int d^2\theta d^2\bar\theta \left(\Phi^\dag_i
e^{2 g V^a T^a} \Phi_i + \ti\Phi^\dag_i e^{-2 g V^a T^a}
\ti\Phi_i\right) + \left(\int d^2\theta\  W +
c.c.\right) \ee
We can extract the multiplet of currents from the kinetic terms in the above Lagrangian. The current correlators can then be computed and their results can be found in \cite{Meade:2008wd,McGarrie:2010kh,Marques:2009yu}.  We will use the result of these current correlators to determine the gaugino massses, sfermion masses on the visible brane and the (positive parity) bulk hyperscalar soft mass.
\subsection{Gaugino masses}
The current correlator in \refe{gauginosoftmass} may be evaluated using the currents found in \cite{Meade:2008wd,McGarrie:2010kh,Martin:1996zb,Marques:2009yu} for the general messenger sector described above.   The zero mode gaugino mass is found to be 
\be 
m^r_{\gl_{0} }= \frac{\alpha_{r}}{4\pi} \Lambda_{G}\ , \ \ \ \  \Lambda_{G}=\sum_{i=1}^{N}[\frac{d_{r}(i) F}{M}g(x_{i})]
\label{gauginomass}
\ee
The label $r=1,2,3$ refers to the gauge groups $U(1),SU(2),SU(3)$, $d_{r}(i)$ is the Dynkin index of the representation of $\Phi_{i},\tilde{\Phi}_{i}$ and
\be 
g(x)= \frac{(1-x)\log(1-x)+(1+x)\log(1+x)}{ x^2}
 \ee
where  $x_{i}=\frac{F}{\eta_{i}M^2}$.  $g(x)\sim 1$ for small $x$ \cite{Martin:1996zb}.

\subsection{sfermion masses}
Starting from equation \ref{product} we may use the result of \cite{Auzzi:2010mb}, which is given in the appendix \ref{app:hyperscalarsoftmasses}, for the sfermion masses on the visible brane
\be
 m^2_{\tilde{f}}=2\left(\frac{F}{M}\right)^2 \sum_{r}(\frac{\alpha_{r}}{4\pi})^2 c(\tilde{f},r)\sum_{i} d_{r}(i)S(x_{i},y_{i})
\ee
with $y_{i}=m_{v}/\eta_{i}M$, where $c(\tilde{f},r)$ is the quadratic Casimir of the gauge group $r$ for the MSSM scalar of representation $\tilde{f}$.  $S(x,y)$ is given in the appendix.

\subsection{Bulk hyperscalar masses}
The positive parity bulk hyperscalar (linking scalar) mass, is given by 
\be
 m^2_{\tilde{h}}=2\left(\frac{F}{M}\right)^2 \sum_{r}(\frac{\alpha_{r}}{4\pi})^2 c(\tilde{h},r)\sum_{i} d_{r}(i)G(x_{i},y_{i})
\ee
where $G(x,y)$ is given in the appendix.

\begin{flushleft}
\end{flushleft}
\begin{figure}[ht]
\centering
\includegraphics[scale=0.6]{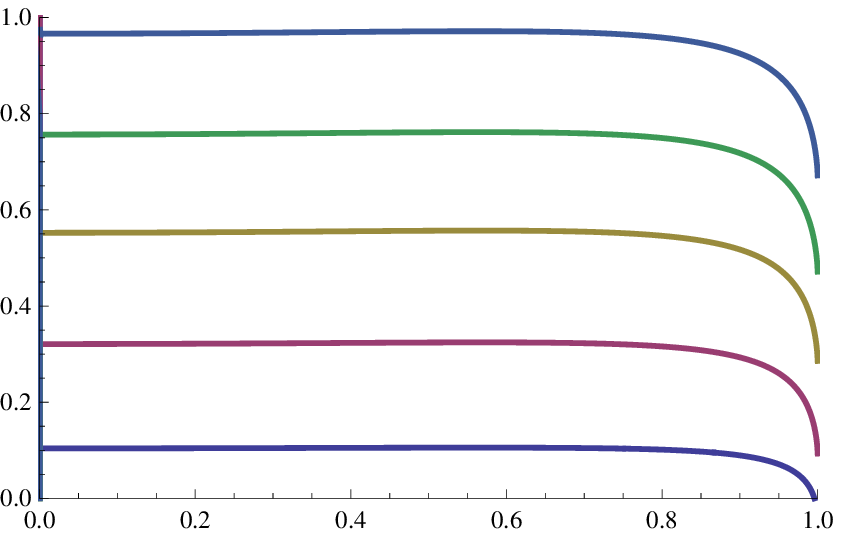}
\includegraphics[scale=0.6]{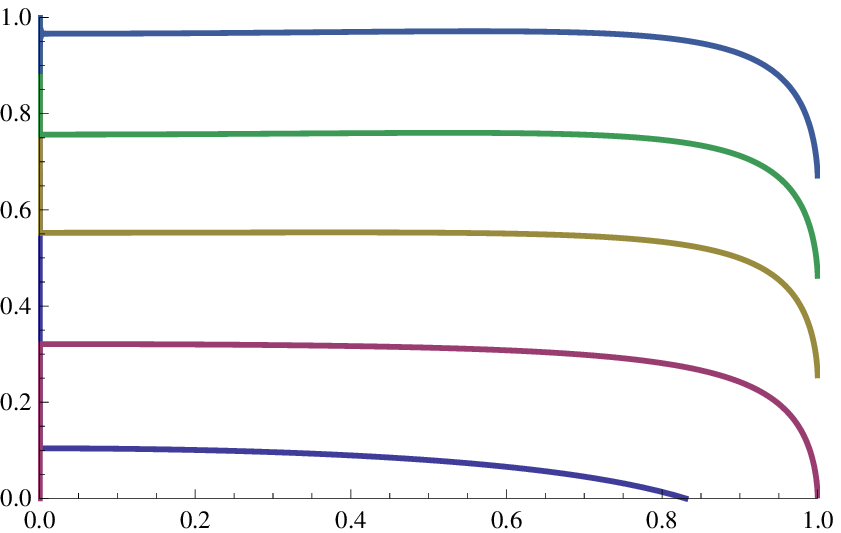}
\caption{On the left is a figure of $S(x,y)$ plotted along the $x$ axis going from bottom to top are $y=1,2.5,5,10,50.$  On the right is a plot of $G(x,y)$ for the same values of $y$. The right plot shows that for small values of $y$, $G(x,y)$ becomes tachyonic as $x$ approaches $1$, far more quickly than $S(x,y)$.}
\label{suppressionplots}
\end{figure}

\begin{flushleft}
\end{flushleft}
\begin{figure}[ht]
\centering
\includegraphics[scale=0.6]{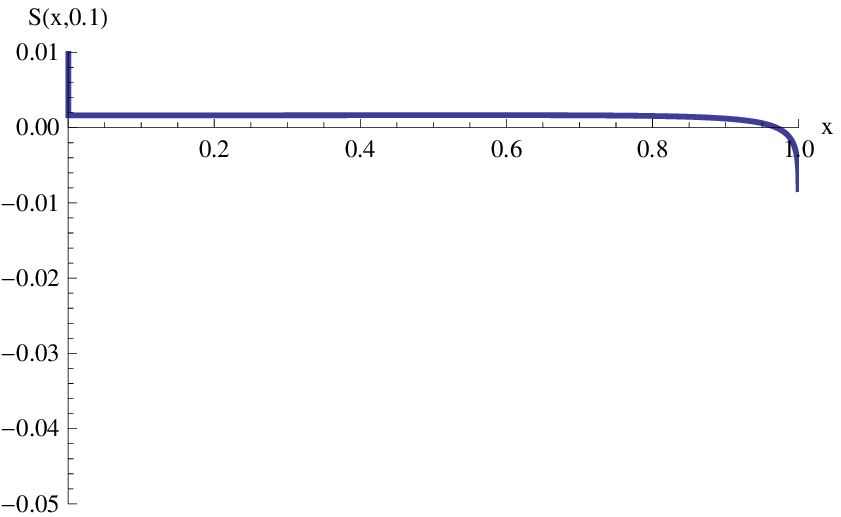}
\includegraphics[scale=0.6]{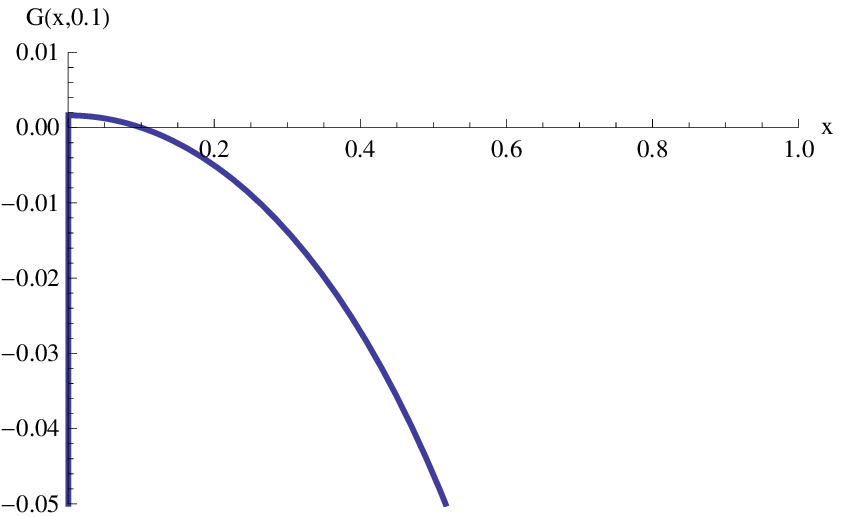}
\caption{On the left is a plot of $S(x,0.1)$ and on the right $G(x,0.1)$ along the x axis.  On the right plot we see the function becoming tachyonic far more quickly than on the left.}
\label{tachy}
\end{figure}

\begin{flushleft}
\end{flushleft}
\begin{figure}[ht]
\centering
\includegraphics[scale=0.6]{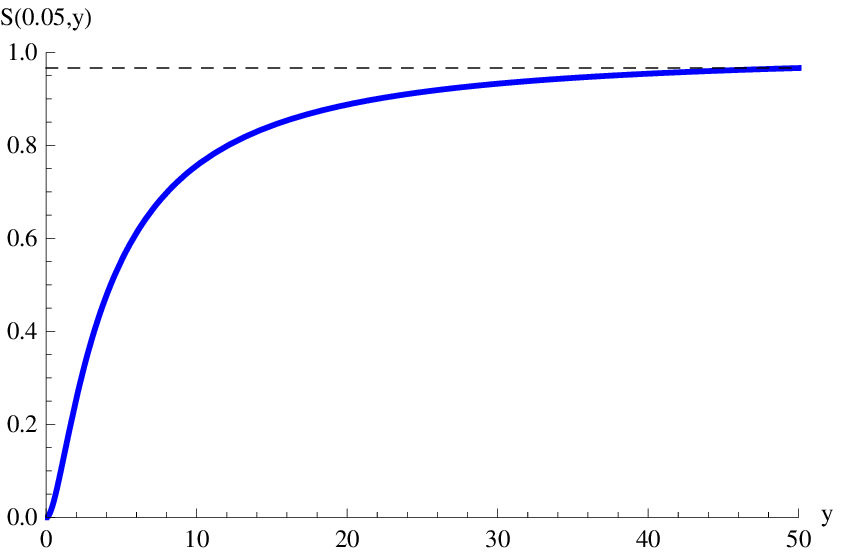}
\includegraphics[scale=0.6]{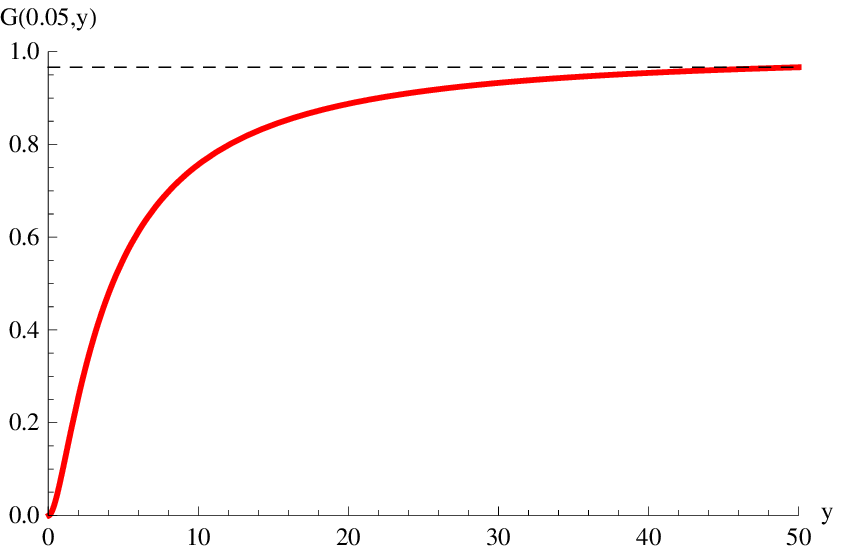}
\caption{On the left is a figure of $S(0.05,y)$ and on the right $G(0.05,y)$.  For small $y$ the functions are screened and scale as $\sim y^2$ and interpolates through the hybrid regime to large $\sim y^0$: the 4d limit.  }
\label{contour}
\end{figure}

In figure \ref{suppressionplots}, \ref{tachy} and \ref{contour}  we compare plots of the functions $S(x,y)$ for sfermion masses and $G(x,y)$ for hyperscalar masses. The plots may become tachyonic for some regimes of the parameter space and both plots have similar behaviour with regimes of strong screening of the masses.  Interestingly, these techniques to obtain analytic results may also be applicable for models with gauge messengers \cite{Intriligator:2008fr,Intriligator:2010be,Matos:2010ie}.

\section{Subleading contributions to general gauge mediation}\label{section:subleading}
The subleading diagram due to the double insertion of the gaugino mass found in \ref{gauginomed} is a three loop diagram.  For the minimal (flat) two state model one obtains (after Wick rotation)
\be
 \delta m_{\tilde{f}}^2= \sum_r c_2(f;r) g^6_{4d} \int \frac{d^4p }{(2\pi)^4}\frac{m^4_{v}}{p^4 (p^2+m^2_{v})^2}(M\tilde{B}_{1/2}(p^2/M^2))^2
\ee
which carries the same momentum dependent form factor as figure \ref{plot2}.  Keeping the full momentum dependence of the mass insertion, this is a three loop contribution and unfortunately cannot be calculated analytically by the same techniques that applied to the two loop leading order contributions of the previous section.  Additionally, there are three loop contributions to sfermion masses from bulk hyperscalar masses, as has been commented on before \cite{Green:2010ww}.  It would certainly be interesting to apply numerical techniques to these superficially subleading diagrams (order $g^6$), keeping their full momentum dependence, to better understand their role in this model.

Very recently a comprehensive examination of the spectrum of the hybrid regime was undertaken \cite{Auzzi:2011gh}, with the model of chapter \ref{chapter3} in mind. Reasonable phenomenological spectra was obtain and in particular, it was commented that the spectrum in the  Hybrid regime and the Gaugino mediated regime have much in common as when the leading order scalar soft mass contribution is suppressed the subleading contribution compensates to give an overall similar outcome.

\section{Summary and conclusion}\label{conclusion5}
In this chapter we have outlined a concrete framework in which higher dimensional models of gauge mediated supersymmetry breaking may obtain a hybrid between gauge and gaugino mediation.  In particular we have shown that the leading order sfermion and bulk hyperscalar masses, being massive two-loop diagrams with zero external momenta, are analytically solvable for this model which allows for a determination of this mass contribution in the hybrid regime that $M\sim m_{v}$ as was first pointed out in \cite{Auzzi:2010mb}.  

These results help to complete the discussion of chapter \ref{chapter2} as they highlight a smooth interpolation between gauge and gaugino mediation, with a new hybrid regime of mediation when $m_v\sim M$.  

In the next chapter, we will move to a more exotic topic:  we introduce more complexity by warping the extra dimensional interval. The resulting complexity of the Kaluza Klein masses and eigenfunctions will mean that we will only be able to make simplifications to the particular limits of gauge or gaugino mediation and the calculable results of the hybrid regime will be lost.  In fact it is also straightforward to see that as the first kk mode of a warped vector superfield is warped towards the IR, a simple hybrid model cannot be constructed.   However, we do not think that this should preclude achieving a soft scalar mass from a warped model that is only partially screened and behaves like a hybrid model, just that this type of result can only be computed numerically rather than analytically.  We believe this is an extremely worthwhile direction for future research outside of this thesis.  Other directions for future research is to apply these techniques and their line of reasoning to models of brane to brane supergravity mediation, in particular because that model also solves problems with flavour changing neutral currents and may generate less suppressed soft masses as usually encountered in the full all order kk mode calculations.

\setcounter{equation}{0}
\chapter{Warped General Gauge Mediation}\label{chapter6}
\section{Introduction}
The framework of general gauge mediation and in particular the decoupling limit $\alpha_{SM}\rightarrow 0$ allows for the application of dualities.  The reasoning behind this is that one may exchange strongly coupled, all order, current correlators with their dual perturbative description.  This has allowed for the implement of Seiberg duality to the GGM construction.  In this chapter we would like to extend this construction through the AdS/CFT programme \cite{Maldacena:1997re}.  Our construction will be general such that it is applicable to model various cases in the literature.  Our key findings are generic such that particular models essentially define the various mass scales as inputs or specify a particular hidden sector to encode into currents or are variations of the construction outlined in this chapter.  In fact there are a large range of models of dynamical supersymmetry breaking with relation to the AdS/CFT \cite{Nomura:2004zs,Nomura:2006fz,Nomura:2006pn,Gherghetta:2010cj,Piai:2010ma,ArkaniHamed:2000ds,Rattazzi:2000hs,Abel:2010uw,Abel:2010vb,Price:2011fk} including conformal sequestering \cite{Schmaltz:2006qs,Abe:2007ki,Craig:2010ip} and models based the duality cascade using Warped throats \cite{Klebanov:2000hb,DeWolfe:2008zy,Simic:2010nv,Hanaki:2010xf,Benini:2009ff}.  Our approach to extracting  MSSM soft masses within the AdS/CFT perspective is rather original in that it focuses on determining the behaviour of global currents, within the AdS background \cite{McGarrie:2010yk} which perhaps has more significant overlap with the AdS/QCD literature \cite{O'Connell:1995wf,Erlich:2005qh,Hirn:2005nr,Harada:2010cn}.

We have in mind some strongly coupled hidden sector with gauge group $G_{hidden}$ with $\tilde{N}$ colours that breaks global $\mathcal{N}=1$ supersymmetry and generates a dynamical supersymmetry breaking scale through 
\be 
M_{susy} \sim M_{Pl} e^{-c/\tilde{g}^2(M_{Pl})}.
\ee
Let us further suppose that $\tilde{g}^2\tilde{N}/16 \pi^2 \gg 1$ and $\tilde{N}$ large.  We may expect that the gauge coupling of the hidden sector evolves slowly between $\Lambda_{IR}$, the scale at which conformal symmetry is broken and $M_{Pl}$ then we may be able to model this approximately conformal field theory (CFT) with an effective AdS dual description between IR and UV branes of an extra dimension. $G_{hidden}$ supplies the background AdS geometry.  Whatever theory $G_{hidden}$ has at energies below $\Lambda_{IR}$ is essentially captured on the IR brane, including supersymmetry breaking, and will be a generic hidden sector with $F$-term and spurion vev $M$.  The IR theory may or may not also be strongly coupled. For instance a theory that was in the conformal window $\frac{3}{2}N_c< N_{f} <3N_C$ with large $N_c$ may, after losing flavours along the flow, at some point move into the magnetic window and admit an IR free magnetic dual on the IR brane.  Similarly the UV theory may have some more complicated structure such as some Calabi Yau.  For our purposes the UV is simply the boundary where fundamental fields such as the MSSM matter ``lives''. In some sense the MSSM matter should be thought of as entirely outside the AdS dual of the hidden sector, which completely decouples in the limit that $\alpha_{SM} \rightarrow 0$.

Next, let us introduce a global symmetry of the hidden sector such as $SU(N_f)\times SU(\tilde{N}_c)$ and then weakly gauge that global symmetry and associate it with the standard model or GUT ``parent'' gauge group.  Through some analogue of ``colour-flavour locking'' bounds states or resonances of the approximate CFT will fill a Kaluza-Klein tower of the weakly gauged global symmetry.  As a result, the vector superfields of the standard model gauge groups will have a full Kaluza-Klein tower and ``live'' in the bulk of the theory. On the four dimensional side, this construction fits within four dimensional general gauge mediation prescription of chapter \ref{chapter1}.  In the AdS side, we see that a full Kaluza-Klein tower of a vector superfield will mediate the effects of supersymmetry breaking from the IR brane up the MSSM fields located on the UV brane. In the four dimensional perspective there is only a massless 4d vector superfield describing the mediation, however due to mixing with the resonances as intermediate states in the two point function \cite{Batell:2007ez}, an effective extra dimension appears. This effect has a QCD analogue of ``vector meson dominance'' \cite{Georgi:1989xy,Bando:1984ej,Komargodski:2010mc}, and should be familiar from the Deconstructed model of chapter \ref{chapter4}.  We see that the decoupling limit $\alpha_{SM}\rightarrow 0$ survives on either side of the duality.  A slightly more subtle point is that we still have not specified the current correlators on the IR brane and one may wish to apply a further Seiberg duality to those.  In any case we should expect a perturbative description of supersymmetry breaking can be encoded into current correlators located on the IR brane, whose effects are mediated by a full Kaluza-Klein tower of vector superfields up through the bulk, to generate soft mass terms for the standard model located on the UV brane.


\begin{figure}[ht]
\centering
\includegraphics[scale=1]{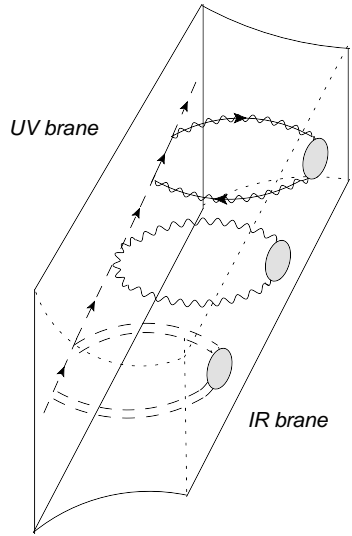}
\caption{A portrait of ``General gauge mediation'' across a warped bulk. Sfermion masses on the visible (UV) brane are generated by propagating the effects of supersymmetry breaking on the hidden (IR) brane. \label{warped figure}} 
\end{figure}
In chapter \ref{chapter3} we showed that a class of $\mathcal{N}=1$ SQCD $SU(N_c)$ theories in the magnetic window have a dual IR-free magnetic description which at low energies generates a deconstructed effective extra dimension. An interesting direction for future research is to see if one could indeed obtain a deconstructed AdS lattice \cite{Son:2003et,Falkowski:2002cm,Erlich:2006hq}  similar to the flat lattice of chapter \ref{chapter3}.

Regardless of the particular UV completion or four dimensional dual from which this model arises, we will simply treat a slice of AdS as an effective description with generic features and make no further comment on this issue.  Within this framework,  we are able to derive general formulas for both gaugino and sfermion masses.   We are also able to accommodate bulk hypermultiplets and again give a general formula for their scalar masses.   A final application of our framework is to calculate the Casimir energy. 

To illustrate this framework we apply our general results in terms of current correlators,  to the case where the hidden sector has a perturbative description in terms of a SUSY breaking spurion coupled to messenger fields charged under the standard model gauge group.  Specifying the hidden sector, specifies the currents and we then extract approximate formulas for the gaugino and sfermion masses of the MSSM, at leading order in $\alpha_i$.  

 Our discussion will focus on the models behaviour with regard to four scales: the AdS warp factor $k$; orbifold length $\ell$; a SUSY breaking $F$-term and hidden sector characteristic mass scale $M$.  Assuming that  $ k\ell \gg 1$, we show that when the $F,k^{2} \ll M^2$, the full Kaluza-Klein tower of the bulk vector multiplet must be considered in the propagation of supersymmetry breaking to the UV brane, whereas in the limit that $F,M^2 \ll k^{2}$, only the zero modes make the dominant contribution to mediation.  We would like to re-emphasise the point that if $k \gg M $ then the phenonomenology is gauge mediated not gaugino mediated.
\subsection{Gauge coupling unification}
Some precursory comments are useful.  The gauge couplings of the standard model will receive contributions from MSSM fields, the bulk fields (kk modes) and from the fields localised on the IR brane that are charged under the standard model.  It has been shown in \cite{Dienes:1998vg,Pomarol:2000hp,Randall:2001gc,Randall:2001gb,Randall:2002qr,Goldberger:2002hb,Goldberger:2002cz,Agashe:2002bx,deBlas:2006fz,Erlich:2006hq,Falkowski:2002cm}, that in certain limits, the contribution from bulk fields is consistent with logarithmic running of the gauge couplings.  That is to say that when evolving the RG equations from $M_z$ up to $k$ the gauge coupling evolution is logarithmic given by 
\be
\frac{1}{g^2_{j}(M_{z})} = \frac{1}{g^2_{j}(k)}  + \frac{b_{j}}{8\pi^2}\log (\frac{k}{M_{z}})
\ee
One \emph{must} still compare this with the other scale $M$ of the hidden sector as to whether one has gaugino mediation or just simply gauge mediation for the soft masses.  For instance if $k\gg M$ we will have unsuppressed scalar masses at leading order, i.e. four dimensional gauge mediation, and logarithmic running up to $E=k$.    Conversely, if $k\ll M$ we will have gaugino mediation and logarithmic running only up to $E=k$ and additional power law contributions between $k$ and $M$. 
These contributions will be of the form 
\be
+ \frac{b_{kk}}{64\pi^2} \frac{\Lambda^2}{k^2}\log{\frac{5M_{kk}}{4\pi k}}
\ee
where $b_{kk}=3N_c-N_c=2N_c$ for the $V$+$\Phi$ kk mode contributions.  To put this another way, if only the massless zero modes contribute significantly to the renormalisation of the gauge coupling \cite{Pomarol:2000hp} then it is likely that only the zero modes contribute significantly to the mediation of supersymmetry breaking and we should expect a gauge mediated not gaugino mediated spectrum. 

Finally, a different and entirely unrelated approach to breaking supersymmetry in $AdS_{5}$ models called Scherk-Schwarz \cite{Scherk:1978ta,Gherghetta:2000kr} or twisted boundary conditions.  These do not fit within the framework of general gauge mediation and are not gauge mediated in any ordinary sense.

\section{Off-shell warped gauge theory}\label{section:Frameworkwggm}
\def\th{\theta}
\def\bth{\bar{\th}}
\def\a{\alpha}
\def\s{\sigma}
\def\vth{\vartheta}
We began in \cite{McGarrie:2010yk} with the $AdS_{5}$ with the metric given by
\be 
\label{metric}
d s^2 = e^{-2\sigma}\eta_{\mu \nu}dx^\mu dx^\nu +dy^2,
\ee
in which  $\eta_{\mu\nu}=\textrm{diag}(-1,1,1,1)$,  $\sigma=k y$ and $1/k$ is the AdS curvature scale with mass dimension one.  From the metric, one can readily read off the  f\"unfbein to be
\be e^a_{\mu}(x,y)=e^{-\sigma}\hat{e}^a_{\mu}(x)=e^{-\sigma }\delta^a_{\mu},\quad e^5_{\mu}=e^a_{5}=0 , \quad e^{\hat{5}}_{5}=1\ . 
\ee
We are interested in describing physics on an interval of this $AdS$ space  given by  $0\le y\le \ell$ with $\ell =\pi R$.    It is helpful to think of the interval as a $\mathbb{Z}_2$ quotient of a periodic $y$ coordinate.  This construction, also known as a warped $S^1/\mathbb{Z}_2$ orbifold, is achieved by replacing $\sigma = k |y|$ in the metric (\ref{metric}) and allowing $y$ to be periodic with range $-\ell <  y\le \ell$.   The $\mathbb{Z}_2$ identification is given by $ y \sim -y$ and has fixed points at $y = 0$ and $y=\ell$ where we shall locate matter on three-branes known as the UV and IR brane respectively.

To build our GGM framework we shall need an off-shell description of the supersymmetric gauge theory living on this space. To obtain the off-shell description for the warped orbifold background one may apply the procedure of ``$\theta$-warping''  mentioned in \cite{Gregoire:2004nn} and developed in \cite{Bagger:2006hm,Xiong2010}. Using this prescription we deform the the superspace coordinates by defining $\vartheta = e^{-\frac{1}{2} \s} \theta$ and supercovariant derivatives by  ${\cal D}_\a = e^{\frac{1}{2} \s} D_\a$.  The warped space action if found by making the replacements   $(\theta, D_\a, d^5x  )\rightarrow (\vartheta, {\cal D}_\a ,   d^5 x \sqrt{-g})$ in the superfields.  After reexpanding in the original $\th$ coordinates one finds
\be
\label{warpact}
S_{warp} = \int d^5 x  \  \frac{1}{2}\Tr \left[ \int d^2 \th  \ W^\a W_\a  + h.c. +  \frac{e^{-2 \s}}{g_5^2} \int d^4 \th\ \left(e^{-2 g_5 V }\nabla_5 e^{2g_5 V}\right)^2  \right] \ ,
\ee
where the superfields have been warped so that they become 
\bea
\label{warpedmultiplets}
V&=&   -\th \s^a  \bth \delta_a^\mu A_\mu + i \bth^2 \th    e^{-\frac{3}{2} \s}\lambda - i \th^2 \bth e^{-\frac{3}{2} \s} \bar{\lambda} + \frac{1}{2} \bth^2 \th^2 e^{-2 \s} D \nonumber \ ,  \\
\Phi &=& \frac{1}{\sqrt{2}} \left(\Sigma +i A_5\right) + \sqrt{2} \th  e^{-\frac{1}{2} \s} \chi + \th^2 e^{-  \s} F \ .
\eea
Expanding in components one finds the kinetic terms for the vector multiplet  are given by 
\be 
 \int d^{5}x \text{Tr}\left[-\frac{1}{2}F^{\mu \nu}F_{\mu \nu} +ie^{-3\sigma}\bar{\gl}\sigma^{\mu}D_{\mu}\gl+\frac{1}{2}e^{-4\sigma}D^2 \right].
\ee  
Mass terms arise in the component expansion of the action (\ref{warpact}) from the term involving   $\partial_5 V$ when the derivative acts on the warp factor.   One finds a Dirac mass for the fermions  and upon integrating out the auxiliary $D$ field, a scalar mass  given by 
\be
\quad m_{\Psi}=\frac{1}{2}\sigma'  \, , \quad  m_{\Sigma}=-4k^2+ 2\sigma'' \, . 
 \ee
We remark that the Abelian version of this theory is related to action written in \cite{Marti:2001iw}, which makes use of a radion superfield.  
  
In what follows we shall expand these fields in terms of their eigen-modes   \cite{Gherghetta:2000qt}  which can be  summarised using the theta-warped superfields by 
\be
V = \frac{1}{\sqrt{2 \ell} }\sum_n V_n(x) f^{(2)}_n (y) \, ,  \quad \Phi = \frac{1}{\sqrt{2 \ell} } \sum_n  \Phi_n(x) g^{(4)}_n(y) \, , 
\ee
 where the even and odd modes  are given by
\bea
 f^{(s)}_{n}(y)&=&\frac{e^{s\sigma/2}}{N_{n}}\left[J_{1}\left(\frac{m_{n}e^{\sigma}}{k}\right)+b \left(m_{n}\right)Y_{1}\left(\frac{m_{n}e^{\sigma}}{k}\right)\right]\, ,  \label{ffunctions}\\ 
 g^{(s)}_{n}(y)&=&\frac{\sigma'}{k} \frac{e^{s\sigma/2}}{N_{n}}\left[J_{0}\left(\frac{m_{n}e^{\sigma}}{k}\right)+b \left(m_{n}\right)Y_{0}\left(\frac{m_{n}e^{\sigma}}{k}\right)\right]      \, , \label{gfunctions}
\eea
and obey orthonormality conditions 
 \be
\frac{1}{2\ell}\int^{\ell}_{-\ell}e^{(2-s)\sigma}f^{(s)}_{n}(y)f^{(s)}_{m}(y)dy=\delta_{nm} \label{ortho}\, ,
\ee
with similar for the odd modes. Orthornormality fixes the normalisation which in the limit $m_n\ll k$ and $k\ell \gg 1$ is given by
\be
N_m \approx \frac{e^{k\ell/2}}{\sqrt{\pi \ell m_n }} \, , 
\ee 
 and boundary conditions at both branes can be used to fix $b(m_n)= -\frac{J_0(m_n/k)}{  Y_0(m_n/k)}$ and deduce the mass spectrum by solving $b(m_n) = b(m_n e^{k \ell})$  which yields  
\be
 m_n \approx \left( n - \frac{1}{4} \right) \pi k e^{-k \ell} \, . 
\ee 
 
   \subsection{Bulk hypermultiplets}
The $\th$-warping technique can be applied to bulk hypermultiplets.  Under the orbifold action the hypermultiplet splits into two $4D$ ${\cal N}=1$ chiral superfields, $H$ of even parity and $H^c$ of odd parity  which transform under the gauge group as 
$H\to e^{-\Lambda}H$ and $H^c\to H^ce^\Lambda$.   Starting with the flat orbifold action and superfields given in \cite{Hebecker:2001ke} and following the warping procedure  one arrives at warped superfields 
\bea
H &=& H^1+\sqrt{2}e^{-\frac{1}{2}\sigma}\theta\psi_L+\theta^2 e^{-\sigma}(F_1+D_5H^2-g_{5}\Sigma H^2) \nonumber \, ,  \\
H^c &=& H^\dagger_2+\sqrt{2}\theta e^{-\frac{1}{2}\sigma} \psi_R+\theta^2 e^{-\sigma}(-F^{\dagger 2}-D_5
H^\dagger_1- g_{5} H^\dagger_1\Sigma)\, , 
\eea
and a warped action
\begin{alignat}{1}
S_{warp}^H=& \int d^5 x e^{-2\sigma}\int d^4\theta [ H^\dagger e^{2g_{5}V}H+H^c e^{-2g_{5}V}H^{c\dagger}] \nonumber \\
&\quad +\int d^5 x e^{-3\sigma} \left( \int d^2 \theta  H^c\nabla_5 H+ \int d^2\bar{\theta} H^{c\dagger}\nabla_5 H^\dagger \right)
\,.
\end{alignat}  
It should be clear from the action that the hypermultiplet will also decouple from the hidden sector in the limit $\alpha_{i}\rightarrow 0$.  In this way hypermultiplets will also follow the prescription of general gauge mediation. Starting from a massless unwarped hypermultiplet and applying $\theta$ warping, the warped hypermultiplet has a mass generated by passing a $\partial_{5}$ through an $e^{-1/2 \sigma}$ factor. This corresponds to the conformal limit $c=1/2$ for the on-shell action of \cite{Gherghetta:2000qt}.\footnote{We exclusively consider hypermultiplets at the conformal limit in this paper.} 
 The positive parity fields in $H$ have eigenfunctions given by \refe{ffunctions}, with $\alpha=1$. Similarly the negative parity fields in $H^c$ are determined by \refe{gfunctions} with $\alpha=0$. In this paper we will compute the soft mass of the zero mode scalar of $H$ which is  given by
\be
H^{(0)}(x,y)=\frac{1}{\sqrt{2\ell} }e^{\sigma}H^{(0)}(x)\; .
\ee
 

\section{Brane localised currents} \label{section:warpcurrents}

In this section we will encode a SUSY breaking sector, localised on the IR brane at $y=\ell$, in terms of current correlators.
 
Since the vector superfield $V$ is of even parity and obeys Neumann type boundary conditions it can couple to matter charged under the gauge group localised on the boundary IR brane.   In general, we expect that the global current multiplet ${\cal J}$ should serve as a source for these interactions however we must take some care to accommodate the effects of the warping.   Starting with the flat space form of these interactions  \cite{McGarrie:2010kh} and applying the $\th$-warping technique produces boundary interactions of the form 
 \begin{equation}
S_{int}=2g_{5}\!\int\!d^5 x e^{-2\sigma} d^{4}\theta \mathcal{J} 
V \delta(y-\ell )
\end{equation}
where the warped current superfield is given by  
\bea
\mathcal{J} &=& J +    ie^{-\frac{1}{2}\sigma} \theta j  -
ie^{-\frac{1}{2}\sigma}\bar{\theta}\bar{j}  -
\theta \sigma^{a}\delta^{\mu}_{a} \bar{\theta} j_{\mu} \nonumber   \\ &  & \quad  + 
\frac{1}{2} e^{-\frac{1}{2}\sigma} \theta^2 \bar{\theta} \bar{\sigma}^{a} \delta^{\mu}_{a}
\partial_{\mu} j  - \frac{1}{2} e^{-\frac{1}{2}\s }\bar{\theta}^2 \theta 
\sigma^{a}\delta^{\mu}_{a} \partial_{\mu} \bar{j} - 
\frac{1}{4} \theta^2 \bar{\theta}^2 \square J \; ,
\eea
and $V$ is given as in (\ref{warpedmultiplets}).   In components these interaction terms read   
\begin{equation}
S_{int} = \int \!d^5 x e^{-4\sigma} g_{5}(JD- \gl j \!-
 \!\bar{\gl} \bar{j}-e^{2\sigma }j^{\mu}A_{\mu})\delta(y- \ell).
\end{equation}

The currents appearing in these expressions are not canonical in the sense they are built out of noncanonically normalised fields due to the warping of the induced metric on the IR brane.   It is helpful to instead work with canonically normalised fields and currents defined by
\be
e^{-2\sigma}J=\hat{J}\, , \quad
e^{-\frac{5}{2}\sigma}j_{\alpha}=\hat{j}_{\alpha}\, , \quad
e^{-2\sigma}j_{\mu}=\hat{j}_{\mu} \, ,   
\ee
so that the interaction terms are given by 
\begin{equation}
S_{int}=\int d^5 x  g_{5}(e^{-2\sigma} \hat{J} D- e^{-3/2\sigma}\lambda \hat{j} \!-
 \! e^{-3/2\sigma}\bar{\lambda} \hat{\bar{j}}-\hat{j}^{\mu}A_{\mu})\delta(y-\ell) \, . 
\end{equation}
With these rescalings two-point functions of canonical currents are given by  the  flat space result but with mass scales accordingly warped down.  In this way we can easily keep track of powers of the warp factor.  
 
 The contribution of the hidden sector can be found by expanding the functional integral in $g_5$  to $O(g_{5}^{2})$   following  \cite{Meade:2008wd}. Upon inserting the relation for the $D= X^3 - D_5 \Sigma$ we find that 
\begin{alignat}{1}\label{E:ChangeL}
\delta \mathcal{L}_{eff}=[&- g_{5}^{2}e^{-3k\ell}\tilde{C}_{1/2}(0) i \lambda \sigma^{\mu} \partial_{\mu} \bar{\lambda}
- g_{5}^{2}\frac {1} {4} \tilde{C}_1(0) F_{\mu\nu} F^{\mu\nu} \nonumber \\
&-g_{5}^{2}e^{-3k\ell}\frac {1}{ 2}(\hat{M} \tilde{B}_{1/2}(0) \lambda \lambda + \hat{M} \tilde{B}_{1/2}(0)\bar{\gl}\bar{\gl})\nonumber 
\\ +& e^{-4k\ell}[\frac {1}{ 2 }g_{5}^{2}\tilde{C}_0 (0)(X^{3}X^{3})+\frac {1}{ 2 }g_{5}^{2}\tilde{C}_0 (0)(D_{5}\Sigma)(D_{5}\Sigma)-g_{5}^{2}\tilde{C}_0 (0)(D_{5}\Sigma)X^{3} ] ]\ .
\end{alignat}
$\hat{M}=e^{-k\ell}M$ is the characteristic mass scale of the hidden brane localised at $y=\ell$. In what follows we shall frequently express our results in terms of the ``supertraced'' set of these current correlators 
\be
 [3\tilde{C}_1(p^2/\hat{M}^2)-4\tilde{C}_{1/2}(p^2/\hat{M}^2)+\tilde{C}_{0}(p^2/\hat{M}^2)]=\Omega \left(\frac{p^2}{\hat{M}^2} \right)\label{Omegawarped1}\, .
\ee
We emphasise that the effect of the induced metric on the IR brane is captured in the canonical current correlators through the warped mass scale $\hat{M}$.
\section{Soft masses and vacuum energy}\label{section:results}
In this section we will give general expressions for the gaugino, sfermion and hyperscalar soft masses and the Casimir energy, in terms of the current correlators located on the IR brane.  In the next section we will explore these general expressions for a generalised messenger sector in specific limits.
\subsection{Gaugino masses}
At $g^2_{5}$ order, we can extract the SUSY breaking contribution to the gaugino masses from the effective Lagrangian as 
\be
\label{gauginosoft}
\mathcal{L}_{\text{soft}}=\frac{\delta(y-\ell)}{2} g^2_{5}e^{-3k\ell}\hat{M}\tilde{B}_{1/2}(0)\lambda\lambda + \text{c.c.} 
\ee
This Majorana mass term for the gaugino  is localised on the boundary which means that  upon performing a KK decomposition this produces a term mixing all KK modes to each other given by 
\be
\mathcal{L}_{\text{soft}}  = \sum_{m, n}\frac{g^2_{4}  }{2}  \frac{1}{2} \hat{M}\tilde{B}_{1/2}(0) \lambda_n \lambda_m f_n^{(2)}(\ell)f_m^{(2)}(\ell) +  \text{c.c.}   
\ee
in which we have expressed the answer in terms of the dimensionless $4D$ gauge coupling  $g^2_{5}= g^2_{4} \ell$.

  In addition to these masses there are also Dirac type Kaluza-Klein masses which mix the positive parity gaugino modes with the negative parity bulk fermion modes.  In general, one must therefore take into account both to understand the gaugino spectrum.     In practice the easiest way to find the gaugino masses is to include the contribution from   \refe{gauginosoft} in the boundary conditions placed on the KK mode expansions as shown in the appendix of  \cite{Marti:2001iw}, similar to the flat case in \cite{ArkaniHamed:2001mi}.   
  
\subsection{Sfermion masses}
\begin{figure}[ht]
\centering
\includegraphics[scale=0.8]{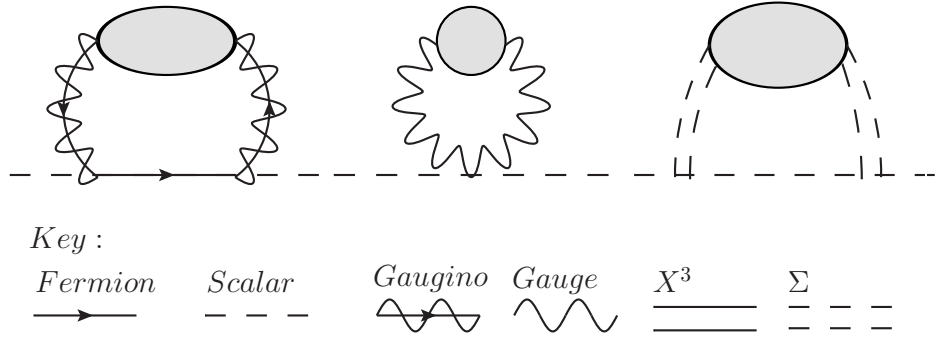}
\caption{The graphical description of the contributions of the two-point functions to the soft sfermion masses. The ``blobs'' represent current correlators localised on the IR brane at $y=\ell$.  The scalar external legs are the sfermions located on the UV brane.  The first diagram represents the current correlator $\braket{j_{\alpha}\bar{j}_{\bar{\alpha}}}$ being mediated by the bulk gaugino $\gl$ from the IR brane to the UV brane at $y=0$.  The second diagram represents mediation of $\braket{j^{\mu} j^{\nu}}$ due to the bulk gauge boson and the final diagram represents mediation of the scalar current correlator $\braket{JJ}$ due to the negative parity bulk scalar $\Sigma$. This is the complete supertraced combination of diagrams for gauge mediation \cite{Meade:2008wd,McGarrie:2010kh}.}
\label{chapter6:figure2}
\end{figure}

The sfermion masses of the MSSM can be determined from the $\tilde{C}_{s}$ current correlators of the SUSY breaking fields on the IR brane and the propagation of this breaking by the vector multiplets in the bulk, up to the UV brane as shown in figure \ref{chapter6:figure2}.    The full momentum dependence of the current correlators should be taken into account as they form part of a loop on the scalar propagator.   The vertex couplings can all be obtained from expanding out a canonical K\"ahler potential for a chiral superfield.    To massage the answers into their final form  one makes use of the  representations   \cite{McGarrie:2010kh,Mirabelli:1997aj}
\be
\delta (0) =\frac{1}{2\ell} \sum_{n} 1   \, , \quad  0 = \sum_n (-1)^n \, .  \label{delta0}
\ee
Judicial use of the later identity allows us to replace factors of $m_n^2$ which occur in the rightmost diagram of  figure \ref{chapter6:figure2} with $p^2$. 
 
 The final result may be written by first defining a positive parity bulk field in the AdS background, propagating from $y=y$ to $y=y'$ \cite{McGarrie:2010kh,Ichinose:2006en,Mirabelli:1997aj,Ichinose:2007zz}
\be
\tilde{G}(y,y')=\frac{1}{2\ell}\sum_{n} \frac{f^{(2)}_{n}(y)f^{(2)}_{n}(y')}{p^2+m^2_{n}} \label{gprop}\, .
\ee
 Then the sfermion mass formula is found to be 
\begin{equation}
m_{\tilde{f}}^2= \sum _r g_{r(5d)}^4  c_2(f;r)E_r
\end{equation}
where
\begin{equation}
E_r= -\! \!\int\! \frac{d^4p}{ (2\pi)^4}\tilde{G}(0,\ell)\tilde{G}(0,\ell) p^2 \Omega^{(r)} \left(\frac{p^2}{\hat{M}^2} \right). \label{P1}
\end{equation}
$r=1,2,3,$ refer to the gauge groups $U(1),SU(2), SU(3)$. $c_2(f;r)$ is the quadratic Casimir for the representation of sfermion $\tilde{f}$, the scalar  in question. 

  As was discussed in \cite{McGarrie:2010kh}, it is interesting to consider the contributions to these masses in different regimes of warping $k$ relative to $M^2, F$.   Since we have that $m_{KK} \approx k^2 e^{-2kl} $ we can see that in the small or intermediate regimes  $k^{2} \ll F,M^2$ or $ F \le k^{2}\ll M^2$ the Kaluza-Klein modes contribute to the mediation of supersymmetry breaking effects across the bulk whereas for large warping, $F ,M^2 \ll k^{2}$, only the zero modes will contribute significantly and we will have an effective $4D$ model with AdS effects.

If  only the zero modes contribute to propagation we may truncate the KK tower. One may then write the zero mode eigenfunctions in terms of gauge boson zero mode eigenfunctions.  The zero mode gauge boson eigenfunctions are flat and one obtains
\begin{equation}
E_r= -\! \!\int\! \frac{d^4p}{ (2\pi)^4}\frac {1}{\ell^{2}} \!  \frac{1}{p^2} \Omega^{(r)}\left(\frac{p^2}{\hat{M}^2} \right). \label{P12}
\end{equation}
Defining $g^2_{(5d)}/\ell=g^2_{(4d)}$, we recover exactly the four-dimensional GGM answer \cite{Meade:2008wd}, with $F$ and $M$ replaced with $\hat{F}$ and $\hat{M}$. This is an effective four-dimensional limit, however as the current correlators are a function of $\hat{M}$, we will still find some suppression due to the warp factor.

\subsection{The subleading contribution to sfermion masses}
A subleading contribution to the scalar soft mass arises from a double mass insertion of the gaugino. In the four dimensional limit $M \ll k$ one finds 
\be
 \delta m_{\tilde{f}}^2= \sum_r c_2(f;r) \frac{g^6_{4d}}{2} \int \frac{d^4p }{(2\pi)^4}\frac{1}{p^4}(\hat{M}\tilde{B}_{1/2}(p^2/M^2))^2.
\ee
In the regime $k\ll M$ one finds 
\be
\delta m_{\tilde{f}}^2= \frac{g^6_{5d}}{16\pi^2 \ell^3}(k\ell)^3 e^{-k\ell} \left(\hat{M}\tilde{B}_{1/2}(0)\right)^2[3/2 + \gamma +\log \Lambda/2 ].
\ee
which is hierarchically suppressed by $e^{-k\ell}$ as expected. This diagram may act as a lower bound on the ratio of gaugino to sfermion masses and when $k\ll M$, the bound is given by 
\be
\frac{m^2_{\lambda} }{m^2_{\tilde{f}}}\lesssim (\frac{4\pi}{\alpha})(k\ell)^2 e^{k\ell}
\ee
and it seems we can have an exponential hierarchy between the two soft mass scales in this limit only.
\subsection{Hypermultiplet scalar masses}
\begin{figure}[ht]
\centering
\includegraphics[scale=0.8]{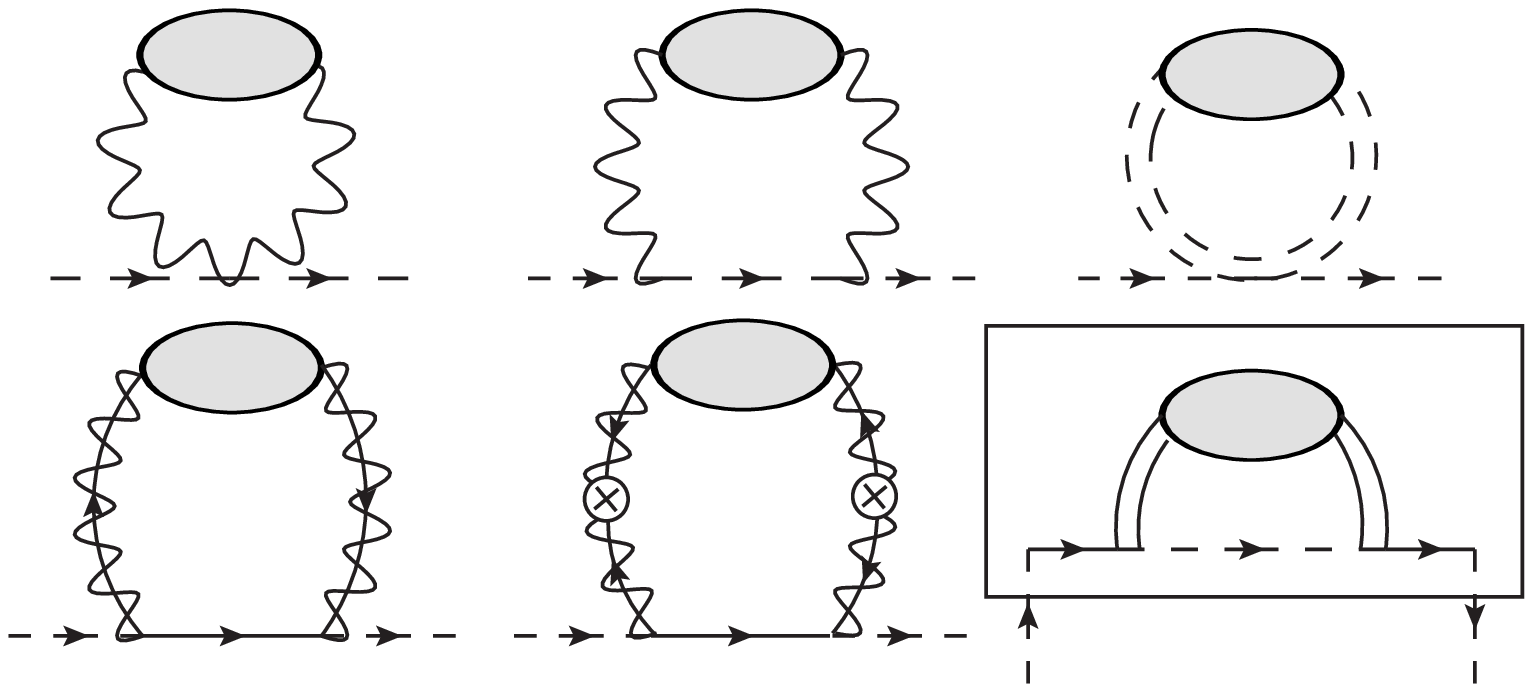}
\caption{The graphical description of the contributions of the two point functions to the hypermultiplet scalar masses. The current correlators (blobs) are located on the IR brane at $y=\ell$.  The external hyperscalar zero mode legs are in the bulk and one must integrate over all possible positions of the external legs.} \label{figure3}
\end{figure}
The supersymmetry breaking masses of a bulk hypermultiplet scalar zero mode may be computed using the general gauge mediation prescription as it decouples from the hidden sector in the limit $\alpha_{i}\rightarrow 0$. The diagrams needed to compute the soft mass are found in figure \ref{figure3}.  When computing the mass, the second diagram vanishes due to transversality. To compute the diagrams, one again takes the current correlators to be brane localised and do not preserve the incoming and outgoing $m_{n}$ eigenmass. The external hypermultiplet legs with mass $m_{n}=0$ eigenfunctions must be specified when computing the diagrams and the vertices to the external hyperscalar must be integrated over all of $y$.  The orthornormality condition \refe{ortho} is used in this integration over $y$.  The right column of diagrams may be collected together by use of \refe{delta0}.  The rectangle on the last diagram signifies that this diagram is completely localised on the IR brane.  The diagram with the symbol $\otimes$, represents a bulk propagator that couples the positive parity fermion $\gl$, to the negative parity fermion $\chi$. 

We find the zero mode hyperscalar soft mass is given by
\begin{equation}
E_r= -\! \!\int\! \frac{d^4p}{ (2\pi)^4}\frac {1}{2\ell^2}\sum_{n} \!  \frac{f^{(2)}_n (\ell)f^{(2)}_n(\ell)}{p^2+m_{n}^{2}}\frac{p^{2}}{p^2+m_{n}^{2}} \Omega^{(r)}\left(\frac{p^2}{\hat{M}^2} \right). \label{P11}
\end{equation}
It should be noted that this equation has only a single sum as the hyperscalar vertices do preserve incoming and outgoing $m_{n}$, unlike for the branes which do not.
The momentum integral is UV divergent as expected \cite{Puchwein:2003jq,Georgi:2000ks}: as we must integrate over all of $y$ for the sewing points of externals hyperscalar legs, there is no finite separation between the bulk hyperscalar and the IR brane to suppress large momentum contributions in the loop.
\subsection{Vacuum energy}
The propagation of supersymmetry in the bulk also produces a nonzero vacuum energy. Each vacuum diagram is generated by forming a closed loop from one end of the current correlator to the other, with the fields that mediate the effects of those current correlators.  There are in fact four diagrams: a closed loop with the gauge boson, gaugino, $\Sigma$ and $X^3$.  The field $X^3$ is non propagating and never leaves the IR brane. The diagram of $X^3$ combines with that of the scalar $\Sigma$ to generate the contribution from the current correlator $C_{0}$. The vacuum energy density is 
\begin{equation}
\mathcal{E}=\sum_{r} \frac{1}{2} g^{2}_{5}  d_{G_{r}} \int \frac{d^{4}p}{(2\pi)^{4}}\tilde{G}(\ell,\ell) p^2\Omega^{(r)}\left(\frac{p^2}{\hat{M}^2} \right). \label{Vacuum1}
\end{equation}
$d_{G}$ is the dimension of the adjoint representation of the gauge group labeled by $r$. The vacuum energy is UV divergent. In the next section we will demonstrate how the Casimir energy may be extracted from this formula, in the $k\ell\gg 1$ limit. 
\section{Generalised messenger sector}\label{section:generalisedmess}
In this section we give a concrete description of matter content of the IR SUSY breaking brane following the construction of \cite{Martin:1996zb}.  We consider two sets of $SU(N)$ vectorlike Chiral superfield messengers $\phi_{i},\tilde{\phi}_{i}$ coupled to a SUSY breaking spurion $X= M + \theta^2 e^{-\sigma} F $. Generalisations to arbitrary hidden sectors are just a straightforward application of the results of  \cite{McGarrie:2010kh,Marques:2009yu}.  The superpotential, which is localised on the IR brane, is 
\be W  =X \eta_{i}\
\phi_i \ti \phi_i . \label{superpotentialchapter6:2}\ee
In principle $\eta_{ij}$ is a generic matrix which may be diagonalised to its eigenvalues $\eta_{i}$ \cite{Martin:1996zb}.  The index $i$ labels the number of messengers from $1$ to $N$.  The messengers on the SUSY breaking brane will couple to the bulk vector superfield as
\be \delta {\cal L} = \int d^2\theta d^2\bar\theta e^{-2\sigma}\left(\phi^\dag_i
e^{2 g_{5} V^a T^a} \phi_i + \ti\phi^\dag_i e^{-2 g_{5} V^a T^a}
\ti\phi_i\right) + \left(\int d^2\theta\ e^{-3\sigma}  W +
c.c.\right) .\label{hiddensectorchapter6} \ee
We can extract the multiplet of currents from the kinetic terms in the above Lagrangian.  It is easiest to work in canonical fields and canonical currents as discussed in section \ref{section:warpcurrents}.  It is also useful to absorb warp factors into the mass scales such that
\be
e^{-k\ell}M=\hat{M}\, , \quad
e^{-2k\ell}F=\hat{F}\, , \quad
e^{-k\ell}k=k^* \, .   
\ee

\subsection{Gaugino masses}
The gaugino soft mass matrix is given by 
\be
\mathcal{L}_{\text{soft}}  = \sum_{m, n}\frac{g^2_{4}  }{2} \frac{1}{2}  \hat{M}\tilde{B}_{1/2}(0) \lambda_n \lambda_m f_n^{(2)}(\ell)f_m^{(2)}(\ell) + \text{c.c.}   
\ee
It should be clear that there is a mass term for coupling all gaugino modes of the Kaluza-Klein tower to all other modes.  Consider just the zero mode portion of this matrix given by 
 \be
 \frac{1}{2} M_{\lambda_0}  \lambda_0 \lambda_0 \equiv  \frac{g^2_{4}}{2}  \frac{1}{2}   \hat{M}\tilde{B}_{1/2}(0)  \lambda_0 \lambda_0 f_0^{(2)}(\ell)f_0^{(2)}(\ell)  =
 \frac{g^2_{4}}{2}   \frac{1}{2}  \hat{M}\tilde{B}_{1/2}(0)  \lambda_0 \lambda_0  
  \ee
 where the   equality is due to the fact that the zero mode for the gauge field is flat, $f^{(2)}_0(y) = 1$.  
      This allows us to establish for the minimal messenger model the scale associated with the gaugino mass 
 \be 
  M^{(r)}_{\lambda_0}  =  \frac{\alpha_r}{4\pi} \Lambda_G\, , \quad \Lambda_{G}=\sum_{i=1}^{N}[e^{-k\ell}\frac{d_{r}(i) F}{M }2 g(x_{i})], 
\ee
 in which have used the result for $ \hat{M}\tilde{B}_{1/2}(0)$ obtained in flat space analogue  \cite{McGarrie:2010kh}  but with the mass  warped accordingly.   The label $r=1,2,3$ refers to the gauge groups $U(1),SU(2),SU(3)$, $d_{a}(i)$ is the Dynkin index of the representation $i$ and
\be 
g(x)= \frac{(1-x)\log(1-x)+(1+x)\log(1+x)}{ x^2}
 \ee
where  $x_{i}=\frac{\hat{F}}{\eta_{i}\hat{M}^2}=\frac{F}{\eta_{i}M^2}$ and $g(x)\sim 1$ for small $x$ \cite{Martin:1996zb}.  

If we wish to consider the full mass matrix we can approximate $f^{(2)}_{n >0}(\ell) = \sqrt{2 k\ell } (-1)^n$  giving rise to 
\be
\frac{1}{2} M_{\lambda_{n,m}}  \lambda_n\lambda_m  =  \frac{g^2_{4}}{2}  (-1)^{(n+m)}  k\ell  \hat{M}\tilde{B}_{1/2}(0)  \lambda_n \lambda_m  \, .
\ee
From this one can see the soft mass contribution associated with the nonzero modes is a factor    $k\ell $ greater than those of the zero mode alone.    
\subsection{Sfermion masses}
The sfermion masses are sensitive to the warping $k^*=k e^{-k\ell}$. Due to the complicated nature of the bulk propagators, we will only comment on the limit when $k\ell$ is large. For the KK modes to contribute, we require that $k$ is small such that $k^2 \ll F,M^2$ or intermediate such that $F \le k^{2}  \ll 
M^2$. We take the exact results of \refe{P1} with \refe{gprop}.  As demonstrated in Appendix \ref{limits},  we approximate the full KK tower of propagators with the eigenfunctions and eigenstates of the heavier modes, which are less localised to the IR brane so will contribute most to the propagation across the bulk.  We then complete a Matsubara summation of the tower of modes. We find the propagator to be
\be
\tilde{G}(0,\ell)\sim \frac{ 2e^{kl/2}(-1)^n}{p \sinh(p/k^*)} ,
\ee
which gives
\be 
E_r \simeq -\! \!\int\! \frac{d^4p}{ (2\pi)^4} \frac{4 e^{k\ell}}{\sinh^2(p/k^*)}  \Omega^{(r)}\left(\frac{p^2}{\hat{M}^2} \right).
\ee
Physically, the tower of KK modes suppresses large momenta contributions from the two point functions on the IR brane, which can be seen from the behaviour of $\sinh^2(p/k^*)$.  In this regime we may expand the current correlators for small momenta as found in \cite{McGarrie:2010kh}, valid when $\frac{p^2}{M^2}\rightarrow 0$:
\be
 \Omega^{(r)}\left(\frac{p^2}{\hat{M}^2} \right) \approx -\frac{1}{(4\pi)^2}\frac{2 d_{r}}{3}x^2 h(x) + O(p^2)
\ee  
with
\be h(x) =\frac{3}{2}[\frac{4+x-2x^2}{x^4}\log(1+x)+\frac{1}{x^2}] + (x\rightarrow -x).
\ee
$h(x)$ for $x<0.8$ can be reasonably approximated by $h(x)=1$ \cite{McGarrie:2010kh}.  In this limit, the function is independent of $p^2$.  Finally, the momentum integral is evaluated and the sfermions masses can then be written as 
\be
 m_{\tilde{f}}^2 \sim 2 C_{\tilde{f}}^r\sum^{3}_r (\frac{\alpha_{r}}{4\pi})^2 \sum_{i}^{N} 4(k^2 \ell)^2 e^{-3k\ell}\zeta(3) d_{r}(i)|\frac{F}{\eta_{i}M^2}|^2 h(x_{i}).
\ee
$C_{\tilde{f}}^r$ is the quadratic Casimir of the $\ti f$ scalar in question, in the gauge group $r$. The sfermion scale $\Lambda_S^2$ is
\be 
\Lambda_S^2 \sim  2 \sum_{i=1}^{N}\left(\frac{\ell k^2 }{|\eta_{i}M |}\right)^2  \left|\frac{F}{M}\right|^2  e^{- 3k\ell}     \zeta(3)
d_{r}(i) h(x_i)\label{warpedLSMGM}.\ee
Next, we may turn to the limit $F,M^2 \ll k^2$.  In this limit only the zero modes contribute significantly to the mediation across the bulk.  We again start from \refe{P1} and keep only the zero modes in the bulk propagators.  One can rewrite this in terms of the zero mode gauge boson eigenfunctions, which are flat \cite{Pomarol:1998sd,Gherghetta:2000kr,Marti:2001iw}.  We arrive at
\begin{equation}
E_r= -\! \!\int\! \frac{d^4p}{ (2\pi)^4}\frac {1}{\ell^{2}}\frac{1}{p^{2}} [3\tilde{C}_1^{(r)}(p^2/\hat{M}^2)-4\tilde{C}_{1/2}^{(r)}(p^2/\hat{M}^2)+\tilde{C}_{0}^{(r)}(p^2/\hat{M}^2)].
\end{equation}
Similarly, one may use the full answer \refe{P1} and expand the current correlators in the limit $\frac{p^2}{M^2}\rightarrow \infty$ as was explored in the appendix of \cite{Mirabelli:1997aj}. The above equation has been evaluated before \cite{Martin:1996zb,Meade:2008wd,Marques:2009yu}, which we rescale by use of $\hat{F}$ and $\hat{M}$. The result is 
\be
 m_{\tilde{f}}^2 = 2 \sum^{3}_r C_{\tilde{f}}^r (\frac{\alpha_{r}}{4\pi})^2 \sum_{i}^{N} e^{-2k\ell} d_{r}(i)|\frac{F}{M}|^2f(x_{i}) .\label{warped4dlimit}
\ee
$f(x_{i})$ is given by 
\be
f(x)=\frac{1+x}{x^2}[\log (1+x)-2\text{Li}_{2}(\frac{x}{[1+x]})+\frac{1}{2}\text{Li}_{2}(\frac{2x}{[1+x]})]+(x\rightarrow -x) .
\ee
For $x<1$, $f(x)\sim 1$. The additional factor of $e^{-2k\ell}$ arises from the dimensionful ratio of $|\frac{\hat{F}}{\hat{M}}|^2$, in the supertrace of $\tilde{C}$ terms. The factors of warping cancel in the ratio $\Lambda^2_{G}/\Lambda^2_{S}$ in this ``$4D$ limit''.  These results show that depending on the ratio of $k$ to $M$, one may obtain an ordinary ``gauge mediated'' result or ``gaugino mediated'' spectrum.  
\subsection{Hyperscalar mass}
We now focus on computing the zero mode hyperscalar soft mass. We start with the exact result found in \refe{P11} and approximate the warped propagator KK mode eigenfunctions by 
\be 
f^{(2)}_{n}(\ell)f^{(2)}_{n}(\ell)\simeq 4 (k \ell) ,
\ee
to find 
\begin{equation}
m_{H_{0}}^2= \sum _r g_{r(5d)}^4 c_2(f;r)D_r
\end{equation}
where, after performing a Matsubara frequency summation,
\be 
D_r \sim \! -\!\int\! \frac{d^4p}{ (2\pi)^4}\frac{2(k\ell)}{\ell^2 k^*}\frac{\coth (p/k^*)+ p/k^*\text{csch}^2(p/k^*)  }{2 p}  \Omega^{(r)}\left(\frac{p^2}{\hat{M}^2} \right). \label{good}
\ee
$D_{r}$ is UV divergent due to the loop in the hypermultiplet diagrams not always being spatially separated by the interval. We would like to extract the $\slashed{D}_{r}$, the $k^*$ dependent part of $D_{r}$. Upon subtracting the UV limit of the integrand, we find
\be
 \slashed{D}_r= -\! \!\int\! \frac{d^4p}{ (2\pi)^4}\frac {k}{\ell k^* }[\frac{\coth(p/k^*)+(p/k^*)\text{csch}^2(p/k^*)-1 }{p}] \Omega^{(r)}\left(\frac{p^2}{\hat{M}^2} \right). 
\ee
We use the expansion of current correlators in the limit $\frac{p^2}{M^2}\rightarrow 0$ found above, to obtain 
\be
 m_{\tilde{H}}^2 \sim \frac{2}{3} C_{\tilde{f}}^r\sum^{3}_r (\frac{\alpha_{r}}{4\pi})^2 \sum_{i}^{N} 4k^2  (k\ell) e^{-2k\ell}\zeta(3) d_{r}(i)|\frac{F}{\eta_{i}M^2}|^2 h(x_{i}).
\ee
This result is less warped than the sfermion soft mass \refe{warpedLSMGM}.  Now we comment on the limit of $M\ll k$.  In this AdS $4D$ limit one obtains the same result as for the sfermion masses \refe{warped4dlimit}.
\subsection{Vacuum energy}
The vacuum energy can be computed starting from \refe{Vacuum1}. The vacuum energy is UV divergent. To obtain the finite part, one may extract the UV limit of the momentum integrand as was carried out for the hyperscalar masses above, or one may use a contour trick in the matsubara summation, outlined in \cite{Mirabelli:1997aj}. Either way leads to the same answer.  For definiteness we use the contour trick and obtain the Casimir energy in the limit of $k\ll M$,
\begin{equation}
\mathcal{E}_{\text{Casimir}}\sim \sum_{r}g^{2}_{5} d_{G_{r}}\int \frac{d^{4}p}{(2\pi)^{4}}\frac{k \ell e^{2k \ell} p}{ e^{2p /k^*}-1} \Omega^{(r)}\left(\frac{p^2}{\hat{M}^2} \right).\label{warpedcasimirenergy}
\end{equation}
Evaluating this for the case of minimal messengers, we find
\be
\mathcal{E}_{\text{Casimir}}\sim  - \frac{1}{2} \sum_{r}\sum_{i}^{N}\frac{g^{2}_{4d (r)}  d_{G_{r}} d_{r}(i) \zeta (5)}{128 \pi^4} (k\ell)k^4 e^{-4k \ell}\left|\frac{F}{\eta_{i}M^2}\right|^2 h(x_{i}) \label{warpedcasimirenergy2}. 
\ee
The warped factor $e^{-4k \ell}$ dominate this result as $k\ell\gg 1$ and we find effectively zero Casimir energy in this regime.  In the $4D$ AdS limit $k\gg M$, we find
\be
\mathcal{E}_{\text{Casimir}} =-\frac{1}{2} \sum_{i}\sum_{r}\frac{4 g^2_{4d(r)} } {(4\pi)^4} d_{g_{r}} d_{r}(i)|\eta_{i} F|^2 e^{-4k\ell} \log (\frac{k}{\eta_{i}M}) 
\ee
which is just as suppressed by the warp factor $e^{-4k \ell}$.  This negative contribution to the Casimir energy is solely from $F$ term breaking of supersymmetry.  It would be interesting to explore the contributions from $D$ term breaking as they have positively signed contributions to the Casimir energy \cite{Mirabelli:1997aj}.  Further, one may include supergravity contributions and explore minimising the total vacuum energy.
\section{Discussion}
General gauge mediation is a powerful model independent framework for gauge mediated supersymmetry breaking.  In this chapter we have shown how the GGM approach can be applied to five-dimensional warped models. 

One must take great care in these warped models, with regard to the ratio $k/M$.  When $k/M \gg 1$  only the zero mode of the vector superfields in the bulk will significantly contribute to the mediation of supersymmetry breaking effects and one will have gauge mediation. We find a natural geometric interpretation of the soft terms being hierarchically small as all soft mass scales arise with at least one factor of $e^{-k\ell}$.  The warp factors cancel in the ratio $\Lambda^2_{G}/\Lambda^2_{S}$ and thus the model behaves like a typical gauge mediated scenario.  This limit closely corresponds to logarithmic running of the standard model gauge coupling up to $k$, as only the zero mode significantly contributes.    These statements appear more obviously in the framework of this chapter as both running gauge couplings and susy breaking soft terms are types of renormalisation.  These results somewhat challenge the common lore, whereby gaugino mediation was assumed: we have shown that gauge mediation appears to be far more likely in this setup, than gaugino mediation.

Next when $k/M \ll 1$, the heavy kk modes that are more uniformly distributed across the bulk will also contribute to the mediation and a gaugino mediated limit will be obtained.   Furthermore if one implements gaugino mediation ($k \ll M$) then one may only use logarithmic running from $M_{z}$ to $k$ and then important power law contributions from the heavy kk modes will be introduced between $k$ and $M$.  In this limit we still expect $\Lambda_G$ to remain of the same order, $\Lambda_S^2$ acquires a suppression by both a dimensionless ratio, as happens in the non-warped case, {\it and} an additional warp factor $e^{-k\ell}$.  In this way it is feasible to obtain not only a hierarchy  between SUSY breaking and Planck scales but also between different soft mass scales.   An interesting question is what happens when $k\sim M$? This regime cannot be computed analytically but we may speculate that it will behave much more similarly to the hybrid mediation of the previous chapter.
We briefly comment on supergravity calculations in the bulk. Generally these corrections will be more suppressed than the gauge mediation contributions as long as there is some difference in scales between $M$ and $M_{planck}$.
 
We believe that this framework can help to make some useful progress in exploring the phenomenology of these models \cite{Bouchart:2011va} in a rigorous way and help to extend the use of certain dualities to increase our understanding of strongly coupled hidden sectors. We hope that this work may be of some assistance to unlocking evidence of supersymmetry and extra dimensions at the LHC.

\setcounter{equation}{0}
\chapter{Conclusion and Outlook}\label{chapter7}
This thesis has emphasised the theoretical construction of higher dimensional models of the mediation of supersymmetry breaking. These models are effective descriptions of an intermediate regime of energy, whose ultimate completion may be string theory.  We have also argued that even entirely four dimensional models may appear to have higher dimensional descriptions, due to the way strongly coupled gauge groups such as QCD, generate composite vector mesons.  In fact we strongly believe that the analogous effect of vector meson dominance found in QCD, will play a role in the mediation of supersymmetry breaking effects from the strongly coupled hidden sector.  The generic signature of which are suppressed scalar mass soft terms, before renormalisation group evolution.  We hope that more emphasise on this prediction is made in future work and is hopefully borne out by future experiments.

Discovering broken supersymmetry at the Large Hadron Collider will open up the possibility to explore a whole new sector of matter and interactions.  It is quite possible this hidden sector that breaks supersymmetry will include a strongly coupled gauge group and may admit a low energy effective description.  If this is the case, understanding this sector will become not just a topic for the theorist but will also become a task for the experimentalist.  From this perspective it will become more obvious that the techniques applied to QCD physics of pions and mesons will become more and more important in the understanding of the hidden sector.  With this in mind our outlook is to apply the tools and knowledge of QCD, such as form factors, spectral sum rules, operator product expansions and comparision with experimentally measurable properties to the models we have outlined in this thesis.  Just as mesons and baryons are made of fundamental quarks, we think it important to develop tools to extract the underlying fundamental degrees of freedom from the composite degrees of freedom we are likely to observe.  Vector mesons and mesinos of the hidden sector as Kaluza-Klein modes of the standard model gauge and gaugino fields would in this case play an important role as a window into the hidden sector.  Finally, refinements of these models will also become important.  The simplest two site model of an extra dimension  in chapter \ref{hybridchapter} whilst easiest to extract results from is probably too simple.  Models based on an AdS metric as appear in chapter \ref{chapter6} add further complication, but it is likely that the model that best fits future experimental results will be more complicated than this.  We imagine that an AdS profile may give the leading order effects, which must be supplemented by a smoother transition of the metric nearer to the fixed points at each end of the interval, for example.   Another direction in this respect is to consider how and in what ways the tools of Seiberg Duality in giving an IR description of the hidden sector overlap with tools from AdS/CFT to give an effective description of a strongly coupled system.  Perhaps this overlap may aid in making more quantative predictions about supersymmetry breaking and also strongly coupled theories in general.

\section{Summary of key results}
\subsection{Gaugino masses}
Lightest gaugino soft mass in terms of the current correlator is given by 
\begin{center}
\begin{tabular}[c]{|c|c|} \hline
Model & Gaugino Mass \\
\hline
4d & $m_{\lambda}=g^2 M\tilde{B}_{1/2}(0)$ \\
5d & $m_{\lambda}=\frac{g^2_{5d}}{\ell} M\tilde{B}_{1/2}(0) $\\
N-site lattice & $m_{\lambda}=\frac{g^2}{N} M\tilde{B}_{1/2}(0)$ \\
Warped & $m_{\lambda}=\frac{g^2_{5d}}{\ell}  \hat{M}\tilde{B}_{1/2}(0)  $
\\ \hline
\end{tabular}
\end{center}

For the particular case where the hidden sector is described by  a set of N chiral superfield messengers coupled to a supersymmetry breaking spurion $X= M +\theta^2 F$
\be
W=X \eta_{i}\Phi_i \tilde{\Phi} _{i}
\ee
The lightest gaugino mass is  given by 
\be 
m^r_{\gl_{0} }= \frac{\alpha_{r}}{4\pi} \Lambda_{G}\ , \ \ \ \  \Lambda_{G}=\sum_{i=1}^{N}[\frac{d_{r}(i) F}{M}g(x_{i})]
\ee
The label $r=1,2,3$ refers to the gauge groups $U(1),SU(2),SU(3)$, $d_{r}(i)$ is the Dynkin index of the representation of $\Phi_{i},\tilde{\Phi}_{i}$ and
\be 
g(x)= \frac{(1-x)\log(1-x)+(1+x)\log(1+x)}{ x^2}
 \ee
where  $x_{i}=\frac{F}{\eta_{i}M^2}$.  $g(x)\sim 1$ for small $x$ \cite{Martin:1996zb}.  For the warped case one must substitute $\hat{F}=e^{-2k\ell}F$ and $\hat{M}=e^{-k\ell}M$.
\subsection{Sfermion masses at leading order}
The brane localised sfermion mass formulas at leading order are given by 
\be
m^2_{\tilde{f}} =\sum_{r}g^4 c_{2}(f;r) E_r
\ee
where we have 
\begin{itemize}
\item 4d: $E_r=-\int \! \! \frac{d^4p}{(2\pi)^4} \frac{1}{p^2} \Omega^{r}(\frac{p^2}{M^2})$ 
\item 5d: $E_r=-\int \! \! \frac{d^4p}{(2\pi)^4 } \sum_{n,\hat{n}}\frac{(-1)^{n+\hat{n}}}{p^2+p^2_{5}} \frac{p^2}{p^2+\hat{p}^2_{5}}\Omega^{r}(\frac{p^2}{M^2})$ 
\item N-site lattice:  $E_r=-\int \! \! \frac{d^4p}{(2\pi)^4 }p^2  (\braket{p^2; 0, N-1})^2\Omega^{r}(\frac{p^2}{M^2})$ 
\item 2-site lattice: $E_r=-\int \! \! \frac{d^4p}{(2\pi)^4 }\frac{1}{p^2}  \left[\frac{m^2_{v}}{p^2+m^2_{v}}\right]^2\Omega^{r}(\frac{p^2}{M^2})$ 
\item Warped: $E_r=-\int \! \! \frac{d^4p}{(2\pi)^4 }p^2 \tilde{G}(0,\ell)\tilde{G}(0,\ell)\Omega^{r}(\frac{p^2}{\hat{M}^2})$ 
\end{itemize}
$\Omega^{r}(\frac{p^2}{\hat{M}^2})$ is the super-traced sum of current correlators defined in section \ref{sec:Frame}.  The propagator $\braket{p^2; 0, N-1}$ is defined in section \ref{section:Frameworkdecon} and $\tilde{G}(y,y')$ in section \ref{section:results}. $C_2(f;r)$ is the quadratic Casimir of the sfermion $\tilde{f}$ in question, in the gauge group $r$.

For the particular case where the hidden sector is described by  a set of N chiral superfield messengers mentioned above the mass formula at leading order are given by 
\be
 m^2_{\tilde{f}}=2\sum_{r=1}^{3}C_2(f;r) (\frac{\alpha_{r} }{4\pi})^2 \Lambda^2_{S}
\ee
\begin{itemize}
\item 4d:  $\Lambda^2_{S}=|\frac{F}{M} |^2 \sum_{i=1}^{N} d_{r}(i) f(x_i)$
\item 5d $(1/\ell <M)$ :  $\Lambda^2_{S}=|\frac{F}{M} |^2 \sum_{i=1}^{N}|\frac{\zeta(3)}{\eta_i M\ell}|^2  d_{r}(i) h(x_i)$
\item N-site lattice $(1/(N a) <M)$:  $\Lambda^2_{S}=\sum_{i=1}^{N}|\frac{F}{\eta_i M^2 a} |^2   d_{r}(i)\frac{1}{128\pi^4 } \frac{2}{3}h(x_i)\mathcal{I}$
\item 2-site lattice: $\Lambda^2_{S}=|\frac{F}{M} |^2\sum_{i=1}^{N} |\frac{d_{r}(i)}{\eta_i M\ell}|^{\rho}\ \ ,\ \  0\leq\rho \leq 2$.
\item Warped 4d $(k \gg M)$: $\Lambda^2_{S}=e^{-2k\ell}|\frac{F}{M} |^2 \sum_{i=1}^{N} d_{r}(i) f(x_i)$
\item Warped 5d $(k \ll M)$:  $\Lambda^2_{S}\sim e^{-3k\ell}|\frac{F}{M} |^2 \sum_{i=1}^{N}(\frac{2k^2 \ell}{|\eta_i M|})^2 d_{r}(i) h(x_i)$.
\end{itemize}
where in the above $h(x_i)$ and $f(x_i)$ may be found in section \ref{sec:general}, $a$ is a lattice spacing and $\mathcal{I}$ may be found in section \ref{latticesfermionmasses}.
\subsection{Sfermion masses at subleading order}
There are also subleading contributions in $\alpha_r$ which are sometimes of the same scale as the leading order contributions.  These contributions arise as a double mass insertion of the gaugino mass and are given by 
\begin{itemize}
\item 4d: $ \delta m^2_{\tilde{f}}= \sum_r c_2(f;r) \frac{g^6}{2}\int \frac{d^4 p}{(2\pi)^4}\frac{1}{p^4}(M \tilde{B}_{1/2} (p^2/M^2))^2$
\item 5d: $\delta m^2_{\tilde{f}}= \sum_r c_2(f;r)\frac{g^6}{2}\int \frac{d^4 p}{(2\pi)^4}p^2 \sum_{n,\hat{n}}\frac{(-1)^{n+\hat{n}}}{p^2+p^2_{5}} \frac{M \tilde{B}_{1/2} (p^2/M^2)}{p^2+\hat{p}^2_{5}}\frac{M \tilde{B}_{1/2} (p^2/M^2)}{p^2+\hat{\hat{p}}^2_{5}}$
\item N-site:  $\delta m^2_{\tilde{f}}= \frac{ \sum_r c_2(f;r)g^6}{2}\!\int \frac{d^4 p}{(2\pi)^4}p^2   \braket{p^2; 0, N\!-\!1}^2  \braket{p^2; N\!-\!1, N\!-\!1}(M \tilde{B}_{1/2} (p^2/M^2))^2$
\item 2-site lattice: $\delta m^2_{\tilde{f}}= \sum_r c_2(f;r)\frac{g^6}{2}\int \frac{d^4 p}{(2\pi)^4}\frac{m^4_v}{p^4(p^2+m^2_{v})^2}(M \tilde{B}_{1/2} (p^2/M^2))^2$
\item Warped: $\delta m^2_{\tilde{f}}= \sum_r c_2(f;r)\frac{g^6}{2}\int \frac{d^4 p}{(2\pi)^4}p^2 (\tilde{G}(0,\ell))^2\tilde{G}(\ell,\ell)(M \tilde{B}_{1/2} (p^2/M^2))^2$
\end{itemize}
For the the particular case where the hidden sector is described by  a set of N chiral superfield messengers mentioned above, this subleading contribution may be \emph{estimated}  to be 
\be
 m^2_{\tilde{f}}=2\sum_{r=1}^{3}C_2(f;r) (\frac{\alpha_{r} }{4\pi})^3 \Lambda^2_G 
\ee
\begin{itemize}
\item 4d:  $\Lambda^2_G=|\sum_{i=1}^{N}[\frac{d_{r}(i) F}{M}g(x_{i})]|^2 $
\item 5d $(1/\ell <M)$ :  $\Lambda^2_G=|\sum_{i=1}^{N}[\frac{d_{r}(i) F}{M}g(x_{i})]|^2 $
\item Warped 5d $(k \ll M)$:  $\Lambda^2_{G}\sim  (kl)^3 e^{-3k\ell}|\sum_{i=1}^{N}[\frac{d_{r}(i) F}{M}g(x_{i})]|^2$.
\end{itemize}

\appendix
\setcounter{chapter}{0}
\renewcommand{\chaptername}{Appendix}
\renewcommand{\theequation}{\Alph{chapter}.\arabic{section}.\arabic{equation}}
\addcontentsline{toc}{chapter}{\numberline{}Appendix}

\setcounter{equation}{0}
\chapter{Notation and Conventions}\label{conventions}
This appendix outlines the notation and conventions of $4d$ $\mathcal{N}=1$.  Useful references are \cite{Srednicki:2007qs,Wess:1992cp,Drees:2004jm,Bilal:2001nv,Lykken:1996xt,Signer:2009dx,Krippendorf:2010ui}.
\section{Poincar\'e group and its Algebra}
The four dimensional metric\footnote{We have chosen the mostly plus signature such that the sign of the determinant does not change when extending by additional dimensions.} is given by 
\be
\eta_{\mu\nu}=\eta^{\mu\nu} =\text{diag}(-1,1,1,1)
\ee
The isometries of this spacetime are the Lorentz transformations and translations:
\be
   x^{\mu} \ \  \mapsto    \ \ x'^{\mu}
    =\underbrace{\Lambda^{\mu}_{\nu}}_{\text{Lorentz}} 
    x^{\nu} \ + \  \underbrace{a^{\mu}}_{\text{translation}}
\ee
whereby $\Lambda^{\mu}_{\nu}=\frac{\partial x'^{\mu}}{\partial x^{\nu}}$. These coordinate transformations leave the metric invariant and hence the Lorentz invariant length element $ds^2 = \eta^{\mu\nu}dx_\mu dx_\nu$.  In general $\text{det}(\Lambda)=\pm 1$ and $+1$ defines the proper orientation. Furthermore, the orthocronous transformations have $\Lambda^{00}\ge 1$. The proper orthocronous subgroup of the Lorentz group is often referred to as $SO(3,1)^\uparrow_{+}$ and the semi-direct\footnote{The Lorentz transformation and translations do not commute.} product of $SO(3,1)^\uparrow_{+}$ and translations form the Poincar\'e group: rotations, boosts and translations.  These continous global transformations of spacetime coordinates are generated by a Lie Algebra:
\begin{alignat}{1}
[P^{\mu} \ , \ P^{\nu} ] \ \ &= \ \ 0 \\
[M^{\mu \nu} \ , \ P^{\sigma} ] \ \ &= \ \ i \, (P^{\mu} \, \eta^{\nu \sigma} \ - \ P^{\nu} \, \eta^{\mu \sigma} )\nonumber \\
[M^{\mu \nu} \ , \ M^{\rho \sigma} ] \ \ &= \ \ i\, (M^{\mu \sigma} \, \eta^{\nu \rho} \ + \ M^{\nu \rho} \, \eta^{\mu \sigma} \ - \ M^{\mu \rho} \, \eta^{\nu \sigma} \ - \ M^{\nu \sigma} \, \eta^{\mu \rho} )\nonumber
\end{alignat}
All these generators are bosonic.  We must find representations of the algebra which will act on a relevant vector space, to generate these transformations.  For pure scalar representations of the group, the algebra is trivially satisfied and  $\text{exp}(0)=1$.  For four vectors $x^{\mu}$, one may choose to satisfy the algebra with 
\be 
(P^{\rho})^{\mu}_{\nu}=i\frac{\partial }{\partial x^{\rho}}\delta^{\mu}_{\nu} ,\ \ \, \ \ \ \  ( M^{\rho \sigma})^{\mu}_{\nu} =i(\eta^{\rho\mu}\delta^{\sigma}_{\nu}-\eta^{\sigma \mu}\delta^{\rho}_\nu)
\ee
Which give the finite transformations 
\be 
T^{\mu}_{\nu}= \text{exp}(-ia^{\rho}P_{\rho}\delta^{\mu}_{\nu}  ) ,\ \ \ \, \ \ \ \ \Lambda^{\mu}_{\nu}= \text{exp}(\frac{1}{2}\omega_{\rho\sigma}M^{\rho \sigma})
\ee
on vector representations of the Poincar\'e group. $a^\rho$ and $\omega_{\rho\sigma}$ are infinitesimal group parameters that determine the type of transformation being carried out in terms of the spacetime coordinates. For scalar fields the transformation is $\phi'(x^\mu)\rightarrow \phi(\Lambda^{-1} x^\mu)=\Lambda^{-1} \phi(x^{\mu})$, which may be written as 
\be 
\Lambda= \text{exp}(\frac{1}{2}\omega_{\rho\sigma}(M^{\rho \sigma})^{\mu \nu} x_{\mu}\partial_{\nu})
\ee
such that $\mathcal{L}^{\mu\nu}=i(M^{\mu\nu})^{\rho \sigma} x_{\rho}\partial_{\sigma}= i(x^\mu \partial^\nu- x^\nu \partial^\mu)$. If the vector space is a four component spinor field $\psi^{A}$, there is a representation of $M_{\mu \nu}$ which satisfies the Poincar\'e algebra
\be 
M^{\mu \nu}=-\mathcal{L}^{\mu\nu} + \Sigma^{\mu\nu}
\ee
where 
\be
\Sigma^{\mu\nu}\equiv  \frac{i}{4}[\gamma^\mu,\gamma^\nu] .
\ee
These gamma matrices satisfy the Dirac algebra 
\be 
\{\gamma^\mu,\gamma^\nu \}= -2\eta^{\mu\nu}  \mathbf{I} .
\ee
The \textbf{Weyl representation} of the Gamma matrices is 

\be
\gamma^\mu=\left(\begin{array}{cc}0 &\sigma^{\mu}_{\alpha \dot{\alpha}}\\ 
\bar{\sigma}^{\mu \dot{\alpha} \alpha }&0
\end{array}\right)
,~~\mbox{and}~~~
\gamma_5 =\left(\begin{array}{cc}
-I & 0\\ 
0 & I
\end{array}\right)\,,
\ee
where $\sigma^\mu_{\alpha \dot{\alpha}}=(1, \vec{\sigma})$ and
$\bar{\sigma}^{\mu \dot{\alpha}
\alpha}=(1,-\vec{\sigma})$. $\alpha,\dot{\alpha}$ are spinor indices
of $\text{SL}(2,C)$.  The Pauli spin matrices, $\vec{\sigma}$, are given by 
\be
\sigma^1=\left(\begin{array}{cc}0 &1\\ 
1&0
\end{array}\right)
\ \ ,\ \ \sigma^2=\left(\begin{array}{cc}0 &-i\\ 
i&0
\end{array}\right)
\ \ , \ \ \sigma^3=\left(\begin{array}{cc}1 &0\\ 
0&-1
\end{array}\right).
\ee 
In the Weyl representation of the Gamma matrices, the Lorentz generator for a Dirac spinor may be decomposed
\be
\Sigma^{\mu\nu}=\left(\begin{array}{cc}(\sigma^{\mu \nu})_{\alpha }^{ \; \beta }&0\\ 
0&(\bar{\sigma}^{\mu \nu })^{\dot{\alpha}}_{\; \dot{\beta}}
\end{array}\right)=\frac{i}{4}\left(\begin{array}{cc}  \sigma^{\mu}\bar{\sigma}^\nu-\sigma^{\nu}\bar{\sigma}^{\mu} &0\\ 
0&\bar{\sigma}^{\mu}\sigma^\nu-\bar{\sigma}^{\nu}\sigma^{\mu}
\end{array}\right)
\ee
Where $\Sigma_L^{\mu \nu}=(\sigma^{\mu \nu})_{\alpha }^{ \; \beta }$ is a representation of the generator that acts on left handed Weyl spinors  and similarly $\Sigma_R^{\mu \nu}=-(\bar{\sigma}^{\mu \nu })^{\dot{\alpha}}_{\; \dot{\beta}}$ acts on the right handed Weyl spinors.  As we can decompose the Dirac spinor into smaller representations that each independently transform under the group $SL(2,C)$ (complexified $SU(2)$) we are at liberty to define a two component notation. The \textbf{two component} spinor  transforms as 
\be
\xi'_\alpha = N_{\alpha}^{\; \beta}\xi_{\beta}.
\ee
We define
\be
\epsilon_{\alpha \beta}=\left(\begin{array}{cc} 0&-1 \\ 1&0
\end{array}\right) \ \ \ \ \epsilon^{\alpha \beta}=\left(\begin{array}{cc} 0&1 \\- 1&0
\end{array}\right).
\ee
These can be used to lift and lower indices without introducing a change of sign. The complex conjugate representation is given by $ \bar{\xi}_{\dot{\alpha}}= (\xi^{\alpha})^*$ or equivalently $ \bar{\xi}_{\dot{\alpha}}= (\xi_{\alpha})^{\dagger}$.  A typical four component spinor in a Weyl basis is 
\be
\psi  = \binom{\lambda^{i}_{\alpha}}{\bar{\chi}^{ i  \dot{\alpha}}} 
\ee
where the operation of Staring lifts/lowers and dots/undots:
\be
(\psi)^* = \binom{\bar{\lambda}^{\dot{\alpha}}}{\chi_{\alpha}} 
\ee
The Dagger operation is 
\be
(\psi)^\dagger = ( \bar{\lambda}_{\dot{\alpha}}  ,   \chi^{\alpha} )
\ee
Dirac conjugate is 
\be
\bar{\psi} =  \psi^{\dagger}
\gamma^{ 0}= ( \bar{\lambda}_{\dot{\alpha}}  ,   \chi^{i\alpha} ) \left(\begin{array}{cc} 0&1 \\ 1&0
\end{array}\right)
=(  \chi^{\alpha},\bar{\lambda}_{\dot{\alpha}})
\ee
\section{The Super-Poincar\'e Algebra}
We have found representations of the Lorentz group which have half integer spin and have further decomposed these representations into Weyl spinors, which also have half-integral spin. However all our representations of the Poincar\'e algebra have integral spin.  It is natural to ask if one may extend the algebra such that some representations of the algebra itself have half-integral spin.  This is called \textbf{the super-Poincar\'e algebra}.  The algebra may be written to include either four or two component spinors as generators.  These generators generate an infinitesimal symmetry of an unphysical superspacetime.  In two component notation, the group parameters may be labeled $\epsilon$ and $\bar{\epsilon}$ and are fermionic.  If these are constant parameters then it is a global symmetry, and when $\epsilon(x^{\mu})$ it is a local superspace.  For global $\mathcal{N}=1$ supersymmetry in two component notation in four dimensions, the algebra is given by 
\begin{alignat}{1}
[P^{\mu} \ , \ P^{\nu} ] \ \ &= \ \ 0 \\
[M^{\mu \nu} \ , \ P^{\sigma} ] \ \ &= \ \ i \, (P^{\mu} \, \eta^{\nu \sigma} \ - \ P^{\nu} \, \eta^{\mu \sigma} )\nonumber \\
[M^{\mu \nu} \ , \ M^{\rho \sigma} ] \ \ &= \ \ i\, (M^{\mu \sigma} \, \eta^{\nu \rho} \ + \ M^{\nu \rho} \, \eta^{\mu \sigma} \ - \ M^{\mu \rho} \, \eta^{\nu \sigma} \ - \ M^{\nu \sigma} \, \eta^{\mu \rho} )\nonumber \\
[Q_{\alpha}\ ,\ M^{\mu\nu}]  \ \ & = (\sigma^{\mu\nu})_{\alpha}^{\; \beta}Q_{\beta}\nonumber \\
[\bar{Q}^{\dot{\alpha}}\ ,\ M^{\mu\nu}]  \ \ & = (\bar{\sigma}^{\mu\nu})^{\dot{\alpha}}_{\; \dot{\beta}}\bar{Q}^{\dot{\beta}}\nonumber \\
[Q_{\alpha}\ ,\ P{\mu}]  \ \ & = 0\nonumber \\
[\bar{Q}^{\dot{\alpha}}\ ,P^{\mu}]  \ \ & = 0 \nonumber \\
\{ Q_{\alpha},Q_{\beta} \}   \ \ & = 0 \nonumber \\
\{ Q_{\alpha},\bar{Q}_{\dot{\alpha}} \}   \ \ & = 2\sigma^\mu_{\alpha \dot{\alpha}} P_{\mu} \nonumber 
\end{alignat}
There is also a global internal $U(1)_{R}$ symmetry with parameter $t$,
\begin{alignat}{1}
[Q_{\alpha},R ] \ \ & =Q_{\alpha}  \\
[\bar{Q}_{\dot{\alpha}},R ] \ \ & =-\bar{Q}_{\dot{\alpha}} \nonumber  
\end{alignat}
The Supercharges are defined by 
\be
Q_{\alpha}=\partial_{\alpha} +i\sigma^{\mu}_{\alpha\dot{\alpha}}\bar{\theta}^{\dot{\alpha}}\partial_{\mu} \ \  , \ \ \bar{Q}_{\dot{\alpha}}=-\partial_{\dot{\alpha}}- i \theta^{\alpha} \sigma^{\mu}_{\alpha\dot{\alpha}}\bar{\theta}^{\dot{\alpha}} \partial_{\mu}
\ee
where we use the up/down  and down/up system 
\be 
{\theta}^2 =\theta^{\alpha} \theta_{\alpha} \ \ , \ \ \bar{\theta}^2 =\bar{\theta}_{\dot{\alpha}} \bar{\theta}^{\dot{\alpha}}.
\ee
It is useful to define the y-space when defining constraint conditions on superfields
\be
y^{\mu}=x^{\mu}- i\theta^{\alpha} \sigma^{\mu}_{\alpha \dot{\alpha}}\bar{\theta}^{\dot{\alpha}}\ \ , \ \ \bar{y}^{\mu}=x^{\mu}+ i\theta^{\alpha} \sigma^{\mu}_{\alpha \dot{\alpha}}\bar{\theta}^{\dot{\alpha}}.
\ee
The superspace measure is given by $\int d^8 z= \int d^4 x d^2\theta d^2 \bar{\theta}$.
The supercovariant derivatives that commute with both partial derivatives \emph{and} with supercharges are given by 
\be
D_{\alpha}=\partial_{\alpha} -i\sigma^{\mu}_{\alpha\dot{\alpha}}\bar{\theta}^{\dot{\alpha}}\partial_{\mu} \ \  , \ \ \bar{D}_{\dot{\alpha}}=-\partial_{\dot{\alpha}}+ i \theta^{\alpha} \sigma^{\mu}_{\alpha\dot{\alpha}}\bar{\theta}^{\dot{\alpha}} \partial_{\mu},
\ee
which obey the anticommutator 
\be
\{ D_{\alpha},\bar{D}_{\dot{\alpha}}\} = 2\sigma^{\mu}_{\alpha\dot{\alpha}}P_{\mu}.
\ee
An infinitesimal supersymmetry transformation is then given by 
\be 
(x^\mu, \theta, \bar{\theta})\rightarrow (x^{\mu}-i \theta \sigma^{\mu}\bar{\epsilon}+i\epsilon \sigma^{\mu}\bar{\theta },\theta+\epsilon,\bar{\theta }+ \bar{\epsilon})
\ee
Acting on a function this gives $F\rightarrow F+\delta_{\epsilon} F$ such that 
\be 
\delta_{\epsilon} F(x,\theta,\bar{\theta})=i (\epsilon Q + \bar{\epsilon}\bar{Q})F(x,\theta,\bar{\theta})
\ee
\section{Off-Shell Superfields}
Four dimensional $\mathcal{N}=1$ superfields are naturally defined in superspace through the use of superspace constraints.  The left handed chiral superfield, in $y$ coordinates, obeys the constraint 
\be 
\bar{D}_{\dot{\alpha}}\Phi(y)=0
\ee
which implies at most a constant in $\bar{\theta}$ i.e.
\be 
\Phi(y)= \phi(y)+\sqrt{2}\theta \psi(y)+\theta^2 F(y) 
\ee
Use of Taylor series and a shift to $x$ coordinates readily gives
\be 
\Phi(x)=\phi(x)-i\theta \sigma^\mu \bar{\theta}\partial_{\mu}\phi(x)-\frac{1}{4}\theta^2\bar{\theta}^2 \partial^2 \phi(x) + \sqrt{2}\theta \psi(x) +\frac{i}{\sqrt{2}} \theta^2\partial_{\mu} \psi(x)  \sigma^\mu\bar{\theta}+ \theta^2 F(x).
\ee
The super symmetry transformations are 
\bea
\delta_{\epsilon}\phi &=& \sqrt{2}\epsilon \psi ,\\
\delta_{\epsilon}\psi &=& \sqrt{2}\epsilon F- \sqrt{2}i (\sigma^{\mu}\bar{\epsilon}\partial_\mu \phi), \\
\delta_{\epsilon} F\ &=& i \partial_{\mu}(\sqrt{2}\psi \sigma^{\mu}\bar{\epsilon}).
\eea
A right handed Chiral superfield in $\bar{y}$ is similarly defined by 
\be
D_{\alpha}\Phi^{\dagger}(y^{\dagger})=0 .
\ee
A vector superfield is defined as a real superfield 
\be
V= \bar{V} 
\ee
with a gauge transformation
\be
V\rightarrow V+i(\Lambda -\bar{\Lambda}) 
\ee
where $\Lambda$ is a chiral superfield.  This allows us to fix to Wess-Zumino gauge
\be 
V=- (\theta \sigma^{\mu}\bar{\theta})A_{\mu}+i\theta^2 \bar{\theta}\bar{\lambda}-i\bar{\theta}^2\lambda+\frac{1}{2}\theta^2 \bar{\theta}^2 D.
\ee
The Gauge invariant information may be gather into a spinorial super field strength 
\be 
W_\alpha=-\frac{1}{4}\bar{D}^2 D_{\alpha} V
\ee
which is a left handed chiral superfield with 
\be 
\bar{D}_{\dot{\alpha}}W_{\alpha}=0 ,  \ \ \ \ D^{\alpha}W_{\alpha}=\bar{D}_{\dot{\alpha}}\bar{W}^{\dot{\alpha}}
\ee
\be
W_{\alpha}(y)= -i \lambda_{\alpha}(y)  + \theta_{\beta}(\delta^{\beta}_{\alpha}D(y)-i(\sigma^{\mu\nu})_{\alpha}^{\beta}F_{\mu\nu}(y)+\theta^2 \sigma^{\mu}_{\alpha \dot{\alpha}}\partial_{\mu}\bar{\lambda}^{\dot{\alpha}}(y).
\ee
The global symmetry current superfield is defined by the linear constraint,
\be 
D^2 \mathcal{J}= \bar{D}^2 \mathcal{J}=0
\ee
A real superfield fixed by these constained is a real linear multiplet which in $x$ is 
\be {\cal J}^a = J^a + i \theta j^a - i \bar \theta  \bar j^a -
\theta\sigma^\mu\bar \theta j_\mu^a +
\frac{1}{2}\theta\theta\bar\theta\bar \sigma^\mu\partial_\mu j^a -
\frac{1}{2}\bar\theta\bar\theta\theta \sigma^\mu\partial_\mu \bar
j^a - \frac{1}{4}\theta\theta\bar\theta\bar\theta \Box J^a\ee
Noether theorem guarantees that a continuous global symmetry gives rise to a conserved current
\be
\partial^\mu j_\mu=0   , \ \ \ Q= \int d^3 x j^{0}
\ee
With conserved charge $Q$. The charge commutes with supersymmetry $[Q_{\alpha},Q ]=0 $ such that the current algebra may be defined up to Schwinger terms 
\be
[Q_{\alpha},j_{\mu} ] = -2i(\sigma_{\mu\nu})_{\alpha}^{\beta}\partial^{\nu}j_{\beta}.
\ee
\subsection{The Minimal Supersymmetric Standard Model}
\renewcommand{\arraystretch}{1.4}
\begin{table}[t]
\begin{center}
\begin{tabular}{|c|c|c|c|c|}
\hline
Names& LH$\chi$SF
& Bosons & Fermions  & $SU(3)_C ,\, SU(2)_L ,\, U(1)_Y$
\\  \hline\hline
squarks, quarks & $Q$ & $(\tilde{u}_L\>\>\> \tilde{d}_L )$&
 $(u_L\>\>\>d_L)$ & $(\>{\bf 3},\>{\bf 2}\>,\>\frac{1}{6})$
\\
($ 3$ Generations) & $U$
&$\tilde{u}^*_R$ & $u^\dagger_R$ & 
$(\>{\bf \overline 3},\> {\bf 1},\> -\frac{2}{3})$
\\ & $D$ &$\tilde{d}^*_R$ & $d^\dagger_R$ & 
$(\>{\bf \overline 3},\> {\bf 1},\> \frac{1}{3})$
\\  \hline
sleptons, leptons & $L$ &$(\tilde{\nu}\>\>\tilde{e}_L )$&
 $(\nu\>\>\>e_L)$ & $(\>{\bf 1},\>{\bf 2}\>,\>-\frac{1}{2})$
\\
($ 3$ Generations) & $E$
&$\tilde{e}^*_R$ & $e^\dagger_R$ & $(\>{\bf 1},\> {\bf 1},\>1)$
\\  \hline
Higgs, higgsinos &$H_u$ &$(H_u^+\>\>\>H_u^0 )$&
$(\tilde{H}_u^+ \>\>\> \tilde{H}_u^0)$& 
$(\>{\bf 1},\>{\bf 2}\>,\>+\frac{1}{2})$
\\ &$H_d$ & $(H_d^0 \>\>\> H_d^-)$ & $(\tilde{H}_d^0 \>\>\> \tilde{H}_d^-)$& 
$(\>{\bf 1},\>{\bf 2}\>,\>-\frac{1}{2})$
\\   \hline\hline
Names& VSF
& Bosons & Fermions  & $SU(3)_C ,\, SU(2)_L ,\, U(1)_Y$
\\  \hline\hline
Gauge,Gauginos& $G$
&$G^a_{\mu}$ & $\tilde{G}^a$ & 
$(\>{\bf 8},\> {\bf 1},\> 0)$
\\(Gauge Fields) & $W$ &$W^3_{\mu}\ , \ W^{\pm}_{\mu}$ & $\tilde{W}^3_{\mu}\ , \ \tilde{W}^{\pm}_{\mu}$ & 
$(\>{\bf 1},\> {\bf 3},\> 0)$
\\ 
 & $B$ &$B_{\mu}$ & $\tilde{B}$ & 
$(\>{\bf 1},\> {\bf 1},\> 0)$
\\ \hline

\end{tabular}
\caption{Left handed Chiral superfields and Vector superfields of the MSSM and their particle content.\label{MSSM}}
\vspace{-0.6cm}
\end{center}
\end{table}
It is sometimes common to label the transpose conjugates of right handed fields ($U,D,E$) with a ($\bar{\phantom{a}}$). The  ($\tilde{\phantom{a}}$) represents superpartners to the standard model fields.

\setcounter{equation}{0}
\chapter{Non-Abelian Bulk Action}\label{NON}
This appendix reviews the $\mathcal{N}=1$ 5D Non-Abelian bulk action for super-Yang-Mills.  This corresponds to $\mathcal{N}=2$ in the 4D perspective.  We compactify on an orbifold, $S^1/\mathbb{Z}_{2}$, such that super-Yang-Mills becomes a  $\mathcal{N}=1$  positive parity vector multiplet and negative parity chiral multiplet.  This review is closely based on ~\cite{Hebecker:2001ke,Mirabelli:1997aj}.    The $\theta$-warping technique of chapter \ref{chapter6} naturally extends these results to a slice of AdS, which may also be found in the appendix of \cite{Bouchart:2011va}.
\section{The Non-Abelian bulk action}
Starting with the off-shell $\mathcal{N}=1$ pure super-Yang-Mills in components
\be
S_{5D}^{SYM}= \int d^{5} x
~\text{Tr}\left[-\frac{1}{2}(F_{MN})^2-(D_{M}\Sigma)^2-i\bar{\gl}_{i}\gamma^M
D_{M}\gl^{i}+(X^a)^2+g_{5}\, \bar{\gl}_i[\Sigma,\gl^i]\right].
\ee
$M,N$ run over $0,1,2,3,4$, while $\mu, \nu$ run over $0,1,2,3$.
Our conventions on the gauge group generators and the metric are
$\text{Tr}(T^{\cal A} T^{\cal B})=\frac 12 \delta^{{\cal A}{\cal B}}$
and $\eta_{MN}=\text{diag}(-1,1,1,1,1)$. The coupling $1/g^2_{5}$ has
been rescaled inside the covariant derivative, $D_{M}=
\partial_{M}+ig_{5} A_{M}$, where $A_{M}$ is a standard gauge
vector field and $F_{MN}$ its field strength.
The other fields are a real scalar $\Sigma$, an $SU(2)_{R}$ triplet of
real auxiliary fields $X^{a}$, $a=1,2,3$ and a symplectic Majorana
spinor $\gl_{i}$ with $i=1,2$ which form an $SU(2)_R$ doublet. The reality condition is
\be
\gl^i= \epsilon^{ij} C\bar{\gl}_{j}^{T} 
\label{real2}
\ee
where $\epsilon^{12}=1$ and $C$ is the 5d charge conjugation matrix
$C\gamma^M C^{-1}=(\gamma^M)^T$. An explicit realisation of the
Clifford algebra $\{\gamma^M,\gamma^N\}=-2\eta^{MN}$ is
\be
\gamma^M=\left(\,\left(\begin{array}{cc}0&\sigma^\mu_{\alpha \dot{\alpha}}\\ 
\bar{\sigma}^{\mu \dot{\alpha} \alpha }&0
\end{array}\right),
\left(\begin{array}{cc}-i&0\\ 0&i\end{array}\right)\,
\right)\,,~~\mbox{and}~~~
C=\left(\begin{array}{cc}
-\epsilon_{\alpha\beta} & 0\\ 
0 & \epsilon^{\dot\alpha \dot\beta}
\end{array}\right)\,,
\ee
where $\sigma^\mu_{\alpha \dot{\alpha}}=(1,\vec{\sigma})$ and
$\bar{\sigma}^{\mu \dot{\alpha}
\alpha}=(1,-\vec{\sigma})$. $\alpha,\dot{\alpha}$ are spinor indices
of $\text{SL}(2,C)$. For the $SU(2)_{R}$ indices we define
\be
\epsilon_{i j}=\left(\begin{array}{cc} 0&-1 \\ 1&0
\end{array}\right) \ \ \ \ \epsilon^{ ij }=\left(\begin{array}{cc} 0&1 \\- 1&0
\end{array}\right)
\ee
The  superalgebra is given by 
\be
\{Q^i,\bar{Q}^j\}=2\gamma^M P_M \delta^{i,j}
\ee
which are also symplectic Majorana.

This action is supersymmetric under the susy
transformations
\bea
\delta_{\epsilon}A^M &=& i\bar{\epsilon}_i\gamma^M \gl^{i}\\
\delta_{\epsilon}\Sigma &=& i\bar{\epsilon}_i \gl^{i}\\
\delta_{\epsilon} \gl^{i} &=& (\gamma^{MN} F_{MN}-\gamma^M D_{M}\Sigma)
\epsilon^i -i(X^a \sigma^a)^{i}_{~j} \epsilon^j\\
\delta_{\epsilon}X^a &=& \bar{\epsilon}_i(\sigma^a)^{i}_{~j}
\gamma^{M} D_{M}\gl^j -ig_{5} 
[\Sigma,\bar{\epsilon}_i(\sigma^a)^i_{~j}\gl^j]
\eea
with $\gamma^{MN}=\frac{1}{4}[\gamma^M,\gamma^N]$. The symplectic
Majorana spinor supersymmetry parameter is
$\bar{\epsilon}_i=\epsilon_{i}^{\dagger}\gamma^{0}$. To clarify
notation we temporarily display all labels, writing the Dirac spinor
in two component form ${\psi}^{i \, T}= (\psi_{\alpha}^{L i},
\bar{\psi}^{R \dot{\alpha} i})$ and $\bar{\psi}_i= (\psi^{R \alpha}_{
i}, \bar{\psi}^L_{ \dot{\alpha} i})$.  The bar on the two component
spinor denotes the complex conjugate representation of $SL(2,C)$.  
In particular, the reality condition~\eqref{real2} implies that
\be
\lambda^1 = \left(\begin{array}{c}
\lambda_{L \alpha}\\ 
\bar\lambda_{R}^{\dot\alpha}
\end{array}\right)~,~~~
\lambda^2 = \left(\begin{array}{c}
\lambda_{R \alpha}\\ 
-\bar\lambda_{L}^{\dot\alpha}
\end{array}\right)~,~~~
(\bar\lambda_1)^{T} = \left(\begin{array}{c}
\lambda_{R}^{\alpha}\\ 
\bar\lambda_{L \dot\alpha}
\end{array}\right)~,~~~
(\bar\lambda_2)^{T} = \left(\begin{array}{c}
-\lambda_{L}^{\alpha}\\ 
\bar\lambda_{R \dot\alpha}
\end{array}\right)~,
\ee
so the $SU(2)_{R}$ index on a two component spinor is a redundant
label.

Next, using an orbifold $S^{1}\!/\mathbb{Z}_{2}$ the boundaries will
preserve only half of the $\mathcal{N}=2$ symmetries. We choose to
preserve $\epsilon_{L}$ and set $\epsilon_{R}=0$. The conjugate
representations are constrained by the reality condition \refe{real2}.
The susy transformations are
\bea
\delta_{\epsilon_L}A^\mu &=& i\bar{\epsilon}_L\bar{\sigma}^\mu\lambda_L+i\epsilon_L\sigma^\mu
\bar{\lambda}_L\label{delam}\\
\delta_{\epsilon_L}A^5 &=& -\bar{\epsilon}_L\bar{\lambda}_R-\epsilon_L\lambda_R\\
\delta_{\epsilon_L}\Sigma &=& i\bar{\epsilon}_L\bar{\lambda}_R-i\epsilon_L\lambda_R\\
\delta_{\epsilon_L}\lambda_L &=& \sigma^{\mu \nu}F_{\mu \nu}\epsilon_L-iD_5\Sigma\epsilon_L
+iX^3\epsilon_L\\
\delta_{\epsilon_L}\lambda_R &=& i\sigma^\mu F_{5\mu}\bar{\epsilon}_L-\sigma^\mu D_\mu
\Sigma\bar{\epsilon}_L+i(X^1+iX^2)\epsilon_L\\
\delta_{\epsilon_L}(X^1+iX^2) &=& 2\bar{\epsilon}_L\bar{\sigma}^\mu D_\mu \lambda_R-2i
\bar{\epsilon}_LD_5\bar{\lambda}_L+ig_{5}[\Sigma,2\bar{\epsilon}_L\bar{\lambda}_L]\\
\delta_{\epsilon_L} X^3 &=& \bar{\epsilon}_L\bar{\sigma}^\mu D_\mu \lambda_L+i
\bar{\epsilon}_LD_5\bar{\lambda}_R-\epsilon_L\sigma^\mu D_\mu\bar{\lambda}_L
-i\epsilon_LD_5\lambda_R\nonumber\\
&&+ig_{5}[\Sigma,(\bar{\epsilon}_L\bar{\lambda}_R+\epsilon_L\lambda_R)]\,,\label{delx3}
\eea
where $\sigma^{\mu \nu}=\frac{1}{4}(\sigma^{\mu}
\bar{\sigma}^{\nu}-\sigma^{\nu}\bar{\sigma}^{\mu} )$. We have a parity
operator $P$ of full action $\mathbb{P}\phi(y)=P\phi(-y)$ and define
$P\psi_{L}=+\psi_{L}$ $P\psi_{R}=-\psi_{R}$ for all fermionic fields
and susy parameters\footnote{The assignment
$P\partial_{5}=-\partial_{5}$ is also required.}. One can group the
susy variations under the parity assignment and it becomes clear that
the even parity susy variations are those of an off-shell $4d$ vector
multiplet $V(x_{5})$.  Similarly the susy variations of odd parity
form a chiral superfield $\Phi(x_{5})$. We may therefore write a $5d$
$\mathcal{N}=1$ vector multiplet as a $4d$ vector multiplet and a
chiral superfield:
\begin{alignat}{1}
V=&- \theta\sigma^{\mu}\bar{\theta}A_{\mu}+i\bar{\theta}^{2}\theta\gl-
i\theta^{2}\bar{\theta}\bar{\gl}+\frac{1}{2}\bar{\theta}^{2}\theta^{2}D\\
\Phi=& \frac{1}{\sqrt{2}}(\Sigma + i A_{5})+
\sqrt{2}\theta \chi + \theta^{2}F\,,
\label{fields2}
\end{alignat}
where the identifications between $5d$ and $4d$ fields are
\begin{equation}
D=(X^{3}-D_{5}\Sigma) \quad F=(X^{1}+iX^{2})\,,
\end{equation}
and we used $\lambda$ and $\chi$ to indicate $\lambda_{L}$ and
$-i\sqrt{2}\lambda_R$ respectively. The Non-Abelian bulk action in ${\cal N}=1$  4D formalism is
\begin{equation}
S^{SYM}_{5}= \int d^{5}x \left\{\frac{1}{2}\text{Tr} \left[ \int d^{2}\theta
 W^{\alpha}W_{\alpha}+
\int d^{2}\bar{\theta}  \bar{W}_{\dot{\alpha}}\bar{W}^{\dot{\alpha}}\right]  
+ \frac{1}{2 g_{5}^2} \int d^{4}\theta  
\text{Tr}\left[e^{-2g_{5}V}\nabla_5 e^{2g_{5}V}\right]^2\right\}\, .
\label{fullaction}
\end{equation}
 $\nabla_5$ is a ``covariant'' derivative with the respect to the
field $\Phi$~\cite{Hebecker:2001ke}:
\be \nabla_5 e^{2g_{5}V}=\partial_5
e^{2g_{5}V} - g_{5}\Phi^\dagger e^{2g_{5}V} - g_{5} e^{2g_{5} V} \Phi . 
\ee

Let us now focus on 5d hypermultiplets. The bulk supersymmetric action is 
\bea
S_{5D}^{H} &=& \int d^5 x[-(D_MH)^\dagger_i(D^MH^i)-
i\bar{\psi}\gamma^MD_M\psi+ F^{\dagger i}F_i-g_5
\bar{\psi}\Sigma\psi+g_5 H^\dagger_i(\sigma^aX^a)^i_jH^j
\nonumber\\
&& +g_5^2 H^\dagger_i\Sigma^2H^i+ig_5\sqrt{2}\bar{\psi}
\lambda^i\epsilon_{ij} H^j-i\sqrt{2}g_5H^{\dagger}_{i}
\epsilon^{ij}\bar{\gl}_{j}\psi\,].\label{hyperaction3}
\eea
$H_{i}$ are an $SU(2)_{R}$ doublet of scalars. $\psi$ is a Dirac
fermion and $F_{i}$ are a doublet of scalars. With our conventions,
the dimensions of ($H_{i},\psi,F_{i}$) are
($\frac{3}{2},2,\frac{5}{2}$). In general the hypermultiplet matter will be in a representation of the gauge group with Dynkin index defined by $d\delta^{ab}=\text{Tr}[T^a T^b]$. The action is supersymmetric under the susy transformations
\bea
\delta_\epsilon H^i &=& -\sqrt{2}\epsilon^{ij}\bar{\epsilon}_j\psi\label{delh}\\
\delta_\epsilon \psi &=& ig_5\sqrt{2}\gamma^MD_MH^i\epsilon_{ij}\epsilon^j-g_5\sqrt{2}\Sigma
H^i\epsilon_{ij}\epsilon^j+\sqrt{2}F_i\epsilon^i\\
\delta_\epsilon F_i &=& i\sqrt{2}\bar{\epsilon}_i\gamma^MD_M\psi+g_5\sqrt{2}\bar{\epsilon}_i
\Sigma\psi-2ig_5\bar{\epsilon}_i\lambda^j\epsilon_{jk}H^k\,.
\eea
To obtain the $\mathcal{N}=1$ sets due to the boundaries preserving only half the supersymmetry, we again choose to preserve $\epsilon_{L}$ and set $\epsilon_{R}=0$.  The susy variations are
\begin{align}
\delta_{\epsilon_{L}} H^1 &= \sqrt{2}\epsilon_{L}\psi_{L} \\
\delta_{\epsilon_{L}} H^2 &= \sqrt{2}\bar{\epsilon}_{L}\bar{\psi}_{R} \\
\delta_{\epsilon_{L}} \psi_{L\alpha} &= ig_5\sqrt{2}\sigma^{\mu}_{\alpha \dot{\beta}}D_\mu H^2\bar{\epsilon}^{L\dot{\beta}} +g_5\sqrt{2}D_{5}H^2 \epsilon^{L}_{\alpha}- g_5\sqrt{2}\Sigma
H^2\epsilon^{L}_{\alpha}+\sqrt{2}F_1\epsilon^{L}_{\alpha}\\
\delta_{\epsilon_{L}} \bar{\psi}^{R \dot{\alpha}} &= 
ig_5\sqrt{2}\bar{\sigma}^{\mu\dot{\alpha}\beta}D_\mu H^2\epsilon^{L}_{\beta}
-g_5\sqrt{2}D_{5}H^1\bar{\epsilon}^{L \dot{\alpha}}
-g_5\sqrt{2}\Sigma
H^1\bar{\epsilon}^{L\dot{\alpha}}
-\sqrt{2}F_2\bar{\epsilon}^{L\dot{\alpha}}\\
\delta_{\epsilon_{L}} F_1 &= i\sqrt{2}\bar{\epsilon}_{L\dot{\alpha}}\bar{\sigma}^{\mu \dot{\alpha}\beta}D_\mu \psi_{L \beta}
-\sqrt{2}\bar{\epsilon}^{L}_{\dot{\alpha}}D_{5}\bar{\psi}^{R \dot{\alpha}}+g_5\sqrt{2}\bar{\epsilon}^{L}_{\dot{\alpha}}\Sigma \bar{\psi}^{R \dot{\alpha}}-2ig_5\bar{\epsilon}^{L j}_{ \dot{\alpha}}\bar{\lambda}^{R\dot{\alpha}j}\epsilon_{jk}H^k\\
\delta_{\epsilon_{L}} F_2 &= -i\sqrt{2} \epsilon^{L\alpha} 
\sigma^{\mu}_{\alpha\beta}D_\mu \bar{\psi}^{R \dot{\beta}}
-\sqrt{2}\epsilon^{L \alpha}D_{5}\psi^{L}_{\alpha}-g_5\sqrt{2}\epsilon^{L \alpha}\Sigma \psi^{L}_{\alpha} +2ig_5\epsilon^{L \alpha}\lambda^{L \alpha j}\epsilon_{jk}H^k\,.
\end{align}
In the 4d superfield formulation, we again use the parity of the $P\psi_{L}=+\psi_{L}$ and $P\psi_{R}=-\psi_{R}$ to group the susy transformations into a positive and negative parity chiral superfields, $PH=+H$ and $PH^c=-H^c$:
\bea
H &=& H^1+\sqrt{2}\theta\psi_L+\theta^2(F_1+D_5H_{2}-g_5\Sigma H_2)\\
H^c &=& H^\dagger_2+\sqrt{2}\theta\psi_R+\theta^2(-F^{\dagger}_{2}-D_5
H^\dagger_1- g_5 H^\dagger_1\Sigma)\,.
\eea
The gauge transformations are
$H\to e^{-\Lambda}H$ and $H^c\to H^ce^\Lambda$. The $\mathcal{N}=1$ action in 4d language is
\be
S_{5d}^H=\int d^5 x (\int d^4\theta [ H^\dagger e^{2g_5 V}H+H^c e^{-2g_5 V}H^{c\dagger}] + \int d^2 \theta  H^c\nabla_5 H+ \int d^2\bar{\theta}H^{c\dagger}\nabla_5 H^\dagger )
\,.
\ee
These results have a natural extension using the techniques of $\theta$-warping discussed in chapter \ref{chapter6}.
\setcounter{equation}{0}
\chapter{Deriving The Bulk Propagators}\label{propagatorsinthebulk}
In this section we outline some steps in determining the bulk propagator in a $  \mathbb{R}^{1,3}\times S^1 / \mathbb{Z}_2 $ background.
\section{The bulk propagator}
We start with a single field in five dimensions whose Fourier transform is 
\be
\phi(x)= \int d^5 x e^{ip.x} \tilde{\phi}(p)
\ee
To go to  $  \mathbb{R}^{1,3}\times S^1 / \mathbb{Z}_2 $   we use the full action of parity operator $\frac{1}{2}(1+\mathbb{P})$ which is a projector, 
whereby 
\be
\mathbb{P} \phi(x_{\mu},x_{5})=p \phi(x_{\mu},-x_{5})=\pm \phi(x_{\mu},-x_{5})
\ee 
so that we arrive at 
\be
\frac{1}{2}(1+\mathbb{P})\int d^5 x e^{ip.x} = \int d^4 x d y_{5} \frac{1}{2}(1+\mathbb{P}) e^{ip.x + ix_{5}.y_{5}}=\int d^4 x d y_{5} e^{ip.x}\frac{1}{2}(e^{ix_{5}.y_{5}}+pe^{- ix_{5}.y_{5}}) .
\ee
Taking the two point function of the scalar field one arrives at the bulk scalar propagator
\be
\braket{\phi(x,x_{5})\phi(y,y_{5})}=	\int_{p5} \frac{i }{p^2 - (p^5)^2}\,
 e^{-ip\cdot (x-y)} (e^{ip^5(x^5-y^5)}
     + P e^{ip^5(x^5+y^5)}) \ ,
\ee
\begin{equation}
   \int_{p5} = \int \frac{d^4p}{ (2\pi)^4} \frac{1}{ 2\ell}\sum_{p^5} \ ,
\end{equation}
The bulk propagator can then similarly be written in terms of the wave functions as (for a negative parity field)
\begin{equation}
P(q^2,a,b)=\sum_{m,n}  \phi_n^*(a)\phi_m(b) \frac{\delta_{mn}}{q^2+p_n^2}=
\frac{2}{L} \sum_n \sin \left( \frac{-\pi n a}{L}\right) \sin \left( \frac{\pi n b}{L}\right) \frac{1}{q^2+p_n^2}.
\end{equation}
Here $p_n=\frac{n\pi}{L}$, and one can convert the sum to an integral using $\Delta n = dp \ L/\pi$ .

One may ask how we arrived at the masses of the kk modes in the propagator.   Well lets look at the fermions. We start with the term 
\be
\int d^5x \bar{\Psi}\gamma^\mu\partial_{\mu}\Psi  
\ee
This is a Dirac four spinor (in 5D it is actually symplectic Majorana as it cannot be Dirac and there is also an $SU(2)_{R}$ index but lets ignore this for now).
We then write this as $\int d^4 x dy_{5}$  and apply orbifold boundary conditions that are like an infinite potential well but with periodic boundary conditions such that some fields reflect off the wall.  The resulting action when expanded is
\be
 \sum_{n} \bar{\gl}_{n}\sigma^\mu\partial_{\mu}\gl_{n} +  \sum_{n}\bar{\psi}_{n}\sigma^\mu\partial_{\mu}\psi_{n} +  \sum_{n} \gl_{n}\partial_{5}\psi_{n} +  \sum_{n} \bar{\gl}_{n}\partial_{5}\bar{\psi}_{n}
\ee
Notice these terms are all the same order in $g$. So to obtain the correct massive propagators we must construct a geometric sum of diagrams.  We will do this separately for each $n$ so there is an overall $\sum_{n}$ outside which we ignore.

So a massless tower of propagators with even parity will  have the form
\begin{equation}
P(q^2,a,b)=\sum_{m,n}  \phi_n^*(a)\phi_m(b) \frac{\delta_{mn}}{q^2}=
\frac{2}{L} \sum_n \cos \left( \frac{-\pi n a}{L}\right) \cos \left( \frac{\pi n b}{L}\right) \frac{1}{q^2}.
\end{equation}

\begin{figure}[cthb]
\begin{center}
\includegraphics[width=10cm]{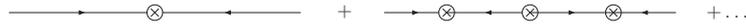}
\caption{Infinite sum of M-insertions.}
\label{M-insertion}
\end{center}
\end{figure}
Now imagine a massless propagor with one mass insertion and then another massless propagator  i.e. an odd number of mass insertions.  Between each mass instertion are propagators that alternate between the $\lambda$ and the $\chi$ fermions, as in figure \ref{M-insertion}
\be
\frac{\slashed{k}}{k^2} M\frac{\slashed{k}}{k^2}+ \frac{\slashed{k}}{k^2} M\frac{\slashed{k}}{k^2} M\frac{\slashed{k}}{k^2}M\frac{\slashed{k}}{k^2}+.....= \frac{\slashed{k}}{k^2}\sum^{\infty}_{i=0}(M\frac{\slashed{k}}{k^2})^{2i+1}
\ee
where we want to interpret the $M=p_{5}$ and $(\slashed{k})^2=k^2$.
\be 
\frac{\slashed{k}}{k^2}\sum^{\infty}_{i=0}(M\frac{\slashed{k}}{k^2})^{2i+1}=\frac{M}{k^2}\sum^{\infty}_{i=0}(\frac{M^2}{k^2})^{i}=\frac{M}{k^2}\frac{1}{1-\frac{M^2}{k^2}}=\frac{M}{k^2-M^2}
\ee
In the even number of mass insertions we have
\be
\frac{\slashed{k}}{k^2}\sum^{\infty}_{i=0}(M\frac{\slashed{k}}{k^2})^{2i}=\frac{\slashed{k}}{k^2}\sum^{\infty}_{i=0}(\frac{M}{k^2})^{i}=\frac{\slashed{k}}{k^2-M^2}.
\ee
Additionally one may need to keep track that at each sewing point we have to do an integral
\be
\int d y_{5} \cos(k_{5}y_{5}) \partial_{5} \sin(k_{5}y_{5}) 
\ee
So there is also a geometric sum of these.  Collecting these ingredients one arrives at the above propagator.  For the particular case that one propagates from  $x_{5}=0$ to $y_{5}=\ell$, we may write
\be
\braket{\phi(x,0)\phi(y,\ell)}=	\int_{p5} \frac{i }{p^2 + (p^5)^2}\,
 e^{-ip\cdot (x-y)} (e^{-ip^5(\ell)}
  + P e^{ip^5(\ell)}) \int_{p5} \frac{i2(-1)^n}{p^2 + (p^5)^2} .
\ee 
 with a Wick rotation implied.
\section{The Contour Trick} \label{sec:contour}
The Matsubara Frequency summation can be applied to the kk propagators.  We will apply it to the integrand
\be
S=\sum_{n}h(k_{5})=\frac{1}{2\ell}\sum_{n}(-1)^{n}\frac{1}{k^2+(k_{5})^2} 
\ee
We would like to remove the sum on $k_{5}$ and carry out an integration on only the $k^2$ momenta.  To do this we exchange the sum of $p_{5}$ for a complex auxiliary function $g(ik_{5})$, that has poles at these point so that when we take the sum of residues, the residues are located at the points where the sum would have been. The sum is then a sum of residues found by integrating the product of $g(ik_{5})h(k_{5})$ along an appropriate closed path.  Typically we will choose 
\bea
g(z)= \frac{\beta}{e^{(\beta z)}-1}  \ \ \
 \text{or}  \ \ \ 
 \frac{\beta}{2}   
 \text{Coth} (\beta z/2)
\eea
where we have  $\beta=2\ell$.  The manipulations are
\be 
S=\sum_{n}h(k_{5})=\frac{1}{2\ell}\sum_{n}(-1)^{n}\frac{2}{k^2+(k_{5})^2} =\sum_{n}\frac{1}{2\ell}\frac{2e^{ik_{5}\ell}}{k^2+(k_{5})^2}
\ee
We apply the residue theorem
\be
\oint \frac{d k_{5}}{2\pi} g(z)h(z)=
\oint \frac{d k_{5} }{2\pi} \frac{1}{2\ell}  \frac{2\ell}{e^{i2k_{5}\ell}-1}  \frac{2 e^{ik_{5}\ell}}{k^2+(k_5)^2}= \sum_{n}\text{Res}[g(z)h(iz)]|_{z=ik_{5}}
\ee
where $k_{5}=\frac{n\pi}{\ell}$. So we construct
\be 
\oint \frac{d k_{5} }{2\pi} g(z)h(z) =\oint \frac{d k_{5} }{2\pi} \frac{1}{2\ell}  \frac{2\ell}{e^{i2k_{5}\ell}-1}  \frac{2 e^{ik_{5}\ell}}{k^2+(k_5)^2}
\ee
Now it is clear that there are poles along the complex $z$ axis, there are also two simple poles when $k_{5}=\pm ik$  We trap these two poles in our contour and exclude the rest, by use of an infinite circle, with two hoops on the simple poles.  Evaluating the residues at these two poles we find
\be
S= \frac{1}{k \text{Sinh} k\ell}
\ee
We can now carry out an integration on the $k^2$ momenta.  This can be applied to each of the bulk propagators. If we had taken $x_{5}=y_{5}=0$ we would have
\be
S=\sum_{n}h(k_{5})=\frac{1}{2\ell}\sum_{n}\frac{1}{k^2+(k_{5})^2} =\frac{1}{k \text{Tanh} k\ell}
\ee

The final one, of a positive parity scalar propagating from a brane to the bulk we start with 
\be
\braket{\phi(x,x_{5})\phi(y,y_{5})}=	\int_{p5} \frac{i }{p^2 - (p^5)^2}\,
 e^{-ip\cdot (x-y)} (e^{ip^5(x^5-y^5)}
     +  e^{ip^5(x^5+y^5)}) \ ,
\ee
and take $x_{5}=0$ and $y_{5}$ as a variable. We finally get, using the contour trick
\be
\frac{1}{2\ell}\sum_{p_{5}=-\infty}^{\infty}\frac{(-1)^n e^{ip_{5}.y_{5}}}{p^2+p^2_{5}}= \frac{Cosh(p|y|-\ell)}{2p \sinh (p\ell)}
\ee 
where one can move the sum on $p_{5}$ or $n$ with no cost in the measure.

\setcounter{equation}{0}
\chapter{Generalised messenger sector in 5D with an orbifold}\label{generalised}

This section extends the results of \cite{Marques:2009yu} to the case of bulk propagation in 5D with an orbifold.  We keep as close as possible to the notation of \cite{Marques:2009yu}.  
\section{Messenger sector currents and correlators}
We consider a messenger sector $\phi_{i},\tilde{\phi}_{i}$ coupled to a SUSY breaking spurion $X$:
\be W = {\cal M}(X)_{ij}\ \phi_i \ti \phi_j =  (m + X \lambda)_{ij}\
\phi_i \ti \phi_j \label{superpotentialappend:2}\ee
$m$ and $\gl$ are generic matrices.  The messengers are in a representation of the gauge group with a Dynkin index $d$, defined by $d\delta^{ab}=\text{Tr}[T^a T^b]$.
The fundamental messengers on the SUSY breaking brane will couple to the bulk vector superfield as
\be \delta {\cal L} = \int d^2\theta d^2\bar\theta \left(\phi^\dag_i
e^{2 g V^a T^a} \phi_i + \ti\phi^\dag_i e^{-2 g V^a T^a}
\ti\phi_i\right) + \left(\int d^2\theta\ W +
c.c.\right) \label{hiddensectorappend} \ee
We can extract the multiplet of currents from the kinetic terms in the above Lagrangian.  We find 
\be {\cal J}^a = J^a + i \theta j^a - i \bar \theta  \bar j^a -
\theta\sigma^\mu\bar \theta j_\mu^a +
\frac{1}{2}\theta\theta\bar\theta\bar \sigma^\mu\partial_\mu j^a -
\frac{1}{2}\bar\theta\bar\theta\theta \sigma^\mu\partial_\mu \bar
j^a - \frac{1}{4}\theta\theta\bar\theta\bar\theta \Box J^a\ee
where
\bea
J^a &=& \phi_i^\dag T^a \phi_i - \ti \phi^\dag_i T^a \ti \phi_i \\
j^a &=& - i \sqrt{2} \left(\phi_i^\dag T^a \psi_i - \ti \phi^\dag_i T^a \ti \psi_i\right) \nn\\
\bar j^a &=& i \sqrt{2} \left(\bar \psi_i T^a \phi_i - \bar{\ti \psi}_i T^a \ti \phi_i\right) \nn\\
j_\mu^a &=& \left(\psi_i \sigma_\mu T^a \bar \psi_i - \ti \psi_i
\sigma_\mu T^a \bar{\ti\psi}_i\right) - i \left( \phi_i^\dag T^a
\partial_\mu \phi_i - \partial_\mu\phi^\dag_i T^a \phi_i -
\ti \phi_i^\dag T^a\partial_\mu \ti \phi_i + \partial_\mu\ti
\phi^\dag_i T^a \ti \phi_i\right)   \nn
\label{supercurrent}\eea

(repeated indices are summed)
\be
\ti C_0 =\sum_{k,n} 2 d_{kn} B_{kn} \int \frac{d^4q}{(2\pi)^4} \frac{1}{(q^2 + (m_k^+)^2)((p + q)^2 + (m_n^-)^2)} 
\ee
\be 
\ti C_{1/2} = -\sum_{k,n} \frac{2 d_{kn}}{p^2} \sum_\pm  A^\pm_{kn} \int
\frac{d^4q}{(2\pi)^4} \frac{p\cdot q}{((p
+ q)^2 + (m_k^\pm)^2)(q^2 + (m_n^0)^2)} \ee
\begin{align}
\ti C_{1} = -& \sum_{k,n}\frac{2 d_{kn}}{3 p^2} \int \frac{d^4q}{(2\pi)^4}
\delta_{kn}
 \frac{4 q\cdot(p + q) + 8 (m_k^0)^2}{(q^2 + (m_k^0)^2)((p
+ q)^2 + (m_k^0)^2)}
\\ + & \sum_{\pm}  \left(
\frac{(p + q)\cdot (p + 2q)}
{(q^2 + (m_k^\pm)^2)((p + q)^2 + (m_k^\pm)^2)}
- \frac{4}{q^2 +(m_k^\pm)^2}  \right)\nn
\end{align}
\be 
M \ti B_{1/2} = \sum_{k,n}2 d_{kn} \sum_\pm \mp A^\pm_{kn} \int
\frac{d^4q}{(2\pi)^4} \frac{m_n^0}{(q^2 + (m_k^\pm)^2)((p +
q)^2 + (m_n^0)^2)} 
\ee
The $\ti C_a$ may be written as
\bea
 \ti C_0
\ &=& \ \sum_{k,n} 2 d_{kn} B^+_{kn} G_1(m_k^+, m_n^-) \\
-4 \ti C_{1/2}\ &=& -\sum_{k,n}\ 4 d_{kn} \sum_\pm A^\pm_{kn}\left[ (G_0(m_k^\pm) -
G_0(m_n^0))+ G_1 (m_k^\pm, m_n^0) \right.\label{cs}\\&& \ \ \ \ \ \ \ \ \ \
\ \ \ \ \ \ \left.  + ((m_k^\pm)^2 - (m_n^0)^2)\frac{1}{p^2}G_1(m_k^\pm, m_n^0)\right]\nn\\
  3 \ti C_1 \ &=&\sum_{k,n}
\  d_{kn} \delta_{kn} \sum_\pm \left[4 (G_0(m_k^\pm) - G_0 (m_k^0)) + G_1(m_k^\pm, m_k^\pm) \right. \\
&&\ \ \ \ \ \ \ \ \ \ \ \ \  \left.+ 2 G_1(m_k^0, m_k^0) +\ 4 (m_k^\pm)^2 \frac{1}{p^2}G_1 (m_k^\pm, m_k^\pm) -
4 (m_k^0)^2 \frac{1}{p^2}G_1(m_k^0, m_k^0)\right] \nn
 \eea
All the terms proportional to $G_0$ in (\ref{cs}) vanish due to the messenger supertrace formula.  The G functions are
\bea G_0(m) &=& \int \frac{d^4 p}{(2\pi)^4} \frac{1}{p^2 + m^2} \label{G0}\\
G_1(m_1, m_2) &=& \! \! \int\!
\frac{d^4 q}{(2\pi)^4}  \frac{1}{(q^2 + m_1^2)((p + q)^2 +
m_2^2)}\label{G1}\label{G3}
\eea
We want to evaluate these $G$ \refe{G3} functions to write them as the function $b$ defined below.  We will need to multiply by $g^{4}_{4}$
\begin{equation}
   g^{2}_{4} \int \frac{d^d q}{ (2\pi)^d}\frac{1}{ q^2 + m_1^2}\frac{1}{ (p+q)^2 + m_2^2}
 = g^{2}_{4} \int \frac{d^d q}{(2\pi)^d} \int_{0}^{\infty}dt_{1}dt_{2}e^{-t1(q^2 + m_1^2)-t_{2}(p+q)^2 -t_{2}m_2^2}
\label{expan}\end{equation}
Using dimensional regularisation $d=4-2\epsilon$ and $T=t_{1}+t_{2}$ 
\bea
=& g^{2}_{4}\mu^{2\epsilon} \int \frac{d^{4-2\epsilon} q}{(2\pi)^{4-2\epsilon}} \int_{0}^{\infty}dt_{1}dt_{2}e^{-t_{1}(q^2 + m_1^2)-t_{2}(p+q)^2 -t_{2}m_2^2}\\
= & g^{2}_{4}\mu^{2\epsilon} \int \frac{d^{4-2\epsilon} q}{(2\pi)^{4-2\epsilon}} \int_{0}^{\infty}dt_{1}dt_{2}e^{-T(q\pm \frac{pt_{2}}{T})^{2}\pm p^{2}\frac{t^{2}_{2}}{T}-t_{2}p^{2}-t_{1}m^{2}_{1}-t_{2}m^{2}_{2}}
\eea
The integral on momentum can be carried out by changing to $p'$ in the exponent and then $d^D p =d^D p'$.  We will keep the minus sign in the exponent.  We use the identity
\be
 \int \frac{d^{4-2\epsilon} q}{(2\pi)^{4-2\epsilon}} e^{-tq^{2}}=\frac{1}{(4\pi)^{2-\epsilon}}T^{\epsilon-2}
\ee
to rewrite the equation as ($x=t_{1}/T$  $(1-x)=t_{2}/T$),
\bea
& g^{2}_{4}\mu^{2\epsilon} \int_{0}^{\infty}dT\int_{0}^{1} dx T \frac{1}{(4\pi)^{2-\epsilon}} T^{\epsilon-2} e^{-p^{2}Tx(1-x)-xm^{2}_{1}T-(1-x)m^{2}_{2}T}
\\= &\frac{g^{2}_{4}\mu^{2\epsilon}}{(4\pi)^{2-\epsilon}}\Gamma(\epsilon)\int_{0}^{1} dx[p^{2}x(1-x)+xm^{2}_{1}+(1-x)m^{2}_{2}]^{-\epsilon}
\eea
One takes the exponential log of everything then expand as a Taylor series.  One finds
\be 
\frac{g^{2}_{4}}{(4\pi)^{2}} \{\frac{1}{\epsilon}-\ln 4 \pi -\gamma+  \ln \mu^{2}- \int_{0}^{1} dx\ln [p^{2}x(1-x)+xm^{2}_{1}+(1-x)m^{2}_{2}] \}+O(\epsilon^{2})
\ee

We then define the function $b(k^2,m_1^2,m_2^2)$ by 
\begin{equation}
 G_1(m_1, m_2)=   \int \frac{d^4 q}{(2\pi)^4}\frac{1}{ q^2 + m_1^2}\frac{1}{ (p + q)^2 + m_2^2}
 = \frac{1}{ (4\pi)^2}\big\{ \frac{1}{ \epsilon} - \gamma - b(p^2,m_1^2,m_2^2) + 
        {\cal O}(\epsilon)\big\}
\label{expandsc}\end{equation}
for $d = 4-\epsilon$.  We can write $b$ more
explicitly as
\begin{eqnarray}
b(p^2,m_1^2,m_2^2) &=& \int^1_0 dx \log\left(x(1-x)p^2 + xm_1^2 + (1-x)
  m_2^2\right) \\ 
 & = & A \log\left[\frac{(A + B_1)(A+ B_2)}{ (A-B_1)(A-B_2)}\right]
        + B_2 \log m_1^2 + B_1\log m_2^2 - 2  \ , \nonumber
\label{valofb}\end{eqnarray}
where 
\begin{equation}
 A = \left[\frac{p^4 + 2p^2(m_1^2+m_2^2) + (m_1^2-m_2^2)^2}{ 4p^4}\right]^{1/2}
\label{Avalue}\end{equation}
and
\begin{equation}
 B_1 = \frac{p^2 + m_1^2 - m_2^2}{ 2p^2}\ , \qquad 
 B_2 = \frac{p^2 + m_2^2 - m_1^2}{ 2p^2}\ . 
\label{Bvalue}\end{equation}
As the divergent terms will cancel we can ignore the Euler $\gamma$ term and $\epsilon$ and focus on the $b$ functions.  We can rewrite the integral as
\be
G_1(m_1, m_2) = -\frac{1}{(4\pi)^2}b(p^{2},m_{1}^{2},m_{2}^{2})+ g(\epsilon, \gamma) \label{G32}
\ee
Where we can safely ignore the functions $g(\epsilon, \gamma)$ as they will cancel.  Putting this together we can write
\be 
[3\ti C_{1}-4\ti C_{1/2}+ \ti C_{0} ]=\Omega 
 \ee
 where
 \bea
\Omega &=& -[\sum_{k,n}  2 d_{kn} B^{+}_{kn} b(p^{2},m_{k}^{+ 2},m_{n}^{- 2})]\nn \\
&+&\! \!  \sum_{k,n} 4 d_{kn} \sum_\pm A^\pm_{kn}[b(p^{2},m_{k}^{\pm 2},m_{n}^{0 2})+\frac{1}{p^{2}}((m^{\pm}_{k})^{2}-(m^{0}_{n})^{2}) b(p^{2},m_{k}^{\pm 2},m_{n}^{0 2})]\nn \\
&-&\sum_{k,n} d_{kn} \delta_{kn} \sum_\pm[b(p^{2},m_{k}^{\pm 2},m_{k}^{\pm 2})+2b(p^{2},m_{k}^{0 2},m_{k}^{0  2})  \\
&&\ \ \ \  +4\frac{1}{p^{2}}(m^{\pm}_{k})^{2}b(p^{2},m_{k}^{\pm 2},m_{k}^{\pm  2})-4\frac{1}{p^{2}}(m^{0}_{k})^{2}b(p^{2},m_{k}^{0 2},m_{k}^{0  2})]\nn \label{useful}
\eea
We may construct a dictionary of 
\be 
[3\tilde{C}_1^{(r)}(p^2/M^2)-4\tilde{C}_{1/2}^{(r)}(p^2/M^2)+\tilde{C}_0^{(r)}(p^2/M^2)]= \frac{1}{(4\pi)^2}\Xi
\ee expressions depending on the hidden sector,  such that they may be used both in 4d and various 5d models.  The factors of two and $\pi$ on the right hand side can simply be taken out to convert $g^2 \rightarrow \alpha$ when going from \refe{mainresult} to \refe{alpha}
We now need to find limits of the b function, which we outline in the next subsection.  

The Majorana gaugino mass matrix couples \emph{every} Kaluza-Klein mode to \emph{every} other mode with the same coefficient.  Each entry can be determined by use of the $b$ function and is given as
\be M^{\ti g} = g^2 M \tilde{B}_{1/2}(0) =\frac{\alpha_r}{4 \pi}
\Lambda_{G} \ , \ \ \ \ \Lambda_{G} = 2 \sum_{k,n = 1}^N \sum_\pm \pm\ d_{kn}\ A_{kn}^\pm\ m^0_n 
 \frac{(m_k^\pm)^2 \log ((m_k^\pm)^2/(m_n^0)^2)}{(m_k^\pm)^2 - (m_n^0)^2}. \label{45}\ee
$k,n$ are messenger indices running from $1$ to $N$, the number of messengers, while $d_{kn}$ is nonzero and equal to $d_{k}$ or $d_{n}$ only if $\phi_{n}$ and $\tilde{\phi}_{k}$ are in the same representation. To find the mass eigenstates, one must include the Dirac masses resulting from the Kaluza-Klein tower.

\subsection{Expansions of b}
We may expand \refe{valofb}  under different limits to get practical expressions which one may then substitute into \refe{useful}.
In the large $\ell$ limit the integral over $p$ in~\refe{mainresult} receives a sizeable contribution only from the region of small momenta ($p<1/\ell$), while the remaining part of the region of integration is exponentially suppressed. So we can expand the function $b$ in~\refe{valofb}, which is the building block for the full integrand~\eqref{useful}, for small momenta. In the regime $1/\ell^2\ll F,M^2$, this expansion gives
\be 
b(p^2,m^2_1,m^2_2)\approx -1 +\frac{m^2_1\log m^2_1-m^2_2 \log m^2_2}{m^2_1-m^2_2}+
\frac{p^2}{2(m^2_1-m^2_2)^{3}}[m^4_{1}-m^4_{2}-2m^2_{1}m^2_{2}\log \frac{m^2_{1}}{m^2_{2}}]\label{nopoles}~.
\ee
Even if it is not immediately obvious, there are no poles in this equation \refe{nopoles} when $F\rightarrow 0$ (i.e. $m_{1}\rightarrow m_{2}$).  Using this limit we can obtain the expression for when both masses are equal:
\be\label{lellem}
 b(p^2,m^2,m^2)\approx \log m^2 +\frac{p^2}{6m^2}+O(\frac{p^4}{m^4}),
\ee
which shows that there are no discontinuities when $F$ changes from $F\gg 1/\ell^2$ to $F\leq 1/\ell^2$. \label{sec:cterms}
\subsection{Minimal GMSB}
When the superpotential is of the form 
\be
W= X\Phi_{i}\bar{\Phi}_{i} 
\ee
where $X=M+\theta^{2}F$, then the model only has three masses: ($M^{2}, M^{2}_{+}=M^{2}(1+x),M^{2}_{-}=M^{2}(1-x)$).  The limit of small $\ell$, with flat space, the 5d model will return to the 4d result:
 \be 
 \Xi \approx -d\frac{4M^{4}}{p^{4}}[x^{2}\log(\frac{p^2}{M^2})+(x^2+3x+2)\log(1+x)-x^2+(x\rightarrow-x)]+O(p^{-6}).
 \ee
The intermediate limit $F\leq 1/\ell^2 \ll M^2$ may be found by expanding the large $\ell$ limit and taking the first order in $x=F/M^2$we find
 \be \Xi \approx- d  
\frac{2 F^2}{3 M^4}
 \ee 
In the large $\ell$ limit where $1/\ell^2 \ll F,M^2$ we find

\be 
\Xi \approx-d[\frac{4+x-2x^2}{x^2}\log(1+x)+1 +(x\rightarrow -x)]+O(p^{4}),
 \ee
In agreement with \cite{Mirabelli:1997aj}.  In this limit we may also make the identification
\be
\frac{3}{2} \Xi=-d x^2 h(x)
\ee
Where $h(x)$ is defined in section \ref{sfermionmasses}

 \subsection{Generalised Messenger Sector}
 When the superpotential of the hidden sector is 
 \be W = {\cal M}(X)_{ij}\ \phi_i \ti \phi_j =  (m + X \lambda)_{ij}\
\phi_i \ti \phi_j \ee
We have three cases. The limit of small $\ell$, with flat space, the 5d model will return to the 4d result\cite{Marques:2009yu}.
In the large $\ell$ limit such that $1/\ell^2 \ll F,M^2$, we can use \refe{nopoles} and \refe{lellem} and $\sum_{n}A_{nn}=\sum_{n}A_{kn}=1$ and $B^{+}_{kn}=B^{-}_{nk}$ to obtain \refe{LambdaS2}.
\chapter{Regularising and renormalising the Casimir energy in Deconstruction}
\section{The Casimir energy from Deconstruction}
The integral we need to extract the finite part from is 
\be 
\sum_{k=0}^{N-1} f(\frac{k}{N})= \! \!\int\! \frac{d^4p}{ (2\pi)^4} [\sum_{k=0}^{N-1} \frac{2(ap)^2 \cos^2 \frac{k\pi}{2N} }{2^{\delta_{k0}}(ap)^2+4\sin^2 \frac{k \pi}{2N} }].
\ee
We follow the steps outlined in \cite{Bauer:2003mh}.  This integral is UV divergent.  We would like to extract from it the lattice dependent finite part that determines Casimir energy.  We will subtract from it the continuum limit of this function\footnote{There is a subtlety here associated with the difference in normalisation of the zero mode with respect to the rest of the Kaluza Klein tower. As a result the $s=0$ overcounts the zero mode piece. However as the zero modes are massless they will actually contribute nothing to the regularised answer and mat be ignored.}:
\be
N \int_{0}^{\infty} ds f(s) = N \int_{0}^{\infty} ds  \int\! \frac{d^4p}{ (2\pi)^4} [ \frac{2(ap)^2 \cos^2 \frac{s\pi}{2} }{2^{\delta_{s0}}(ap)^2+4\sin^2 \frac{s \pi}{2} }]
\ee
Using 
\be
 \! \!\int\! \frac{d^4y}{ (2\pi)^4}\frac{y^2}{(y^2+\Delta)^\alpha}= \frac{1}{(4\pi)^{d/2}}\frac{d}{2}\frac{\Gamma(\alpha-d/2-1)}{\Gamma(\alpha)}(\Delta)^{1+d/2-\alpha},
\ee
we set $\alpha=1$ and use $d=4-2\epsilon$ to obtain
 \be
\frac{2N}{a^4} \int_{0}^{\infty}  \! \!ds \cos^2 (\frac{s \pi }{2}) \frac{\Delta^2}{(4\pi)^{2}}(2-\epsilon)\Gamma(\epsilon-2)e^{-\epsilon \log(4\pi)-\epsilon \log(\Delta)}.
\ee
We may use
\be
\Gamma(\epsilon-2) =\frac{1}{2\epsilon}+(\frac{3}{4}-\frac{\gamma}{2})+O(\epsilon)
\ee
to give
\be
N \int_{0}^{\infty} ds f(s) = \frac{2N}{a^4} \int_{0}^{\infty} \! \! ds \cos^2 (\frac{s \pi }{2}) \frac{\Delta^2}{(4\pi)^{2}}[\frac{1}{\epsilon}+ (1-\gamma)-\log (4\pi)-\log (\Delta)+ O(\epsilon)]
\ee
Defining 
\be
\sum_{k=0}^{N-1} f(\frac{k}{N})-N  \int_{0}^{\infty} ds f(s)=  \frac{2}{(4\pi)^2\ell^4} \mathcal{S}(N)
\ee
\begin{align}
\mathcal{S}(N)=-[N^4 & \sum^{N-1}_{k=1}\cos^2 (\frac{k\pi}{N})(\Delta(k/N))^2\log (\Delta(k/N)) \\&-N^5\!\!\int_{0}^{\infty}\! \!\!ds \cos^2(\frac{s\pi}{2})(\Delta(s))^2\log (\Delta(s))]
\end{align}
where
\be
\Delta(s)=(a m(s))^2= 4\sin^2 \frac{s \pi}{2}
\ee
In the limit that $ N\rightarrow \infty$ , the mass eigenstates will return to that of a contiuum $S^1/\mathbb{Z}_{2}$ namely $m_{k}=\frac{n\pi}{\ell}$.  We use the Abel-Plana formula \cite{Saharian:2000xx,Bauer:2003mh}
\be
\sum_{k=0}^{N-1} f(\frac{k}{N})-N  \int_{0}^{\infty} ds f(s)=  \frac{1}{2}f(0)+ i\int_{0}^{\infty}dn \frac{f(+in)-f(-in)}{\exp(2\pi n)-1}
\ee
to extract the continuum limit of the Casimir energy:
\be
\lim_{N\rightarrow \infty} S(N)\rightarrow \int \frac{d^4y}{(4\pi)^2} \frac{y}{e^y-1}=3\zeta(5).
\ee

\chapter{Evaluation of the hyperscalar soft mass integrals}\label{app:hyperscalarsoftmasses}
In this appendix we will evaluate some integrals relating to the leading order bulk hyperscalar soft mass with a generalised messenger sector of section  \ref{section:general}.  This is a different calculation to the computation of the brane localised scalar soft mass found in \cite{Auzzi:2010mb}, however the techniques are the same and we will use and review those techniques here. The original and more complete references are \cite{Martin:1996zb,Ghinculov:1994sd,vanderBij:1983bw,Marques:2009yu}.
\section{Analytic evaluation of the scalars masses at two loops}
First we define the notation
\be
\langle m_{11}, \ldots, m_{1 n_1} | m_{21} , \ldots , m_{2 n_2} | m_{31},
 \ldots, m_{3 n_3} \rangle
\ee
\be
= \int \frac{d^d k}{\pi^{d/2}}  \frac{d^d q}{\pi^{d/2}} \prod_{i=1}^{n_1}
\prod_{j=1}^{n_2} \prod_{l=1}^{n_3} \frac{1}{k^2+m_{1i}^2}
 \frac{1}{q^2+m_{2j}^2}  \frac{1}{(k-q)^2+m_{3l}^2} \, . \nonumber
\ee
The hyperscalar soft mass for the minimal model is given by \refe{Thekeyequation} and is a sum of three terms.  The first term is the four dimensional soft mass result given by taking $y\rightarrow \infty$ in $S(x,y)$, which gives
\be 
S(x,\infty)=\frac{s_{0}}{2x^2}
\ee
with $s_{0}$ defined below. The second term is

\be
(+ g^4 /(4 \pi)^d) \left( - \langle m_+|m_+|m_v \rangle  -  \langle m_-|m_-|m_v  \rangle \right.
\ee
\be \left.
-4 \langle m_f|m_f|m_v  \rangle  -2  \langle m_+|m_-|m_v  \rangle + 4  \langle m_+|m_f|m_v  \rangle  \right.
\nonumber\ee
\be
\left.  +4 \langle m_-|m_f|m_v  \rangle -4 m_+^2 \langle m_+|m_+|0,m_v  \rangle \right.
\nonumber\ee
\be \left. - 4 m_-^2  \langle m_-|m_-|0,m_v \rangle +  8 m_f^2 \langle m_f|m_f|0,m_v  \rangle \right.
\nonumber\ee
\be \left. +4(m_+^2 -m_f^2)   \langle m_+|m_f|0,m_v \rangle  +
 4(m_-^2 -m_f^2)   \langle m_-|m_f|0,m_v  \rangle
\right) \, .
\nonumber\ee
This result is obtained by removing one massless and one $m_{v}$ entry in each term in \cite{Auzzi:2010mb}. Using the regulator $m_{\epsilon}$ for the massless propagator  and starting from this result, one applies partial fractions 
\be 
\frac{1}{[(p+k)^2-m^2_{1}][(p+k)^2-m^2_{2}]}=\frac{1}{m^2_{1}-m^2_{2}}\left[\frac{1}{(p+k)^2-m^2_{1}} -\frac{1}{(p+k)^2-m^2_{2}}\right]
\ee
to the last 5 terms such that all integrals are of the same form. 
The third term is given by evaluating 
\be
(-m^2_{v} g^4 /(4 \pi)^d) \left( - \langle m_+|m_+|m_v,m_v \rangle  -  \langle m_-|m_-|m_v,m_v  \rangle \right.
\ee
\be \left.
-4 \langle m_f|m_f|m_v,m_v  \rangle  -2  \langle m_+|m_-|m_v,m_v  \rangle + 4  \langle m_+|m_f|m_v,m_v  \rangle  \right.
\nonumber\ee
\be
\left.  +4 \langle m_-|m_f|m_v,m_v  \rangle -4 m_+^2 \langle m_+|m_+|0,m_v,m_v  \rangle \right.
\nonumber\ee
\be \left. - 4 m_-^2  \langle m_-|m_-|0,m_v,m_v  \rangle +  8 m_f^2 \langle m_f|m_f|0,m_v,m_v  \rangle \right.
\nonumber\ee
\be \left. +4(m_+^2 -m_f^2)   \langle m_+|m_f|0,m_v,m_v  \rangle  +
 4(m_-^2 -m_f^2)   \langle m_-|m_f|0,m_v,m_v  \rangle
\right) \, .
\nonumber\ee
This result is obtained by removing a massless entry in each term in \cite{Auzzi:2010mb}. The symbolic manipulations are straightforward applications Mathematica by repeated use of ``Rules''.  We apply 
\be
 \langle m_a | m_b | 0, m_v, m_v  \rangle = \frac{  \langle m_a | m_b | 0 \rangle -   \langle m_a | m_b | m_v \rangle }{m_v^4}
-\frac{ \langle m_a | m_b | m_v , m_v  \rangle}{m_v^2}  \, ,
\ee
and then apply  ($d=4-2 \epsilon$),
\be
 \langle  m_0|m_1|m_2  \rangle=
\frac{1}{-1+2 \epsilon} \left(
m_0^2    \langle  m_0,m_0|m_1|m_2  \rangle  \right.
\ee
\be \left. + m_1^2   \langle m_1,m_1|m_0|m_2  \rangle+
m_2^2  \langle  m_2,m_2|m_0|m_1  \rangle
\right) \, .\nonumber
\ee
This reduces to just the first two terms on the right hand side when $m_2=m_{\epsilon}=0$.  One may also use 
\be 
\langle m_0 | m_1 | m_2, m_2  \rangle= \langle m_2 , m_2 | m_0| m_1 \rangle.
\ee
All the terms are now expressed in terms of the basic object
\be
\langle m_0,m_0|m_1|m_2 \rangle=
\frac{1}{2 \epsilon^2}+\frac{1/2-\gamma - \log m_0^2}{\epsilon}
\ee
\be 
+\gamma^2-\gamma +\frac{\pi^2}{12}
+(2 \gamma -1) \log m_0^2 + \log^2 m_0^2 -\frac{1}{2}+ h(a,b) \, .
\ee
The function $h$ is given by the integral  \cite{vanderBij:1983bw}:
\be
h(a,b)= \int_0^1 dx  \left( 1+ {\rm Li}_2 (1-\mu^2) -\frac{\mu^2}{1-\mu^2}  \log \mu^2 \right) \, 
\ee
The dilogarithm is defined as ${\rm Li}_2(x)=-\int_0^1 \frac{dt}{t}\log(1-xt)$  with
\be
\mu^2=\frac{a x + b(1-x)}{x(1-x)} \, \ \ , \ \   a=m_1^2/m_0^2   \ \ , \ \ b=m_2^2/m_0^2.
\ee
It is useful to first evaluate the terms with massless propagators, whereby the function $h$  simplifies to  $h(0,b)=1+\rm{Li}_2 (1-b)$ and has a symmetry $h(b,0)$=$h(0,b)$.   Then for the terms with entirely massive propagators, the analytic expression for $h$ is used to obtain the plots:
\be
h(a,b)=1-\frac{\log a \log b}{2} -\frac{a+b-1}{\sqrt{\Delta}} \left( {\rm Li}_2 \left (-\frac{u_2}{v_1} \right) + {\rm Li}_2 \left (-\frac{v_2}{u_1} \right)
\right.
\ee
\be
\left. + \frac{1}{4} \log^2 \frac{u_2}{v_1}  +  \frac{1}{4} \log^2 \frac{v_2}{u_1} +    \frac{1}{4} \log^2 \frac{u_1}{v_1} -    \frac{1}{4} \log^2 \frac{u_2}{v_2} + \frac{\pi^2}{6} \right) \, ,
\ee
where
\be
\Delta= 1-2 (a+b) +(a-b)^2 \, ,  \qquad u_{1,2}= \frac{1+b-a \pm \sqrt{\Delta}}{2} \, ,
\ee
\be v_{1,2}=\frac{1-b+ a \pm \sqrt{\Delta} }{2} \, .
\ee
The final result is that 
\be
m^2_{\tilde{h}}= 4(\frac{\alpha}{4\pi})^2(\frac{F}{M})^2 G(x,y)
\ee
where 
\be
G(x,y)= \frac{1}{2x^2} (s_{0}+ \frac{t_{1}+t_{2}}{y^2}+t_{3}+y^2 t_{4} )+ (x \rightarrow -x)
\ee
\be
s_0=2(1+x) \left( \log (1+x) -2 {\rm Li}_2  \left( \frac{ x}{1+x}\right)
+\frac{1}{2} {\rm Li}_2 \left( \frac{2 x}{1+x}\right) \right)  ,
\ee
\be
t_{1}=-4x^2-2x(1+x) \log (1+x)-x^2 {\rm Li}_2 \left(x^2\right) \nonumber
\ee
\bea
t_{2}&=-8h\left(1,y^2\right) +8(1+x)^2 h\left(1,\frac{y^2}{1+x}\right)
\nonumber
\\ &-4x h\left(1+x,y^2\right) -4x(1+x)h\left(\frac{1}{1+x},\frac{y^2}{1+x}\right) \nonumber
\eea
\bea
t_{3}&=2h\left(1,y^2 \right)+(1+x)h\left(1,\frac{y^2}{1+x}\right)-2h\left(1+x,y^2 \right)+(1+x)h\left(\frac{1-x}{1+x},\frac{y^2}{1+x}\right) \nonumber \\&  -2 h\left(\frac{1}{y^2},\frac{1}{y^2}\right) -2(1+x) h\left(\frac{1}{1+x},\frac{y^2}{1+x}\right)+ 2(1+x) h\left(\frac{1+x}{y^2},\frac{1+x}{y^2} \right)-2x h\left(\frac{1+x}{y^2},\frac{1}{y^2}\right)\nonumber
\eea
\be
t_{4}= 2h\left(\frac{1}{y^2},\frac{1}{y^2}\right)-4h\left(\frac{1+x}{y^2},\frac{1}{y^2}\right)+h\left(\frac{1+x}{y^2},\frac{1+x}{y^2}\right)+h\left(\frac{1+x}{y^2},\frac{1-x}{y^2} \right) .\nonumber
\ee
As a consistency check we also computed the result of  \cite{Auzzi:2010mb} after applying further rules presented in that paper, which gives an analytic expression for $s(x,y)$:
\be
s(x,y)=\frac{1}{2x^2}\left(s_0 +\frac{s_1+ s_2}{y^2} + s_3 +s_4 +s_5 \right)
 + \, (x \rightarrow -x)\, , 
\ee
where
\be
s_0=2(1+x) \left( \log (1+x) -2 {\rm Li}_2  \left( \frac{ x}{1+x}\right)
+\frac{1}{2} {\rm Li}_2 \left( \frac{2 x}{1+x}\right) \right)   \, , 
\ee

\be
s_1=- 4 x^2   - 2 x(1+x) \log^2(1+x) - x^2 \, {\rm Li}_2(x^2) \, ,
\nonumber\ee
\be
s_2=8 \left(1+x\right)^2 h\left(\frac{y^2}{1+x},1\right)-4 x \left(1+x\right) h\left(\frac{y^2}{1+x},\frac{1}{1+x}\right)
\nonumber\ee
  \be
   -4 x    h\left(y^2,1+x \right)-8 h\left(y^2,1\right) \, ,
\nonumber\ee
\be
s_3= -2 h\left(\frac{1}{y^2},\frac{1}{y^2}\right)
-2 x \,   h\left(\frac{1+x}{y^2},\frac{1}{y^2}\right) +
2(1+ x) h\left(\frac{1+x}{y^2},\frac{1+x}{y^2}\right) \, ,
\nonumber \ee
\be
s_4=(1+x) \left(  2  h\left(\frac{y^2}{1+x},\frac{1}{1+x}\right)
 - h\left(\frac{y^2}{1+x},1\right)- h\left(\frac{y^2}{1+x},\frac{1-x}{1+x}\right)  \right) \, ,
\nonumber \ee
\be
s_5= 2 h\left(y^2,1+x\right)-2 h\left(y^2,1\right) \, . \nonumber
\ee
 \chapter{Limits on the bulk propagator} \label{limits}
In chapter \ref{chapter6} we used certain approximations of the bulk wavefunctions and masses of the Kaluza-Klein modes to sum up the effects of the KK tower.  In this appendix we flesh out the details of those approximations.  
\section{Limits on the bulk propagator in AdS}
In principle, all Kaluza-Klein modes may propagate supersymmetry breaking across the interval.  To discern which modes contribute most, it is important to determine the limiting behaviour of the bulk eigenfunctions, as they change with mass eigenstates $m_{n}$   \cite{Ichinose:2006en}.   This may also be carried out in Poincar\'e co-ordinates, with similar results.
First we focus on the positive parity eigenfunctions 
\be 
f^{(2)}_{n}(y)=\frac{e^{\sigma}}{N_{n}}\left[J_{1}\left(\frac{m_{n}e^{\sigma}}{k}\right)+b(m_{n})Y_{1}\left(\frac{m_{n}e^{\sigma}}{k}\right)\right].
\ee
Let us first evaluate the function at $y=\ell$ in the mass regime $m_{n}\gg k$ and take  $x=m_{n}e^{\sigma}/k$.  In this regime $m_{n}\simeq n\pi k^*$ with very large $n$. Using the identities valid for large $x$,
\be
J_{1}(x)\simeq\left(\frac{2}{\pi x}\right)^{1/2}\cos (x-3\pi/4)=  \left(\frac{2}{\pi x}\right)^{1/2}\sin (x-\pi/4)
\ee
\be
Y_{1}(x))\simeq\left(\frac{2}{\pi x}\right)^{1/2}\sin (x-3\pi/4)= - \left(\frac{2}{\pi x}\right)^{1/2}\cos (x-\pi/4)
\ee
and taking $b(m_{n})\sim 1$ one obtains 
\be 
f^{(2)}_{n}(\ell)\simeq-\frac{e^{k\ell}}{N_{n}}\left(\frac{2 k}{\pi m_{n} e^{k\ell}}\right)^{1/2} \sqrt{2}(-1)^n.
\ee
Next we may look at $y=0$ in the same mass regime $m_{n}>>k$, where now $x=m_{n}/k$.  Similar manipulations result in 
\be 
f^{(2)}_{n}(0)\simeq -\frac{1}{N_{n}} \left(\frac{2 k}{\pi m_{n}}\right)^{1/2}\sqrt{2}\cos(m_{n}/k).
\ee
Taking $\cos(m_{n}/k)$ to be order $1$ we may define the eigenfunction from the IR brane to UV brane as 
\be
 f^{(2)}_{n}(0)f^{(2)}_{n}(\ell)\simeq\frac{2}{N_{n}} \left(\frac{2 k}{\pi m_{n}}\right)^{1/2} \frac{e^{ k\ell}}{N_{n}}\left(\frac{2 k}{\pi m_{n} e^{k\ell}}\right)^{1/2}(-1)^n
\ee
with 
\be
N_{n}\simeq \frac{1}{\sqrt{m_{n}e^{-k\ell}\pi \ell}}.
\ee
Some simplifications finally result in
\be
 f^{(2)}_{n}(0)f^{(2)}_{n}(\ell)\simeq 4(k\ell)(-1)^n e^{-k\ell/2}.
\ee
An eigenfunction for the derivative of the negative parity fields can be computed similarly,
\be
\partial_5g^{(4)}_{n}(0) \partial_5g^{(4)}_{n}(\ell) \simeq f^{(2)}_{n}(0)f^{(2)}_{n}(\ell) m^2_{n}\simeq 4 m^2_{n}(k\ell)(-1)^n e^{3 k\ell/2}.
\ee
Now we turn to the regime $m_{n}\ll k$.  These states are highly localise at the IR brane and cannot propagate significantly across the bulk. To see this, we evaluate $f^{(2)}_n(y)$ at $y=0$
\be
 f^{(2)}_{n}(0)\simeq \frac{x}{2N_{m}}\simeq 0,
\ee
where $x=m_{n}/k\ll 1$, for which we used
\be
 J_{1}(x)\simeq x/2 \  , \ \ \ \    Y_{1}(x)\simeq -\frac{2}{\pi x}.
\ee
This is exponentially suppressed and we see that there is very little probability to be located near the UV brane for $m_{n}\ll k$.  Conversely, at y=$\ell$ we use the mass spectrum $m_{n}\simeq (n-1/4)\pi k^*$ and find 
\be
 f^{(2)}_{n}(\ell)\simeq -\frac{e^{ k\ell/2}}{N_{n}}\left(\frac{2 k}{m_{n}\pi}\right)^{1/2}.
\ee
This demonstrates that the light modes are significantly localised to the IR brane not to propagate supersymmetry breaking effects across the interval.  
\section{Matsubara frequency summation}
An accurate determination of the full summation of propagators of the Kaluza-Klein tower should be done numerically.  We make a simplifying assumption that we may carry out a Matsubara frequency summation of the KK modes \cite{McGarrie:2010yk}, by approximating the whole KK tower with the states of $m_{n}\gg k$, which have masses $m_{n}\simeq n\pi k^*$.  The resulting summations  give
\be 
\frac{k^*}{2} \sum_{m_{n}}\frac{(-1)^n}{p^2+m^2_{n}}\sim \frac{1}{2}\frac{1}{p \sinh (p/k^*)}
\ee
and
\be 
\frac{k^*}{2} \sum_{m_{n}}\frac{1}{p^2+m^2_{n}}\sim \frac{1}{2} \frac{1}{p \tanh (p/k^*)}
\ee
such that 
\be
\sum_{n}\frac{k^*}{2\ell k^*} \frac{f^{(2)}(0)f^{(2)}(\ell)}{p^2+m^2_{n}}\sim  \frac{2  e^{ k \ell/2 }}{p\sinh (p/k^*)}.
\ee
Similarly, using a contour pulling argument detailed in \cite{Mirabelli:1997aj}, we may convert
\be 
\frac{k^*}{2} \sum_{m_{n}}\frac{1}{p^2+m^2_{n}}\sim \frac{1}{p}\frac{1}{e^{2p/k^*}-1}+ \int^{\infty}_{-\infty} \frac{dp_{5}}{(2\pi)}.
\ee
The second term is independent of $k^*$ and is divergent.  The first term is the finite part that is relevant for computing the Casimir energy. To summarise these results, we emphasise that as most kk states are localised towards the IR brane for $m_n\ll k$, only the zero modes significantly contribute to both the running gauge coupling and to gauge mediated supersymmetry breaking.  Only if $k\ll M$, where $M$ is the characteristic mass scale of the hidden sector, will the most heavy kk modes, where $m_n\gg k $ start to contribute to cause the gaugino mediated limit.


\addcontentsline{toc}{chapter}{\numberline{}Bibliography}

\bibliographystyle{utphys}
\bibliography{thesispostviva.bbl}


\end{document}